\documentclass[lineno,authoryear]{FLO_v1}%

\usepackage{graphicx}
\usepackage{upgreek}
\usepackage{multicol,multirow}
\usepackage{amsmath,amssymb,amsfonts}
\usepackage{mathrsfs}
\usepackage{amsthm}
\usepackage[figuresright]{rotating}
\usepackage{appendix}
\usepackage[authoryear]{natbib}
\usepackage{ifpdf}
\usepackage[T1]{fontenc}
\usepackage{newtxtext}
\usepackage{newtxmath}
\usepackage{textcomp}
\usepackage{xcolor}

\usepackage{siunitx}
\usepackage{float}
\usepackage{bm}

\usepackage{subfigmat}
\usepackage{placeins}  
\usepackage{afterpage}  

\usepackage[colorlinks,allcolors=blue]{hyperref}
\definecolor{jourcolor}{cmyk}{1,0.57,0.01,0.38}
\hypersetup{
    colorlinks,%
    citecolor=jourcolor,%
    filecolor=jourcolor,%
    linkcolor=jourcolor,%
    urlcolor=jourcolor
}

\theoremstyle{definition}

\articletype{RESEARCH ARTICLE}

\DOI{10.1017/flo.2025.1}

\Year{2025}

\Vol{1}

\Price{}

\art-id{FLO2000049}

\citearticle{Lemarquand, H., \& Lugrin, M., \& Content, C., \& Caillaud, C., Esquieu, S., Sipp, D.}

\begin{document}



\title[Transition mechanisms in hypersonic wind-tunnel nozzles: a methodological approach using global linear stability analysis]{Transition mechanisms in hypersonic wind-tunnel nozzles: a methodological approach using global linear stability analysis}

\author[H. Lemarquand, M. Lugrin, C. Content, C. Caillaud, S. Esquieu and D. Sipp]{Hugo Lemarquand$^{1\ast}${{\href{https://orcid.org/0009-0005-7917-3231}{\includegraphics{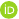}}}}, Mathieu Lugrin$^{1}${{\href{https://orcid.org/0000-0003-1901-5483}{\includegraphics{orcid_logo}}}}, Cédric Content$^{2}${{\href{https://orcid.org/0000-0002-5762-0900}{\includegraphics{orcid_logo}}}}, Clément Caillaud$^{3}${{\href{https://orcid.org/0009-0002-2120-972X}{\includegraphics{orcid_logo}}}}, Sébastien Esquieu$^{3}${{\href{https://orcid.org/0009-0007-9429-3789}{\includegraphics{orcid_logo}}}} and Denis Sipp$^{4}${{\href{https://orcid.org/0000-0002-2808-3886}{\includegraphics{orcid_logo}}}}}




\address[1]{DAAA ONERA, Institut Polytechnique de Paris, 92190, Meudon, France}
\address[2]{DAAA ONERA, Institut Polytechnique de Paris, 92320, Châtillon, France}
\address[3]{CEA-CESTA, 15 Avenue des Sablières, 33114, Le Barp, France}
\address[4]{ONERA, Institut Polytechnique de Paris, 91120, Palaiseau, France}

\corres{*}{Corresponding author. E-mail:
\emaillink{hugo.lemarquand@onera.fr}}


\date{\textbf{Received:} XX 2025; \textbf{Revised:} XX XX 2025; \textbf{Accepted:} XX XX 2025}


\abstract{Base-flow computations and stability analyses are performed for a hypersonic wind tunnel nozzle at a Mach number of 6. Isothermal and adiabatic wall boundary conditions are investigated, and moderate stagnation conditions are used to provide representative scenarios to study the transition in quiet hypersonic wind tunnel facilities. Under these conditions, the studied nozzle shows a small flow separation at the convergent inlet. 
Global stability analysis reveals that this recirculation bubble may trigger a classical three-dimensional stationary unstable global mode. Resolvent analysis reveals Görtler, first and second Mack modes affecting the divergent part of the nozzle, along with a Kelvin-Helmholtz instability induced by the bubble.
The present study also highlights the key impact of perturbations located in the convergent inlet on the development of instabilities further downstream in the divergent outlet, helping understand the need and efficacy of a suction lip upstream of the nozzle throat to mitigate instabilities in the divergent nozzle. Detailed knowledge of all these mechanisms is essential for understanding flows in quiet hypersonic wind tunnel nozzles and, consequently, represents a key step toward the optimisation of such nozzles.}

\maketitle


\vspace{-0.5em}
\begin{boxtext}

\textbf{\mathversion{bold}Impact Statement}

Nozzle turbulent boundary layers are a significant source of acoustic noise in conventional hypersonic wind tunnels. They
must be avoided when designing "quiet" (low-noise) hypersonic nozzles. 
The few existing quiet hypersonic wind tunnels are currently limited in Mach number and transition Reynolds number due to this transition, and thus do not fully cover the atmospheric flight conditions required for hypersonic vehicle studies.
Consequently, understanding and defining accurate metrics representing the instability potential of a given nozzle design is a crucial point for the optimisation of future higher Reynolds number quiet nozzles. 
In this article, we provide a spatial high-order tool that computes the base-flow of a given nozzle and characterises all possible instabilities, such as globally unstable modes or resolvent modes affecting separated regions and attached boundary layers. 
For a specific nozzle design, we analyse the various instability mechanisms at play to provide guidelines for the interpretation of such results.
\end{boxtext}


\vspace{-1.0em}
\section{Introduction}

    During the design phase of hypersonic vehicles, wind tunnel testing plays a critical role in validating key factors, such as aerodynamic performance. These tests often focus on investigating the laminar-to-turbulent transition of the boundary layer developing on the test model. Accurate prediction of this transition is crucial as the turbulent boundary layer leads to a significant increase in wall heat flux. For example,
    \citet{van1956problem}
    demonstrated that for a flat plate at Mach 6, the heat flux increases by approximately a factor of five when the boundary layer becomes turbulent. Nowadays, the vast majority of current hypersonic wind tunnels are not sufficiently representative of the atmospheric flight conditions encountered in real-life scenarios, 
    particularly due to limitations in enthalpy levels 
    \citep{gu2020capabilities}.
    Some facilities allow for the accurate reproduction of similarity parameters such as Mach number, Reynolds number, and enthalpy conditions, but fail to replicate the free-stream turbulence level. Such fluctuations are known to drastically alter the laminar-to-turbulent transition of the boundary layer on the test model due to the receptivity process \citep{morkovin1969many}. The receptivity mechanism describes how external disturbances penetrate the boundary layer and establish the initial amplitudes and scales of the instabilities. Consequently, the level of fluctuation in the test section (also referred to as the "noise level") becomes another new significant similarity parameter. There can be various sources of these disturbances in wind tunnel facilities \citep{schneider2008development, morkovin1957transition}, but above Mach $2.5$ the primary source of noise is the eddy-Mach-wave radiation from the turbulent boundary layer that develops on the nozzle wall \citep{laufer1961aerodynamic}. This noise is around 1 to 2 orders of magnitude above flight levels \citep{schneider2008development}. 
    \citet{chazot2019need} illustrate this significant discrepancy in their Fig.\ 2, by comparing the noise levels in conventional wind tunnels (VKI H3) to those in quiet wind tunnels (BAM6QT at Purdue University) which achieve noise levels comparable to atmospheric flight conditions (note: flight-level atmospheric fluctuations are $\leq 0.05 \%$ \citep{schneider2001effects}).
    Practically, the effect of noise levels on the boundary layer transition in conventional hypersonic wind tunnels is illustrated by experimental studies of \citet{durant2015mach} (Fig.\ 12) and \citet{mckiernan2021boundary} (Figs.\ 12 and 13). The comparison of TSP (Temperature Sensitive Paint) measurements reveals an early transition on the model in the conventional wind tunnel setup compared to the low-noise configuration. Thus, one of the main issues in quiet hypersonic wind tunnels is controlling the laminar-to-turbulent transition of the boundary layer on the nozzle wall. Such a control helps to ensure a quiet freestream environment in the test section in order to closely reproduce atmospheric flight conditions. 
    Currently, such quiet wind tunnel, in which the laminarity of the boundary layers along the wall is maintained to significantly reduce the noise levels, do exist for Reynolds numbers up to \SI{20e6}{\per\meter} at Mach 6 (for example, BAM6QT at Purdue) \citep{schneider2008development}. However, such facilities are rare, nonexistent in Europe \citep{chazot2019need}, and remain limited in both Mach and Reynolds numbers \citep{schneider2008development, beckwith1986design}. 

    \vspace{-0.5em}
    \begin{figure}[H]
        \centering
        \includegraphics[width=0.9\textwidth]{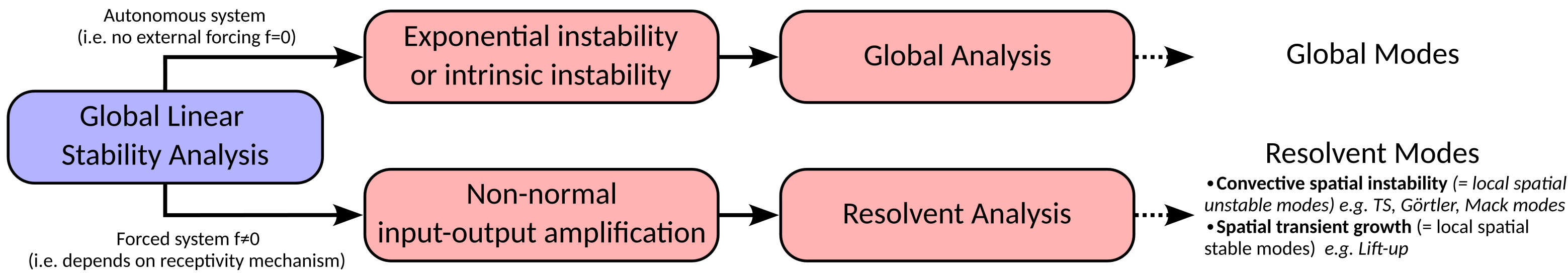}
        \caption{Stability terminology used throughout this study.}
        \label{fig:Terminology}
    \end{figure}\vspace{-1.0em}

\begin{figure}
        \centering
        \includegraphics[width=1.0\textwidth]{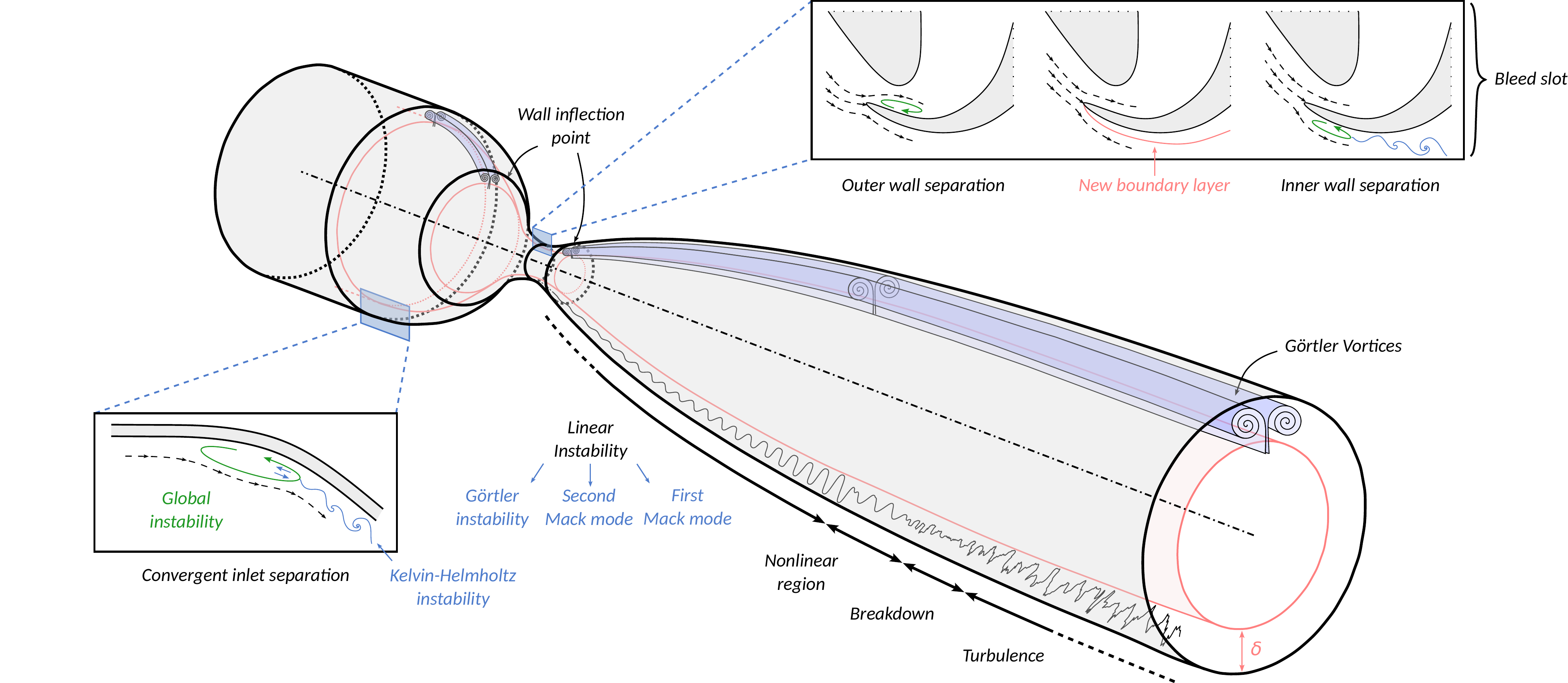}
        \caption{Schematic of all instabilities within a hypersonic nozzle.}
        \label{fig:SchemaInstabNozzle}
    \end{figure}

    As hypersonic wind tunnel nozzles are studied (including the settling chamber, convergent, and divergent sections), the boundary layer along the wall evolves from subsonic to supersonic regimes, where various instabilities can develop.
    Following the classical global linear stability analysis terminology (see figure \ref{fig:Terminology}), 
    two families of instabilities can develop in a nozzle: exponential or intrinsic instabilities (oscillatory flow, driven by global modes) and non-normal input-output amplification (noise-amplifier flow, driven by resolvent modes) \citep{huerre1990local}. 
    Global modes arise within the flow itself and do not rely on a receptivity mechanism, i.e.\ no external forcing is introduced in the Navier-Stokes equations. 
    On the other hand, resolvent modes depend on a receptivity mechanism and therefore require external disturbances to project sufficient energy on the optimal and sub-optimal forcing structures.
    A summary of all the potential instabilities that can develop in hypersonic nozzle flows is provided below and summarised in figure \ref{fig:SchemaInstabNozzle}. 
    First, 
    Görtler instability can arise over the concave walls of the nozzle \citep{saric1994gortler}.
    Görtler instability is known to be significant and is considered in wind tunnel nozzle design \citep{schneiderENMethode, lakebrink2018optimization, chen1985instabilities}.
    This instability occurs across all Mach numbers, from incompressible to hypersonic flow conditions, and results from a centrifugal instability mechanism related to the concave curvature of the streamlines \citep{rayleigh1917dynamics, xu2024gortler}. \citet{sipp2000three} have shown that unstable regions that can develop this centrifugal instability are characterised by a negative Rayleigh discriminant ($\Delta < 0$). The Görtler instability is characterised by counter-rotating vortices oriented in the streamwise direction \citep{gortler1941instabilitat}, and these streamwise elongated structures are stationary or at low-frequency \citep{wu2011excitation, xu2024gortler}. This mechanism is sensitive to the disturbance environment 
    \citep{xu2024gortler}. This instability has been experimentally observed in several wind tunnels at NASA Langley, as reported by \citet{beckwith1981gortler} and \citet{beckwith1984nozzle}, showing oil-flow streaks visible on the wall in the convergent section and at the divergent exit (see Figs. 3, 4, and 9 in \citet{schneider2008development} for precise pictures).
    %
    A second class of important instabilities is composed of the first and second Mack modes. Stability analysis have shown that such instabilities can grow within hypersonic wind tunnel nozzles \citep{schneiderENMethode, schneider2001shakedown, lakebrink2018optimization}.
    The oblique (i.e.\ three-dimensional) first Mack mode corresponds to a viscous inflectional mode, in which fluctuations are localised around the generalised inflection point.
    %
    In addition to this viscous first mode, \citet{mack1984boundary} found that multiple inviscid instability modes can exist in the supersonic boundary layer, referred to as the second Mack mode and higher-order modes. The second Mack mode corresponds to a two-dimensional trapped acoustic wave close to the wall 
    and become unstable above $M\approx4$ \citep{mack1984boundary}.
    %
    Lastly, recirculation bubbles can form at the convergent inlet \citep{schneider1998design} or at the suction lip upstream of the nozzle throat \citep{taskinoglu2005numerical, benay2004design}. 
    These bubbles can lead to new instabilities including 
    global modes
    (self-sustained in time instabilities that grow exponentially over time) 
    \citep{marquet2009direct, hildebrand2018simulation} and convective instabilities such as the Kelvin-Helmholtz instability which arises from the separated shear layer \citep{barbagallo2012closed}.
    The figure \ref{fig:SchemaInstabNozzle} offers a summary of all these aforementioned instabilities.

    The present work intends to characterise and identify all linear instabilities that can develop in 
    hypersonic wind tunnel nozzles.
    Therefore, the objective of this study is not to study in detail the laminar-to-turbulent transition process of the boundary layer, i.e.\ we will not focus on the nonlinear interactions leading to the transition. 
    Instead, we aim to understand how and where linear instabilities develop in the boundary layer, allowing future studies to focus on reducing their growth to delay the transition. Since we are focusing on the early stages of the transition, before nonlinear interactions become dominant, we will therefore use methods based on linear stability theories. 
    Thus, the turbulent boundary layer and all its implications for the flow will not be studied here, note that DNS studies of turbulent boundary layer on hypersonic wind tunnel nozzles have, for example, been carried out by: \citet{hildebrand2022direct, duan2019characterization}.
    As for now, most instabilities in hypersonic nozzles, presented in the previous paragraph, are studied using the LST (Local Stability Theory) \citep{schmid2002stability} and the PSE (Parabolised Stability Equations) \citep{herbert1997parabolized}.
    However, such methods are only valid for weakly non-parallel flows.
    Thus, relying on PSE or LST methods to inform an optimisation process, aimed at computing optimal nozzle designs to delay transition, may yield results that do not guarantee the optimality of the computed design due to an incomplete physical representation.
    Therefore, to address more general configurations, comprehensive methods, such as the resolvent analysis, are available. These methods are increasingly used as they became computationally affordable in the recent years. 
    By construction, the Resolvent framework is particularly suited to account for the non-modal growth of instabilities resulting from the non-normality of the Navier-Stokes operator \citep{schmid2007nonmodal, sipp2013characterization}.
    As this global linear stability framework relies solely on the assumption of linearity, the non-parallel effect are fully accounted for, allowing to accurately probe the effects of flow separation or strong gradients along regions of high wall curvature.
    Moreover, this global linear stability framework offers a convenient access to the gradients of the instabilities amplification with respect to various quantities, a key ingredient for future optimisation work. 
    For instance, this method gives access to the regions of the flow where small modifications to the base-flow have the greatest impact on its stability \citep{marquet2008sensitivity, metto2013, poulain2024adjoint}.

    This paper is organised as follows. First the flow configuration and the nozzle geometry are presented (\S \hspace*{0.01em} \ref{section:Flow_Configuration}). Then, the theoretical aspects and tools
    used to study both global and resolvent modes using a Global Linear Stability framework (GSA) are introduced (\S \hspace*{0.01em} \ref{section:Methods}). 
    In \S \hspace*{0.01em} \ref{section:baseflow}, base-flows for the isothermal and adiabatic walls are described. 
    \S \hspace*{0.01em} \ref{section:Global_Stability_Analysis} focuses on the study of global modes in the nozzle for both the isothermal and adiabatic cases. 
    In \S \hspace*{0.01em} \ref{section:Convective_Instabilities}, resolvent modes are studied and identified in the nozzle flow for the isothermal case, followed by a comparison of resolvent modes between the two cases in \S \hspace*{0.01em} \ref{section:Comp_isoT_adiaT}.
    Finally, in \S \hspace*{0.01em} \ref{section:forcing_field_restriction} we will take advantage of the global linear stability tool to demonstrate, through a simple model, the effectiveness of using suction lips upstream of the throat to design quiet wind tunnels.


\vspace{-1.0em}
\section{Flow configuration}\label{section:Flow_Configuration}

    The nozzle geometry and flow conditions selected to study the linear instabilities that can develop within the boundary layer of hypersonic wind tunnel nozzles are detailed in this section.
    This nozzle was selected because the flow it produces contains all the instabilities representative of hypersonic wind tunnel flows, e.g.\ Görtler, first and second Mack modes. Moreover, it can exhibit, under certain stagnation conditions, a boundary layer separation at the convergent inlet leading to a recirculation bubble. 
    Investigating the effect of such separated regions on instabilities amplification is important for understanding the transition in the nozzle.

\begin{figure}
        \centering
        \includegraphics[width=\textwidth]{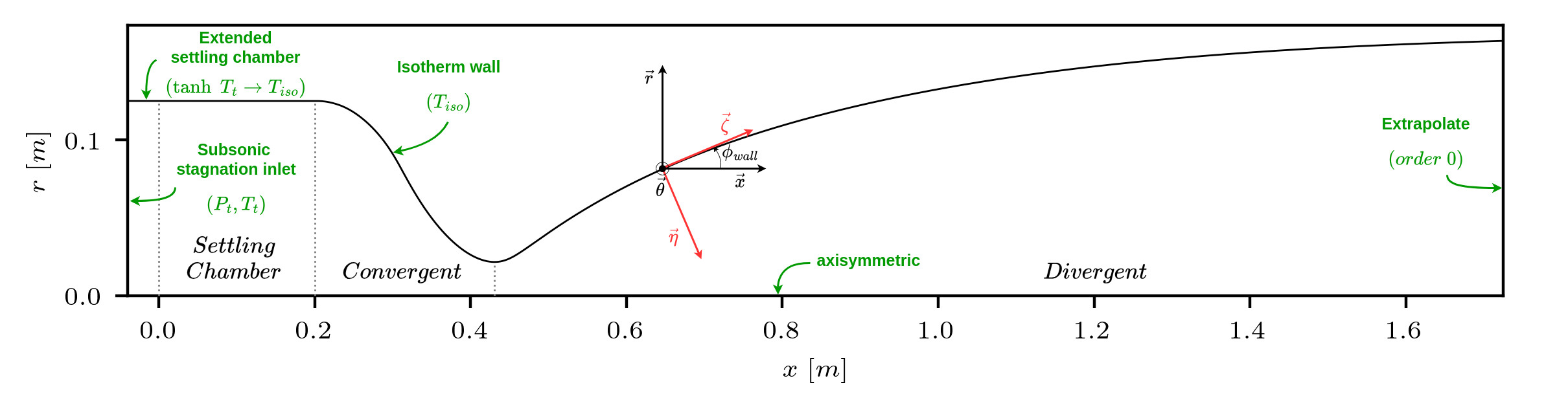}
        \caption{Numerical setup for base-flow computation in the isothermal case.}
        \label{fig:Baseflow_setup}
    \end{figure}

    The geometry of the hypersonic wind tunnel nozzle used in this study is shown in figure \ref{fig:Baseflow_setup}. It is an axisymmetric nozzle from ONERA with a design Mach number of 6. The nozzle consists of three parts: the settling chamber, the convergent, and the divergent. Note that the settling chamber length has been increased to accommodate the inlet boundary condition, the dimensions used in this study are therefore not exactly representative of the real configuration. Nevertheless, this part can be seen as a simplified model of a real settling chamber commonly used in this type of wind tunnel (no filter, no supply line, no heater, etc.). 
    Figure \ref{fig:Baseflow_setup} also illustrates the two coordinate systems used in this study. The first is the cylindrical coordinate system $(\Vec{e_x},\Vec{e_r},\Vec{e_{\theta}})$ used for base-flow and stability computations. The second is a local coordinate system $(\Vec{e_{\zeta}},\Vec{e_{\eta}},-\Vec{e_{\theta}})$ defined with respect to the wall normal and will be used to study the boundary layer and its disturbances.


    
    The inlet boundary condition is set with a stagnation pressure of $P_t = 4 \; \text{bar}$ and a stagnation temperature of $T_t = 550 \; \text{K}$. These stagnation conditions represent the lower Reynolds number and temperature range used in this wind tunnel.
    However, this choice of stagnation conditions provides representative scenarios for the study of transition while offering a reduced computational cost.
    
    Given the short run time in this type of blowdown-to-vacuum facility and that the nozzle wall is made of steel, the wall temperature is assumed to remain isothermal $T_{iso} = 300 \; \text{K}$. Just before the settling chamber, a zone is added to ensure a smooth variation between the inlet boundary condition at $T_t$ and the wall temperature at $T_{iso}$ using a $\tanh{}$ function for the wall boundary condition, thus avoiding a temperature bump that is numerically problematic (temperature discontinuity at the upper-left corner). Note that the size of the "\textit{extended settling chamber}" shown in figure \ref{fig:Baseflow_setup} has been enlarged for clarity, the length is fixed at 1\% of the settling chamber length. 
    The base-flow is also computed under adiabatic conditions. Although this adiabatic condition may not reflect the physical reality of this flow, but it will provide an interesting comparison of wall temperature effects during the stability study. 
    All these conditions are summarised in table \ref{tab:Boundary_freestream_conditions}.

    %
    {\renewcommand{\arraystretch}{1.35} 
    {\setlength{\tabcolsep}{0.35cm} 
        \begin{table}
            \vspace{-1.0em}
            \caption{\label{tab:Boundary_freestream_conditions} Boundary conditions and free-stream conditions in the test section.}
            \centering
            \begin{tabular}{ccccc}
                \hline
                $P_t$ $[\text{bar}]$ & $T_t$ $[\text{K}]$ &  $T_{iso}$ $[\text{K}]$ & $M_{\infty}$ $[-]$ & $Re_{\infty}$ $[\text{m}^{-1}]$  \\
                \hline
                $4$ & $550$ & $300$ &$6.0$ & $2.8 \cdot 10^{6}$ \\
                \hline
            \end{tabular}
    \end{table}}}


\vspace{-1.0em}
\section{Methods}\label{section:Methods}

    A description of the methods used for the axisymmetric nozzle is provided in this section. \S \hspace*{0.01em} \ref{subsection:NS} presents the governing equations in cylindrical coordinates, the base-flow equation and the linearisation of these equations. \S \hspace*{0.01em} \ref{subsection:Global_Stability_Oscillator} and \ref{subsection:Resolvent_Amplifier} outline the global linear stability equation for exponential or intrinsic instabilities (global analysis) and non-normal input-output amplification (resolvent analysis). And \S \hspace*{0.01em} \ref{subsection:Numerical_methods_strategy} introduces the numerical strategy used to compute the base-flow and solve the eigenvalue problem.


    \subsection{Governing equations and linear dynamics}\label{subsection:NS}

        As we are studying axisymmetric wind tunnel nozzles, the cylindrical coordinate system $(\Vec{e_x},\Vec{e_r},\Vec{e_{\theta}})$ is chosen throughout this study. 
        The flow is governed by the compressible Navier-Stokes equations written in conservative form:
        %
        %
        \begin{equation} 
            \frac{\partial q}{\partial t} = R(q),
            \label{eq:NSequation} 
        \end{equation}
        with $q = (\rho, \rho u_x, \rho u_r, \rho u_{\theta}, \rho E)^T$ the state vector of conservative variables (density, momentum and total energy) and $R(q)$ is the discrete residual obtained after spatial discretisation (see \S \hspace*{0.01em} \ref{subsubsection:SolverDiscretisation}). For more details on this equation refer to appendix \ref{appendix:NavierStokesEquation}. Note that all variables used are made non-dimensional using length $\nu_{\infty}/U_{\infty}$, density $\rho_{\infty}$, velocity $U_{\infty}$, and temperature $T_{\infty}$. All are calculated from the reference Mach number $M_{\infty}$, the stagnation pressure $P_t$ and temperature $T_t$ (table \ref{tab:Boundary_freestream_conditions}), and using isentropic flow relations and the Sutherland's law \citep{sutherland1893lii}. 
        
        To study the linear dynamics of this system, a small 3D disturbance $q'(x,r,\theta,t)$ is added to the 2D base-flow $\overline{q}(x,r)$ which is a steady solution of equation (\ref{eq:NSequation}) such as $R(\overline{q})=0$, and can be seen as an equilibrium point. Since the disturbance is assumed to be small compared to the base-flow, $q = \overline{q} + \epsilon q'$ with $\lVert \epsilon q' \rVert \ll \lVert \overline{q} \rVert$, the non-linear equation can then be linearised around the base-flow:
        \begin{equation} 
            \frac{\partial q'}{\partial t} = \bm{\mathcal{J}} q',
            \label{eq:equationlinear} 
        \end{equation}
        where $\bm{\mathcal{J}}$ is the Jacobian operator around $\overline{q}$ such as $\bm{\mathcal{J}} = \left.\frac{\partial R(q)}{\partial q}\right|_{\overline{q}}$.

        Two families of instabilities can develop in a flow: exponential or intrinsic instabilities (oscillatory flow, driven by global modes) and non-normal input-output amplification (noise-amplifier flow, driven by resolvent modes) \citep{huerre1990local}.
        Even though it is generally known that resolvent modes seem to dominate in wind tunnel nozzle flows, in some cases it can be interesting to also consider global modes, especially when recirculation bubbles are present \citep{marquet2009direct}, as in our case. Two different mathematical approaches are implemented to study these two families of instabilities and are presented in the following two sections. 

    \subsection{Global analysis
    }\label{subsection:Global_Stability_Oscillator}

        
        Global modes are self-sustained in time and are intrinsic, i.e.\ they arise within the flow and are therefore less sensitive to receptivity mechanisms. Thus, the flow does not need external forcing, i.e.\ the autonomous system is studied.
        As the flow is axisymmetric, the small disturbance can be expressed in Fourier space (harmonic in time and space) by: 
        %
        %
        \begin{equation} 
            q'(x,r,\theta,t) = \hat{q}(x,r) \, e^{\lambda t + i m \theta},
            \label{eq:disturbance_globalmode} 
        \end{equation}
        %
        %
        %
        Note that since an axisymmetric nozzle is used in this study, physically only discrete $m \in \mathbb{Z}$ values are allowed.
        By injecting the small disturbance (\ref{eq:disturbance_globalmode}) into the linearised system (\ref{eq:equationlinear}), the following eigenvalue problem is obtained:
        \begin{equation} 
            \bm{\mathcal{J}}(m)\hat{q} = \lambda \hat{q},
            \label{eq:globalstability} 
        \end{equation}
        where $\lambda = \sigma + i \omega$ is the complex eigenvalue, $\Re(\lambda) = \sigma$ is the growth rate of the eigenfunction, $\Im(\lambda) = \omega$ denotes the associated frequency and $m$ is the azimuthal wavenumber. Therefore, if the real part of the eigenvalue $\lambda$ is positive, then the fixed point is globally unstable. 
        Note that the Jacobian operator $\bm{\mathcal{J}}(m)$ depends on the azimuthal number $m$ to analytically account for the $\theta$-direction thereby extending the flow stability analysis to 3D, for more details refer to \S \hspace*{0.01em} \ref{subsubsection:Extension3Dstability}.
        The
        global analysis
        of the base-flow is therefore assessed by computing the global modes of the eigenvalue problem (\ref{eq:globalstability}). Thus, the asymptotic stability of the system is assessed by examining the most unstable eigenvalues of the Jacobian operator $\bm{\mathcal{J}}$ \citep{jackson1987finite}.

    \subsection{Resolvent analysis}\label{subsection:Resolvent_Amplifier}

        The second family of instabilities is termed non-normal input-output amplification (selective noise amplifiers). Even if the flow is globally stable, the flow may exhibit nonmodal amplification of convective instabilities at specific frequencies due to the non-normality of the Jacobian operator $\bm{\mathcal{J}}$ \citep{schmid2007nonmodal, sipp2013characterization}. These instabilities depend on the receptivity mechanism and therefore need to be forced in order to be amplified.
        A small 3D forcing perturbation term $f'(x,r,\theta,t)$ (e.g., pressure and/or temperature fluctuations, wall roughness, etc.) is thus added to the Navier-Stokes equation (\ref{eq:NSequation}), the new linearised equation is:
        \begin{equation} 
            \frac{\partial q'}{\partial t} = \bm{\mathcal{J}} q' + f'.
            \label{eq:equationlinear_forcing} 
        \end{equation}
        As the flow is axisymmetric, $\theta$-direction is supposed to be homogenous. Hence these small disturbances can be expressed in Fourier modes of azimuthal wavenumber $m$. 
        As the forcing is considered harmonic at frequency $\omega$, disturbances can then be expressed as:
        \begin{equation} 
            \left\{\begin{aligned}
    		q'(x,r,\theta,t) &= \check{q}(x,r) \, e^{i (m\theta + \omega t)} + c.c. \\
    		f'(x,r,\theta,t) &= \check{f}(x,r) \, e^{i (m\theta + \omega t)} + c.c.
    	\end{aligned}\right.            
            \label{eq:equation_disturbances} 
        \end{equation}
        %
        Equation (\ref{eq:equationlinear_forcing}) leads to the following input/output problem:
        \begin{equation} 
            \check{q} = \bm{\mathcal{R}}(\omega,m) \check{f},
            \label{eq:resolvent_equation_reponse} 
        \end{equation}
        where $\bm{\mathcal{R}} = (i \omega I - \bm{\mathcal{J}})^{-1}$ is the Resolvent operator. This operator represents a linear transfer function between the incoming forcing $\check{f}$ (input) and the flow response $\check{q}$ (output). Then, for a given frequency $\omega$ and azimuthal wavenumber $m$, the most amplified resolvent mode is found by solving an optimisation problem defined as a ratio between two energies:
        \begin{equation} 
            \mu^2 = \sup_{\check{f}} \frac{\lVert \check{q} \rVert_{E}^{2}}{\lVert \check{f} \rVert_{F}^{2}} = \sup_{\check{f}} \frac{\lVert \bm{\mathcal{R}} \check{f} \rVert_{E}^{2}}{\lVert \check{f} \rVert_{F}^{2}},
            \label{eq:resolvent_equation} 
        \end{equation}
        where $\mu^2$ is the gain, $\lVert \cdot \rVert_{E}$ and $\lVert \cdot \rVert_{F}$ are the norms used to evaluate the energy amplitude of the response and the forcing respectively. Using a discrete scalar product norm, these norms can be expressed with their Hermitian matrix $Q_E$ and $Q_F$ such as $\lVert \check{q} \rVert_{E}^{2} = \check{q}^{*} Q_E \check{q}$ and $\lVert \check{f} \rVert_{F}^{2} = \check{f}^{*} Q_F \check{f}$, where $\check{q}^{*}$ is the conjugate transpose of $\check{q}$. 
        The Chu energy norm is used for $Q_E$ and $Q_F$ to account for the compressibility effects in the energy measure \citep{chu1965energy, hanifi1996transient}.
        These matrices are written with conservative variables and are block diagonal (see appendix \ref{appendix:ChuDefinition} for details). Moreover, in order to restrict the forcing to a specific area or specific component, we introduce a prolongation matrix $P$ and define the forcing $\check{f}$ and forcing energy $Q_F$ only on the restricted domain/component. 
        Equation (\ref{eq:resolvent_equation}) can be recast as,
        \begin{equation} 
            \mu^2 = \sup_{\check{f}} \frac
            {\check{f}^{*} P^{*} \bm{\mathcal{R}}^{*} Q_E \bm{\mathcal{R}} P \check{f}}
            {\check{f}^{*} Q_F \check{f}}.
            \label{eq:resolvent_equation_recast} 
        \end{equation}
        The resulting optimisation problem (\ref{eq:resolvent_equation_recast}) can be viewed as a Rayleigh quotient and can be recast as a generalised Hermitian eigenvalue problem:
        \begin{equation} 
           P^*\bm{\mathcal{R}}^{*} Q_E \bm{\mathcal{R}} P \check{f}_{opt,i} = \mu_i^2 Q_F \check{f}_{opt,i},
            \label{eq:eigenvalueproblem_resolvent} 
        \end{equation}
        where $\mu_i^2, \; i \in  [\![0,\dots]\!],$ are the eigenvalues sorted by energy such that $\mu_i^2 \geq \mu_{i+1}^2$. The $\check{f}_{opt,i}$ are the optimal forcing, while 
        $\check{q}_{opt,i}=\bm{\mathcal{R}} P \check{f}_{opt,i}$ 
        are optimal responses. These two sets of vectors define orthonormal sets of the forcing and response spaces: $\check{f}_{opt,i}^*Q_F \check{f}_{opt,j}=\check{q}_{opt,i}^*Q_E \check{q}_{opt,j}=\delta_{ij}$.
        This eigenvalue problem (\ref{eq:eigenvalueproblem_resolvent}) allows to map the linear system amplification and thus characterise the receptivity of the base-flow by solving it for a range of frequencies $\omega$ and azimuthal wavenumber $m$. More details on the global linear stability of dynamical systems are available in \citet{schmid2007nonmodal} and \citet{sipp2010dynamics}.

    \subsection{Numerical methods and strategy}\label{subsection:Numerical_methods_strategy}

        \subsubsection{Solver and finite volume discretisation}\label{subsubsection:SolverDiscretisation}

            The open-source CFD code BROADCAST \citep{poulain2023broadcast} is used in this study. This toolbox provides all the tools and numerical methods needed to compute the base-flow and the discrete operators, as well as to perform global linear stability analysis of two-dimensional Cartesian or axisymmetric flows. 
    
            After two dimensional discretisation of the domain into a structured $N_x \times N_r$ mesh, the equation (\ref{eq:NSequation}) is numerically solved using this high-order finite volume solver. This toolbox uses a 7th order Flux-Extrapolated MUSCL scheme for the inviscid fluxes \citep{cinnella2016high} (validated by \citet{sciacovelli} for hypersonic flows), a fourth order accuracy scheme with a five-point compact stencil for the viscous fluxes \citep{shen2009high} and Algorithmic Differentiation (AD) with the software TAPENADE \citep{hascoet2013tapenade} in order to extract the discrete direct and adjoint global linear operators (of arbitrary order). The AD tool accurately computes the $n$-order derivatives of the various operators required for accurate stability analysis (global modes, resolvent modes). A description of the methods used in the BROADCAST toolbox, as well as their validation, can be found in \citet{poulain2023broadcast}, and has been extended and validated for axisymmetrical flows in \citet{caillaud2024separation}. A reminder of the extension to 3D stability of the base-flow, as well as the definition of the 3D residual in axisymmetric flows, can be found in appendix \ref{appendix:Extension_3D}.

        \subsubsection{Methods for solving eigenvalue problems for global and resolvent modes}\label{subsubsection:MethodSolveEigenvalueProblem}

            The dimension of the various discrete operators presented in the previous sections (Jacobian $\bm{\mathcal{J}}$, resolvent $\bm{\mathcal{R}}$, etc.) depends on the mesh size ($N_x \times N_r$) and the number of conservative variables ($n_c=5$), thus $N_{dof}=(N_x \times N_r \times nc)^2$. Since we are working in a discrete framework with a finite stencil (with the schemes used (see \S \hspace*{0.01em} \ref{subsubsection:SolverDiscretisation}) $stencil = (2 N_{gh} + 1)^2$ and $N_{gh} = (7+1)/2$, with $N_{gh}$ the number of ghost cells ($gh$)) these matrices are very sparse, and the number of non-zero entries is approximately $nnz = N_x \times N_r \times stencil \times n_c^2$ \citep{metto2013}. Therefore, all linear systems involve sparse matrices, and in this study, they are solved using PETSc/SLEPc 
            \citep{balay2019petsc, hernandez2005slepc}. 
            Since the BROADCAST code is written in Python, the petsc4py and slepc4py interfaces are used \citep{dalcin2011parallel}.
            This study involves solving linear systems with large sparse matrices and requires LU decompositions, so the direct sparse LU solver MUMPS is used \citep{amestoy2001fully}. Note that, even with sparse algorithms, these operations require a large amount of memory usage. The memory needed to perform the LU factorisation scales as $(nnz)^{4/3}$ \citep{l2012multifrontal, fer2023scaling}, where $nnz$ is the number of non-zero entries in the matrix.

            
            For global modes, the eigenvalue problem (equation (\ref{eq:globalstability})) is solved using a shift-invert strategy by looking in the complex plane at the closest eigenvalues to a target. Eigenvalues and eigenvectors are computed using an Arnoldi algorithm \citep{hernandez2007krylov}. More details on how this eigenvalue problem is solved in BROADCAST can be found in \citet{poulain2023broadcast}.

            For resolvent modes, the numerical resolution of the Generalised Hermitian Eigenvalue Problem (GHEP) (equation (\ref{eq:eigenvalueproblem_resolvent})) is again solved using SLEPc, which implements various Krylov methods 
            \citep{hernandez2007krylov}. More details on the resolvent algorithm and the GHEP resolution in BROADCAST can be found in \citet{poulain2023broadcast}.

        \subsubsection{Mesh, flow initialisation and resolution}


            A structured mesh of the nozzle is obtained using an in-house meshing tool. Two meshes are used for this study: a coarse structured mesh to quickly eliminate transients from initialisation, and a fine grid ($N_x \times N_r = 5780 \times 400$) refined enough to accurately capture the various instabilities. To ensure mesh convergence, the number of cells per wavelength was estimated for different instabilities based on the density structures at the maximum of Chu energy. For the second Mack mode, the highest-frequency resolvent mode in the flow, the estimation yields approximately $(50\times12)$ for the response and $(50\times15)$ for the forcing. \citet{poulain2023broadcast} recommended approximately around $12$ cells per wavelength using a 7th order FE MUSCL scheme to correctly capture the optimal gain with the numerical schemes used in our study. Given that the second Mack mode is the highest-frequency instability in this flow, the streamwise resolution of our mesh exceeds the required precision for all instabilities in the divergent section. In the radial direction, the resolution is close to the recommended values from \citet{poulain2023broadcast}. In the convergent section of the nozzle, the low-frequency Kelvin-Helmholtz instability must be captured. The number of cells per wavelength for this instability is estimated at around $14$ in the streamwise direction, which is enough with the numerical schemes used in this study. These results confirm that the mesh provides adequate resolution in both the convergent and divergent sections of the nozzle to accurately study the development of the linear instabilities in this flow. Note that the estimated memory required to perform the LU decomposition of the Jacobian operator on the refined mesh is $nnz \approx 1.3 \times 10^9$, therefore the memory requirement is approximately $\approx 1.4 \; \text{TB}$. Consequently, global linear stability analysis demands substantial memory resources.
            
            The boundary conditions applied in the numerical simulation for the isothermal case are illustrated in figure \ref{fig:Baseflow_setup}. For the adiabatic case, the "\textit{extended settling chamber}" and the wall are set with an adiabatic wall boundary condition. The flow in the centre of the nozzle is initialised using quasi-1D isentropic relations. And, the near-wall region is initialised with a boundary layer computed using the compressible boundary layer solver CLICET \citep{aupoix2010couches}.
    
            The laminar flow in the nozzle (called fixed point or base-flow $\overline{q}$) is computed with the following steps. First, a preliminary flow is computed on a coarse structured mesh ($N_x \times N_r = 1600 \times 400$) using "implicit iterations" (Symmetric Gauss-Seidel (LU-SGS) \citep{yoon1986multigrid}) to eliminate transients due to the initialisation. Second, once the residuals have decreased sufficiently, the solution is interpolated onto a fine grid ($N_x \times N_r = 5780 \times 400$). After again a few implicit iterations, the base-flow is obtained using an iterative pseudo-Newton method. The state vector $q^n$ is updated with $q^{n+1} = q^n + \delta q^n$, and $\delta q^n$ is computed using a pseudo transient continuation method \citep{crivellini2011implicit} to ease convergence:
            \begin{equation} 
                \left( \frac{I}{\Delta t} + \bm{\mathcal{J}}(q^n) \right) \delta q^n = - R(q^n)
                \label{eq:Newton_steps} 
            \end{equation}
            
            The base-flow obtained is converged with a decrease close to $12$ orders of magnitude of the $L^2$ norm of the residuals, i.e.\ close to machine precision. 



            


\vspace{-1.0em}
\section{Characterisation of the base-flow}\label{section:baseflow}

    In this section, the characterisation of the base-flow in the hypersonic nozzle for both isothermal and adiabatic cases is performed. A comparison between these two cases is also provided to highlight the differences in the flow induced by these two boundary conditions.

    \begin{figure}
        \centering
        \includegraphics[width=0.925\textwidth]{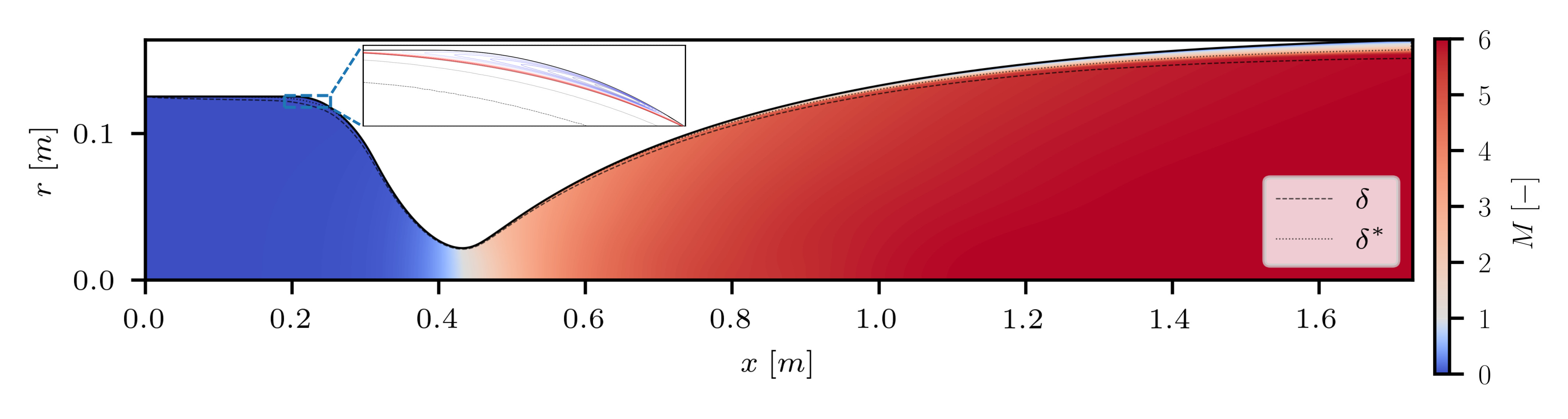}
        \caption{Base-flow Mach number field for the isothermal case. Zoom: recirculation bubble at the convergent inlet.}
        \label{fig:PlotMach_bubble}
    \end{figure}
    \begin{figure}
        \vspace{-1.5em}
        \centering
        \includegraphics[width=0.925\textwidth]{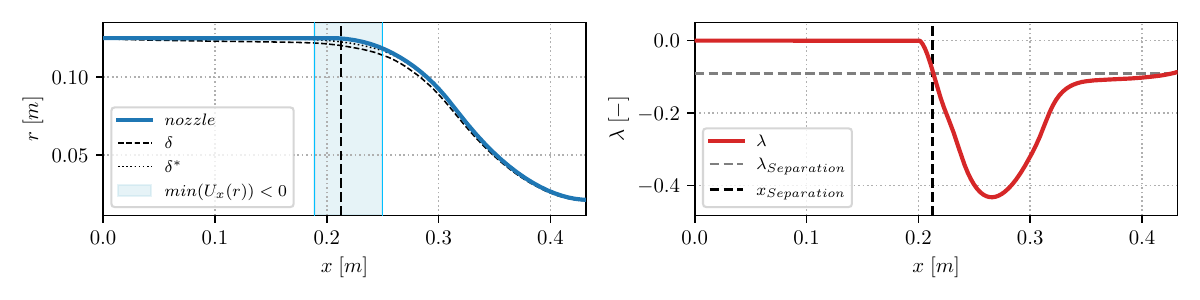}
        \caption{Estimation of the boundary layer separation using the Rott-Crabtree gradient parameter $\lambda$.}
        \label{fig:Rott_Crabtree}
    \end{figure}

    The evolution of the Mach number for the isothermal base-flow is shown in figure \ref{fig:PlotMach_bubble}. A boundary layer separation is observed at the convergent inlet for both isothermal and adiabatic cases, leading to a recirculation bubble in this region. This flow feature is important
    because recirculation bubbles can trigger global modes and resolvent modes such as Kelvin-Helmholtz instability. For example, the low-frequency instability measured in the NASA Langley Mach 6 quiet nozzle is suspected to come from this kind of separation \citep{schneider1998design, wilkinson1997review}. 
    Note that the presence of this recirculation bubble here is likely because this wind tunnel was designed to operate at higher stagnation pressures than those used in this study (see table \ref{tab:Boundary_freestream_conditions}).

    A comparatively straightforward criterion, known as the Rott-Crabtree method, provides a convenient means for estimating whether boundary layer separation is likely to occur at the convergent inlet.
    This method was used, for example, to properly design a Mach-6 quiet tunnel convergent \citep{schneider1998design, schneider1991quiet}.
    The Rott-Crabtree method provides access to the gradient parameter $\lambda$, which must be greater than $-0.09$ to avoid boundary layer separation. Figure \ref{fig:Rott_Crabtree} shows that this criterion accurately predicts the boundary layer separation at the convergent inlet, as $\lambda$ drops below $-0.09$ exactly where a negative streamwise velocity is detected in the flow due to the recirculation bubble. Thus, this straightforward criterion provides a quick estimation of the separation; however, a simulation is usually required to validate the result, as in our case.

        

    \vspace{-1.0em}
    \subsection{Comparison between isothermal and adiabatic base-flow}

    \begin{figure}
        \centering
        \includegraphics[width=0.80\textwidth]{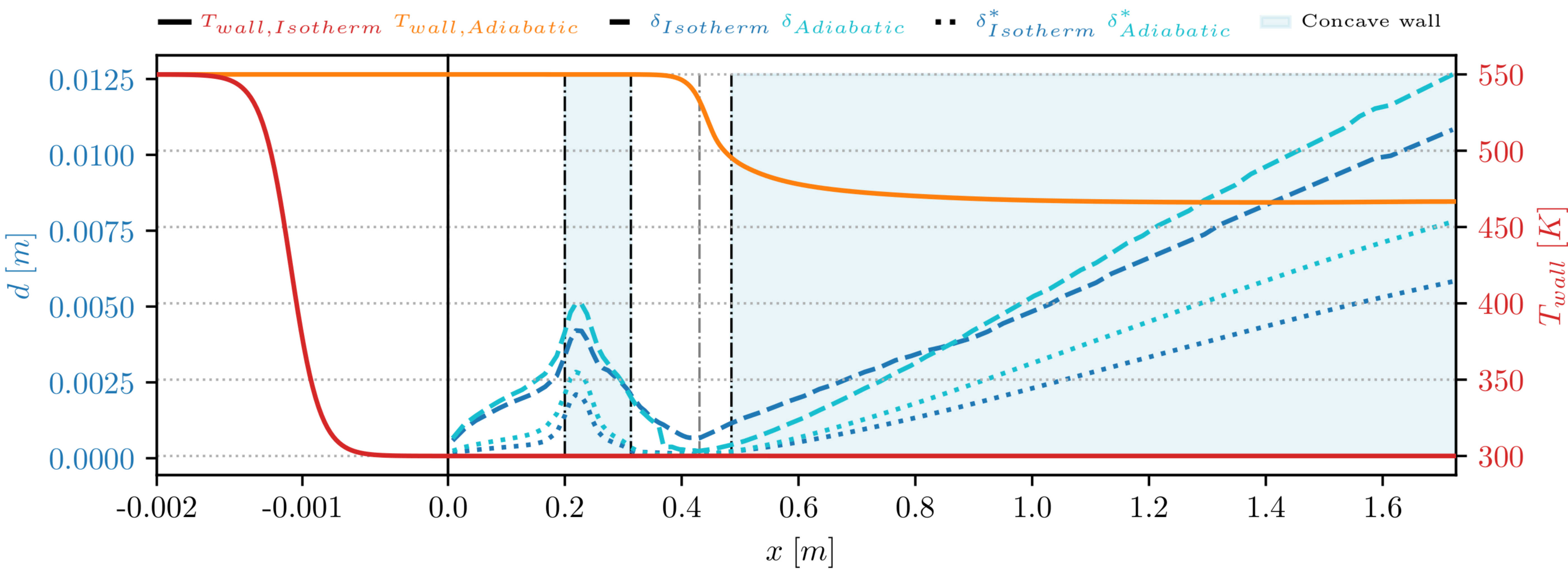}
        \caption{Wall temperature and boundary layer thickness comparison between the isothermal and adiabatic cases. On the left, zoom on the "\textit{extended settling chamber}" which allows for a smooth transition between $T_t$ and $T_{iso}$ using a $\tanh$ function in the isothermal case.}
        \label{fig:Baseflow_wallT_BoundaryLayer}
    \end{figure}
    \begin{figure}
        \vspace{-1.0em}
        \centering
        \subfigure[At the convergent inlet ($x = 0.20 \, \text{m}$).]{\includegraphics[width=0.32\textwidth]{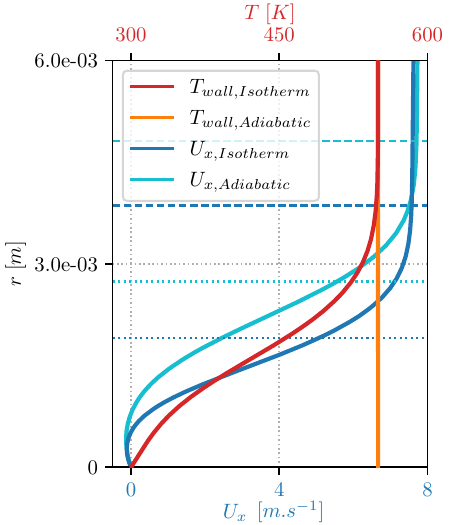}}
        \subfigure[At the outlet ($x = 1.726 \, \text{m}$).]{\includegraphics[width=0.32\textwidth]{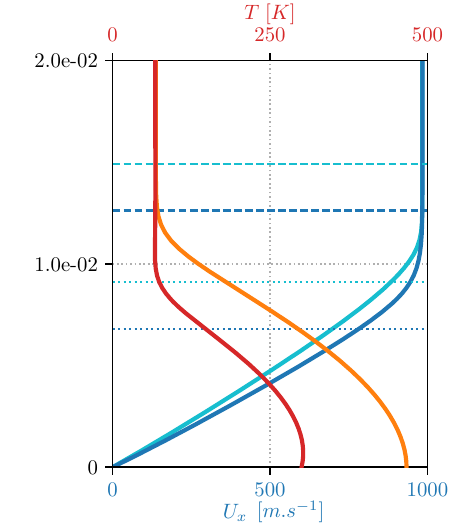}}
        \caption{Wall-normal profiles of streamwise velocity and temperature at different positions in the nozzle, comparison between the isothermal and adiabatic cases. ({\includegraphics[width=\baselineskip]{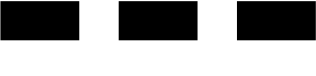}}): boundary layer thickness $\delta$. ({\includegraphics[width=\baselineskip]{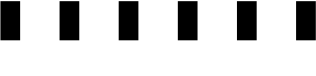}}): displacement thickness $\delta^*$.}
        \label{fig:BL_comp}
    \end{figure}

    A comparison of the solutions obtained for the two base-flows, under isothermal and adiabatic boundary conditions, is presented in figures \ref{fig:Baseflow_wallT_BoundaryLayer} and \ref{fig:BL_comp}. Note that the boundary layer thicknesses are estimated using the BLX\_like method detailed by \citet{begou2018prevision}, a method based on the maximum of a diagnostic function which depends on the distance to the wall and the vorticity vector modulus.
    The wall temperature in figure \ref{fig:Baseflow_wallT_BoundaryLayer}, along with the temperature profiles in figure \ref{fig:BL_comp}, shows that $T_{wall,adiabatic}$ is higher than the isothermal temperature along the whole length of the nozzle. The "decrease" in wall temperature observed in the adiabatic case corresponds to the recovery temperature ($T_{recovery} = T_{\infty} \left( 1 + r\frac{\gamma-1}{2}M^2\right)$ with $r \approx Pr^{1/2} \approx 0.85$ in laminar regions). These differences implies that the boundary layer profiles, and therefore the characteristic thicknesses, vary significantly between the two cases. As a result, instabilities will not scale and develop in the same way, which will be interesting for stability comparison. Additionally, as shown in figures \ref{fig:Baseflow_wallT_BoundaryLayer} and \ref{fig:BL_comp} (a), the size of the recirculation bubble is affected by the wall boundary condition, therefore the global modes developing within are also likely to differ. In summary, the two wall boundary conditions used in this study will allow us to compare the development of global 
    and resolvent modes in the nozzle, given that the recirculation bubble and the boundary layer profiles obtained are different in these two cases.


\vspace{-1.0em}
\section{Global analysis}\label{section:Global_Stability_Analysis}

    \begin{figure}
        \centering
        \subfigure[Isothermal case.]
        {\includegraphics[width=\textwidth]{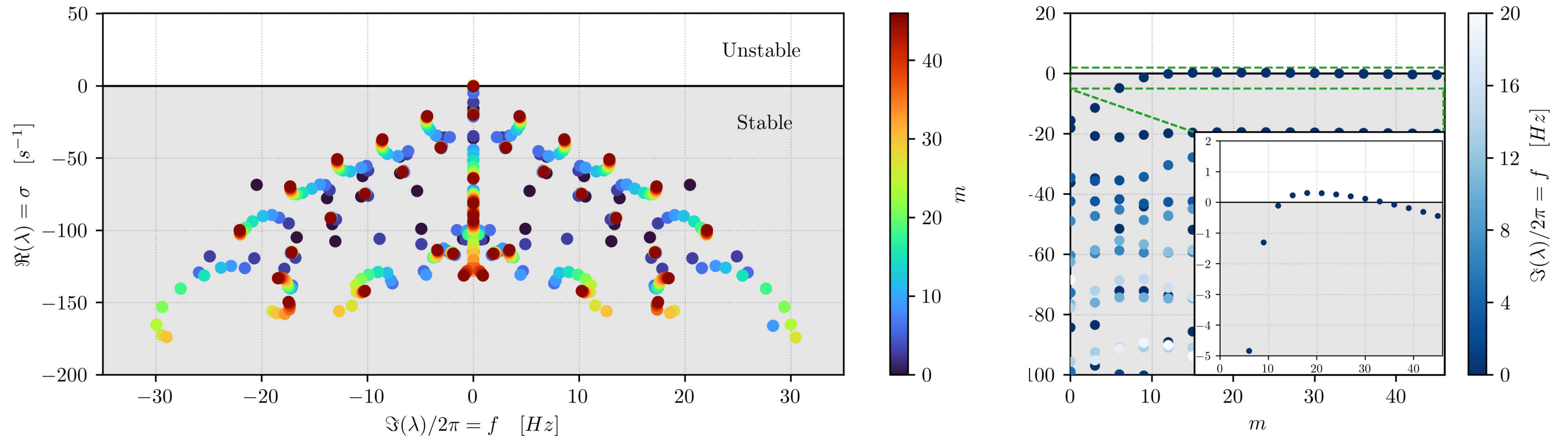}}
        \subfigure[Adiabatic case.]
        {\includegraphics[width=\textwidth]{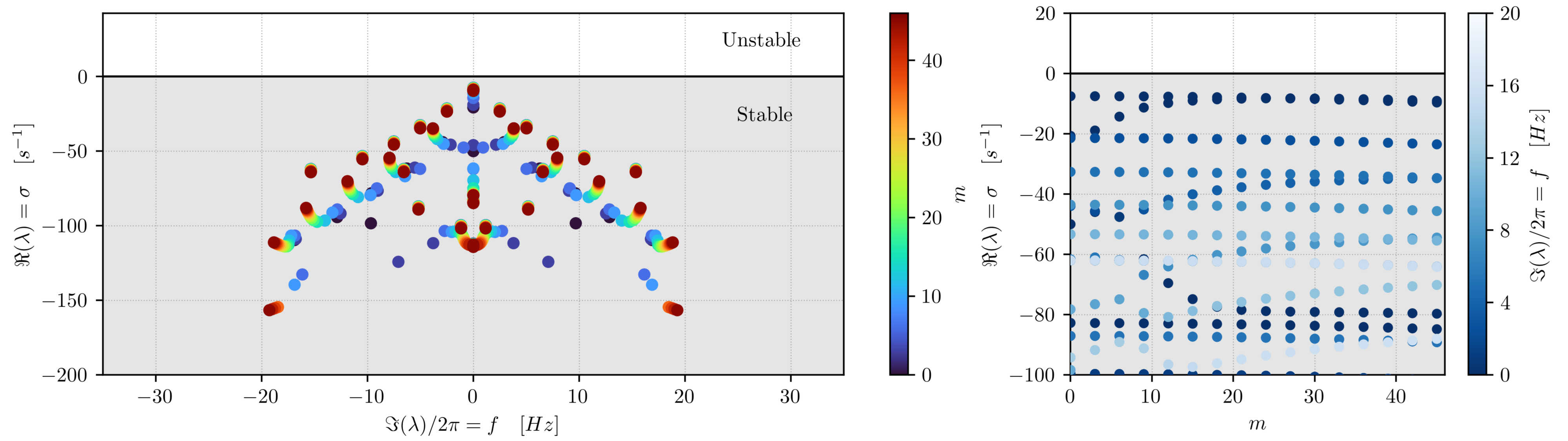}}
        \caption{Discrete eigenvalue spectrum of the nozzle flow. The left plot shows the eigenvalue in the complex $(f,\sigma)$ plane with respect to the azimuthal wavenumber $m$ (with matching colour). The right plot shows the same values, but with the axes of the graph changed to $(m,\sigma)$ and zooming in on the values close to the global mode.}
        \label{fig:EigenvalueSpectrum}
    \end{figure}
    \begin{figure}
        \centering
        \includegraphics[width=0.9\textwidth]{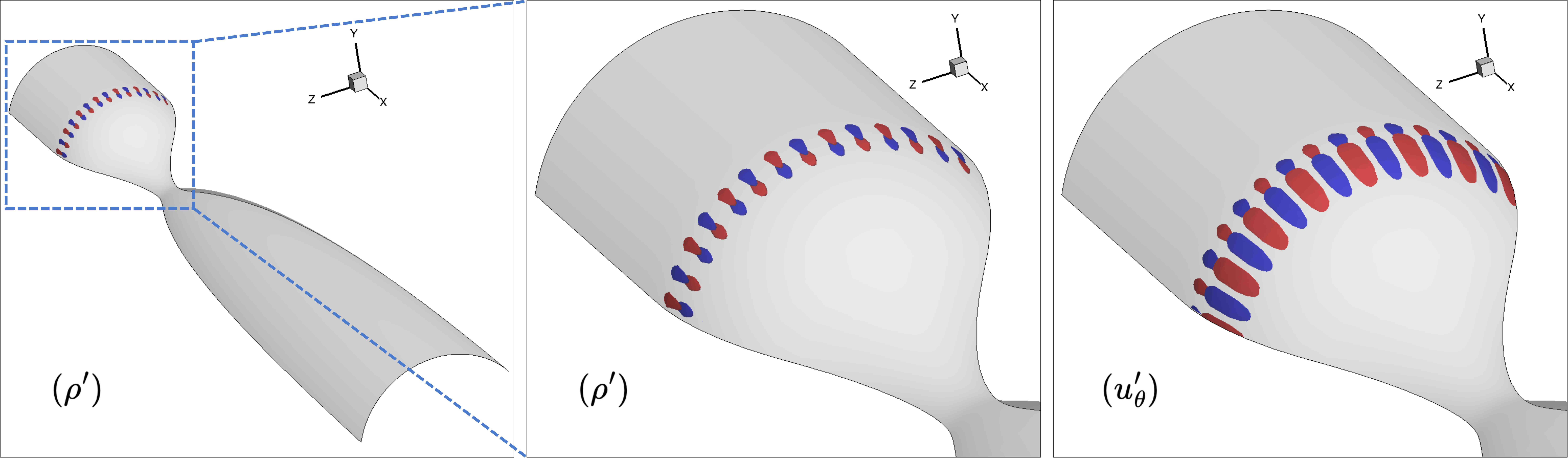}
        \caption{Example of disturbance $\rho'$ and $u_{\theta}'$ associated with a globally unstable mode in the isothermal case, $(f,m) = (0\; \text{Hz}, \, 21)$. Red and blue: iso-surfaces at $\pm 55\%$ and $\pm 20\%$ of the maximum absolute values $\rho'$ and $u_{\theta}'$. One-third view of the nozzle.}
        \label{fig:stabglob_eigenfunc_example}
    \end{figure}

    In this section, the
    global analysis
    of the nozzle flow is performed (see \S \hspace*{0.01em} \ref{subsection:Global_Stability_Oscillator}), to verify the presence or absence of global modes induced by the recirculation bubble in our flow, as it is well known that such flow features can lead to global modes \citep{theofilis2000origins,marquet2009direct}.
    To investigate this, we study the eigenvalues of the Jacobian operator, equation (\ref{eq:globalstability}). 

    Global unstable modes are identified in the isothermal case for a few azimuthal wavenumbers around $m \in  [\![12, 33]\!]$ at zero frequency (figure \ref{fig:EigenvalueSpectrum}). Since $m > 0$, these global modes will lead to a steady three-dimensionalisation of the base-flow.
    An example of the eigenfunction associated with one of these unstable modes is shown in figure \ref{fig:stabglob_eigenfunc_example}, the chosen mode is the most amplified and is referred to as the leading global modes according to \citet{marquet2009direct, jackson1987finite}. As expected, these modes are localised within the recirculation bubble. The spatial structure of the leading global mode obtained is consistent with other studies of separated flows
    \citep{marquet2009direct, barkley2002three, gallaire2007three}. 
    These modes, observed in the isothermal case, are relatively problematic because they can alter the laminar flow topology and therefore the boundary layer transition process.

    The ability of the method used to capture global modes is not necessarily relevant for boundary layer separation at the convergent inlet, as this issue is well addressed in the state-of-the-art of wind tunnel nozzle design \citep{schneider2001shakedown}, and can be controlled by properly designing the convergent under the stagnation conditions used, for example by using the Rott-Crabtree criterion \citep{schneider1998design, schneider1991quiet} or by numerical simulation. However, it is crucial to detect these instabilities when designing the suction lip upstream of the nozzle throat, as it can introduce boundary layer separations and thus recirculation bubbles \citep{taskinoglu2005numerical, benay2004design} (see figure \ref{fig:SchemaInstabNozzle}). Due to the three-dimensionalisation of the base-flow, the entire flow field is modified, and consequently, predictions based on linear stability analysis applied to this 2D base-flow could be significantly altered. The boundary layer transition may therefore be inaccurately predicted, potentially occurring early within the nozzle and leading to a noisy flow in the test section \citep{schneider2008development}. 
    Global analysis could enable proper characterisation of these bubbles at the suction lip. However, the topology of global modes at the suction lip, if they exist, could be radically different due to the presence of a transonic region compared to those detected in this study at the convergent inlet in a subsonic and weakly compressible zone.

\vspace{-1.0em}
\section{Resolvent modes with an isothermal wall}\label{section:Convective_Instabilities}

    Having identified the global unstable modes, the resolvent analysis of the nozzle flow can now be conducted to investigate resolvent mode (see \S \hspace*{0.01em} \ref{subsection:Resolvent_Amplifier}).
    Such studies are normally performed on a globally stable flow (i.e.\ the Jacobian does not exhibit unstable eigenvalues) \citep{schmid2007nonmodal}, which is not our case (in the isothermal case). However, it is important to keep in mind that the goal here is not to characterise the exact transition mechanism of this nozzle (which could be altered due to the globally unstable mode), but rather to identify (using global linear stability analysis) all linear instabilities that can develop in hypersonic wind tunnel nozzles. Therefore, we apply resolvent analysis to examine how such flow would behave if there were no global unstable mode. 

    For the numerical setup, the restriction matrix $P$ is used to exclude the area in front of the settling chamber region ($x<0$, in figure \ref{fig:Baseflow_setup}). 
    Thus, the forcing field can be localised anywhere in the settling chamber and/or convergent and/or divergent of the nozzle. On the other hand, the Chu energy matrix $Q_E$ is restricted, in the optimisation problem, to a near-wall region defined by $2\delta$. This restriction of the Chu energy is applied to only study resolvent modes that develop within the boundary layer and thus not focus on resonant acoustic modes that may exist in the settling chamber and convergent regions due to the confined subsonic flow.

    \begin{figure}[t]
        \centering
        \includegraphics[width=\textwidth]{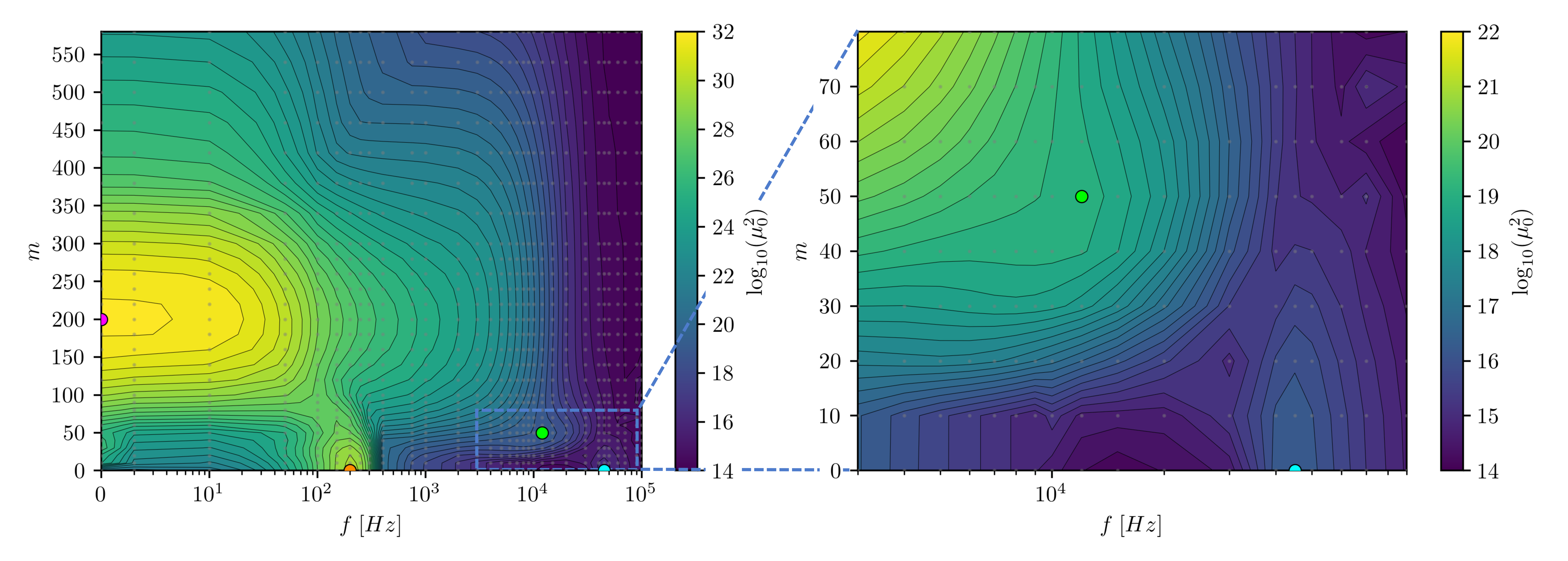}
        \caption{Optimal gain map $\mu_0^2$ in frequency $f$ and azimuthal wavenumber $m$ space for the isothermal case. The right plot shows the optimal gain map zoom on the area near the first and second Mack modes. The grey dots ({\includegraphics[width=0.4\baselineskip]{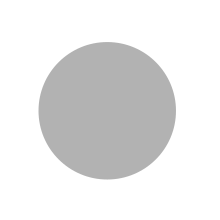}}) show the different computation locations. ({\includegraphics[width=0.4\baselineskip]{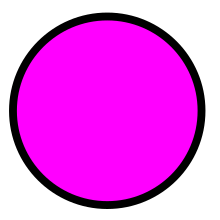}}): Görtler instability peak. ({\includegraphics[width=0.4\baselineskip]{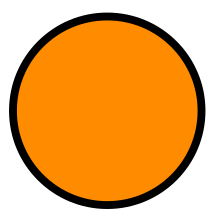}}): Kelvin-Helmholtz instability peak. ({\includegraphics[width=0.4\baselineskip]{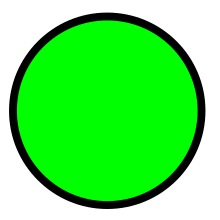}}): first Mack mode peak. ({\includegraphics[width=0.4\baselineskip]{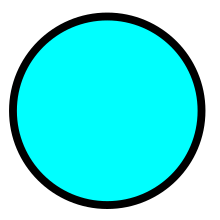}}): second Mack mode peak.}
        \label{fig:OptimalGainMap}
    \end{figure}
    {\renewcommand{\arraystretch}{1.25} 
    {\setlength{\tabcolsep}{0.25cm} 
        \begin{table}[H]
            \caption{\label{tab:Instability_Tiso} Resolvent modes in the isothermal case.}
            \centering
            \begin{tabular}{c|ccc}
                \hline
                Instability & frequency $f$ [Hz] & Wavenumber $m$ & Optimal gain $\log_{10}(\mu_0^2)$ \\ 
                \hline
                Görtler & 0 & 200 & 32.19 \\ 
                Kelvin-Helmholtz & 200 & 0 & 29.70 \\ 
                First Mack mode & 12 000 & 50 & 19.04 \\ 
                Second Mack mode & 45 000 & 0 & 16.30 \\ 
                \hline
            \end{tabular}
    \end{table}}}
    %
    \vspace{-2.0em}
    \begin{figure}[H]
        \centering
        \includegraphics[width=1.0\textwidth]{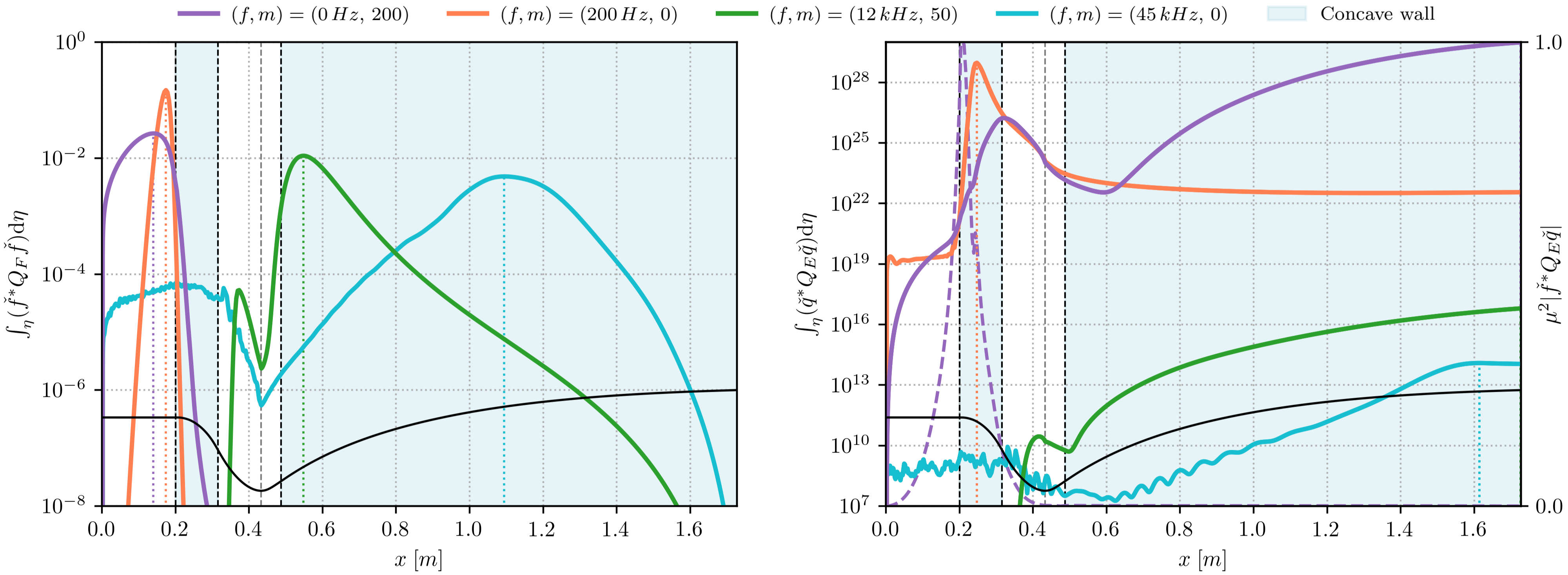}
        \caption{({\includegraphics[width=\baselineskip]{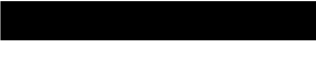}}) Evolution of the quantity $E_{Chu}$ of the optimal forcing (left) and response (right) computed along the gridlines in the $r$-direction by integrating the local Chu energy contribution. ({\includegraphics[width=\baselineskip]{Ligne1.png}}) Evolution of the rate of work done by the forcing onto the response mode $\xi = \mu^2 |\check{f}^* Q_E \check{q}|$ computed along the gridlines in the $r$-direction. $(f,m) = (0 \, \text{Hz}, 200)$: Görtler instability. $(f,m) = (200 \, \text{Hz}, 0)$: Kelvin-Helmholtz instability. $(f,m) = (12 \, \text{kHz}, 50)$: first Mack mode. $(f,m) = (45 \, \text{kHz}, 0)$: second Mack mode.}
        \label{fig:Streamwise_gortler}
    \end{figure}
    \vspace{-1.0em}
    %
    Figure \ref{fig:OptimalGainMap}, shows the evolution of the optimal gain $\mu_0^2 (\omega,m) = \lVert \check{q}_{opt,0} \rVert_{E}^{2} / \lVert \check{f}_{opt,0} \rVert_{F}^{2}$ as a function of frequency $f$ and azimuthal wavenumber $m$ for the isothermal case.
    Note that the optimal gain $\mu_0^2$ are made non-dimensional by dividing them by the square of the reference time $(\nu_{\infty}/U_{\infty}^2)^2$.
    Four regions of maximum gain can be observed,
    indicating four dominant mechanisms in our flow, which are: the Görtler instability, the 
    Kelvin-Helmholtz instability (shear mode due to the recirculation bubble), the first and second Mack modes. 
    Figure \ref{fig:Streamwise_gortler} shows the streamwise evolution of the quantity $E_{Chu}$ (\citet{chu1965energy}, see appendix \ref{appendix:ChuDefinition}), computed along the gridlines in the $r$-direction by integrating the local Chu energy contribution of the optimal forcing and response for the four leading mechanisms. In these figures, the blue areas indicate the two concave wall zones: the first part of the convergent and the second part of the divergent. This figure will be referenced in the following sections for discussion of the four dominant mechanisms.
    In the following sections, we will identify these various resolvent modes in the isothermal case, examining their shapes, locations on the nozzle, and other characteristics.

    %
    %


    \subsection{Görtler instability}\label{subsection:gortler_instability}

        \begin{figure}
            \centering
            \includegraphics[width=0.91\textwidth]{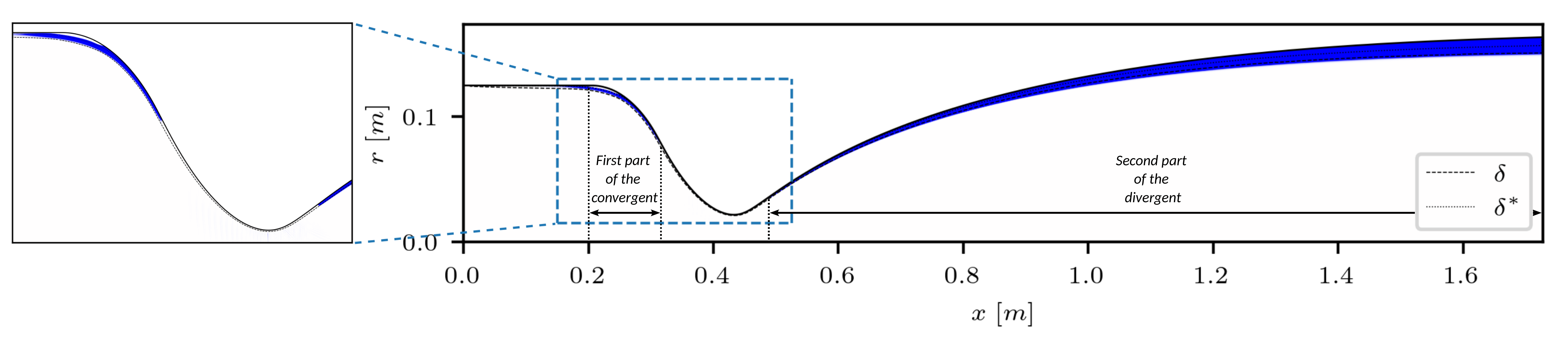}
            \caption{Blue areas indicate regions in the flow where the Rayleigh discriminant is negative ($\Delta < 0$), i.e.\ the unstable zones where centrifugal instabilities can develop \citep{sipp2000three}. Left: zoom near the throat.}
            \label{fig:Rayleigh_discriminant}
        \end{figure}
        %
        \begin{figure}
            \vspace{-1.0em}
            \centering
            \includegraphics[width=\textwidth]{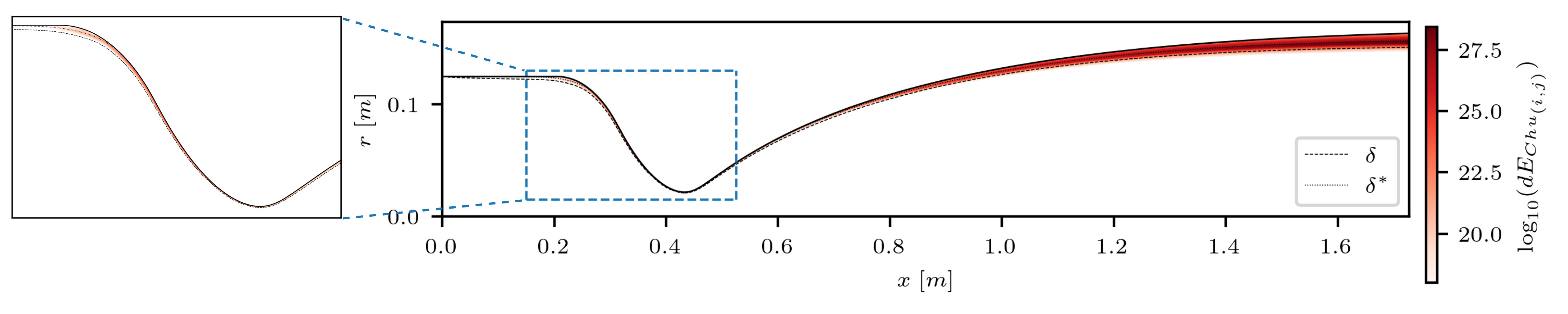}
            \caption{Görtler instability $(f,m) = (0 \, \text{Hz}, 200)$: evolution of the local quantity $E_{Chu_{(i,j)}}$ of the optimal response. Left: zoom near the throat.}
            \label{fig:Echuij_response_gortler}
        \end{figure}

        The first identified mechanism is the Görtler instability, which peaks at zero frequency (thus stationary) and at an azimuthal wavenumber of $m=200$. In our case, this is the dominant mechanism, which is not surprising as the nozzle we are studying is known to be relatively short compared to nozzles found in other hypersonic wind tunnels (refer to Fig.\ 27 of \citet{threadgill2024scaling} for a length comparison of this nozzle with a nozzle having the typical length of a quiet wind tunnel nozzle). The short nozzle length implies a significant wall curvature, which promotes the development of centrifugal instability.

        In order to confirm that the observed instability is indeed the Görtler instability, we compute the centrifugal instability criterion known as the Rayleigh discriminant:
        \begin{equation} 
            \Delta = \frac{2 \, \omega_z \lVert \overline{u} \rVert}{R_{curv}}, \; \; \text{with} \; \; R_{curv} = \frac{\lVert \overline{u} \rVert^3}{(\nabla \psi) \cdot (\overline{u} \cdot \nabla \overline{u})}  
            \label{eq:Rayleigh_discriminant} 
        \end{equation}
        %
        %
        where $\lVert \overline{u} \rVert$ is the norm of the velocity field, $\omega_z = \partial_x u_r - \partial_r u_x$ the vorticity, $\psi$ the streamfunction and $R_{curv}$ the local algebraic radius of curvature. \citet{sipp2000three} have demonstrated a sufficient criterion for centrifugal instability using the Rayleigh discriminant: the flow is unstable and can develop centrifugal instabilities where $\Delta < 0$. According to figure \ref{fig:Rayleigh_discriminant} two regions in the flow can develop centrifugal instabilities, both located near concave wall sections, i.e.\ the first part of the convergent and the end of the divergent. By looking at the evolution of the local Chu energy $E_{Chu_{(i,j)}}$ of the optimal response at $(f,m) = (0 \, \text{Hz}, 200)$, figure \ref{fig:Echuij_response_gortler}, we can see that the growth in energy occurs in these two $\Delta < 0$ zones. 
        Note that the energy between these two zones corresponds to a decay in energy, as shown in figure \ref{fig:Streamwise_gortler}, indicating that the instability takes some distance to be advected and dissipates after passing through a suitable centrifugal instability zone. All this confirms that the centrifugal Görtler instability is observed here in the flow.
        We will now examine in more detail how this instability evolves within the nozzle.

        Figure \ref{fig:Streamwise_gortler} shows that the forcing peak of the Görtler instability is localised in the settling chamber while the response peak is located at the exit of the divergent section. We can also see that the response grows in the two concave parts of the nozzle.
        Interestingly, variations in energetic growth are observed in the concave region of the convergent section (two distinct "bumps").
        However, by computing the localisation of the rate of work done by the forcing onto the response mode, following the Hadamard product formulation as defined by \citet{skene2022sparsifying, qadri2017frequency, houtman2023resolvent, ribeiro2023triglobal}: $\xi = \mu^2 | \langle \check{f}, \check{q} \rangle_{Chu} | = \mu^2 |\check{f}^* Q_E \check{q}|$, our analysis reveals that this quantity is predominantly localised on the first "bump".
        Therefore, the first growth in the concave region of the convergent is attributed to both the base-flow contribution and the rate of work done by the forcing on the response mode, while the second growth is driven solely by the base-flow contribution.
        
\begin{figure}
            \centering
            \includegraphics[width=0.49\textwidth]{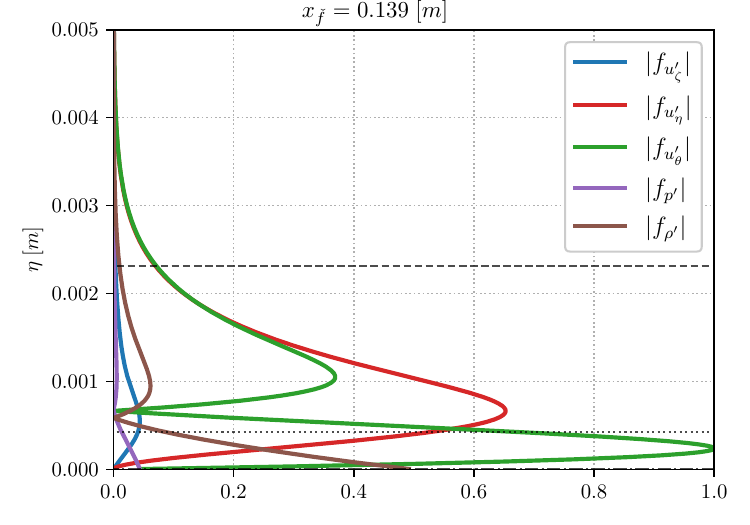}
            \includegraphics[width=0.49\textwidth]{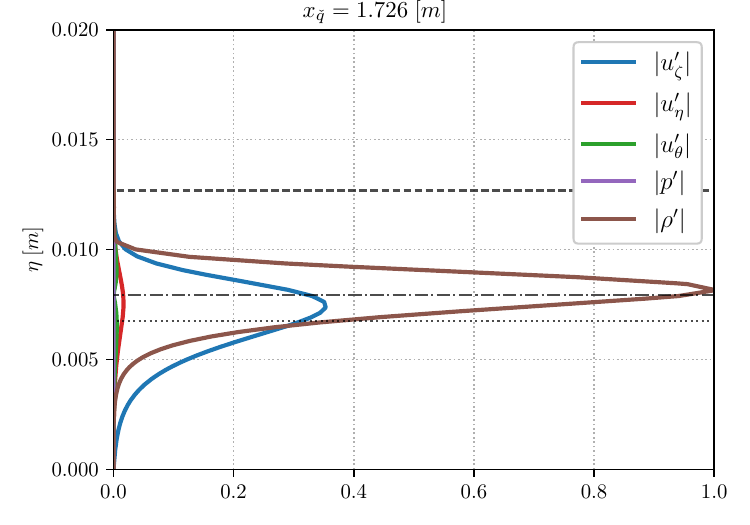}
            \caption{Görtler instability $(f,m) = (0 \, \text{Hz}, 200)$: eigenfunctions at the maximum Chu energy. Left: Forcing. Right: Response. ({\includegraphics[width=\baselineskip]{Ligne1.png}}): boundary layer thickness $\delta$. ({\includegraphics[width=\baselineskip]{Ligne2.png}}): displacement thickness $\delta^*$. ({\includegraphics[width=\baselineskip]{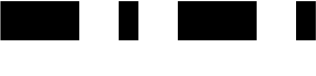}}): generalised inflection point.}
            \label{fig:Eigenfunctions_gortler}
        \end{figure}

        The local eigenfunctions obtained at the maximum Chu energy location are shown in figure \ref{fig:Eigenfunctions_gortler}. These eigenfunctions are consistent with the state-of-the-art of Görtler instabilities studied using linear stability \citep{li2022secondary, hao2023response, cao2023stability, zhao2024investigation}. They are counter-rotating streamwise vortices in the forcing field. While the response mainly exhibits a dominant streamwise structure, but slight counter-rotating streamwise vortices still remain in the response. Note that a density peak is also observed for the response along the generalised inflection point. This density peak will also be observed in the study of the first and second Mack modes, which seems to be a property shared by these different instabilities. This property is observed by \citet{hanifi1996transient} and also noted by \citet{bugeat20193d} in a more recent study.

\begin{figure}
            \centering
            \includegraphics[width=\textwidth]{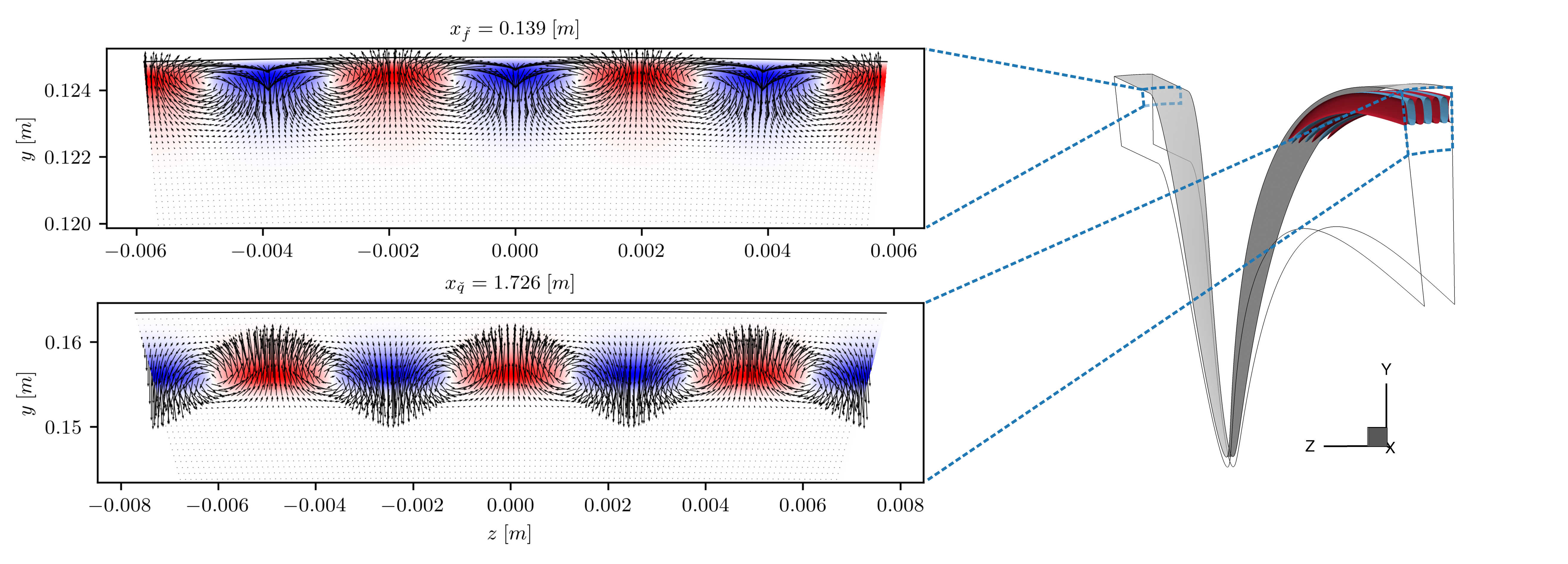}
            \caption{Görtler instability $(f,m) = (0 \, \text{Hz}, 200)$. Left: 2D forcing (top) and response (bottom) at the maximum Chu energy, velocity vector field $(\Vec{e_{\eta}},\Vec{e_{\theta}})$ and plot $f_{u'_{\zeta}}$ (top) or $u'_{\zeta}$ (bottom). Right: 3D response iso-surfaces at $10\%$ of $u'_{\zeta}$.}
            \label{fig:PlotSlice_gortler}
        \end{figure}

        To better visualise these disturbance structures, figure \ref{fig:PlotSlice_gortler} shows cross-sections of the nozzle at the same x-position as figure \ref{fig:Eigenfunctions_gortler}. Counter-rotating streamwise vortices are clearly visible, with the upwash and downwash effects near the wall in the forcing and response fields, respectively. In the response field, these counter-rotating vortices are located farther from the wall compared to the forcing field, due to the thicker boundary layer at that position. Additionally, the response field shows dominant alternating longitudinal structures in the streamwise direction. All of these structures may be responsible for the oil-flow streaks observed by \citet{beckwith1981gortler, beckwith1984nozzle} on the nozzle wall. 

        

        %
        \begin{figure}
            \vspace{-1.0em}
            \centering
            \includegraphics[width=0.67\textwidth]{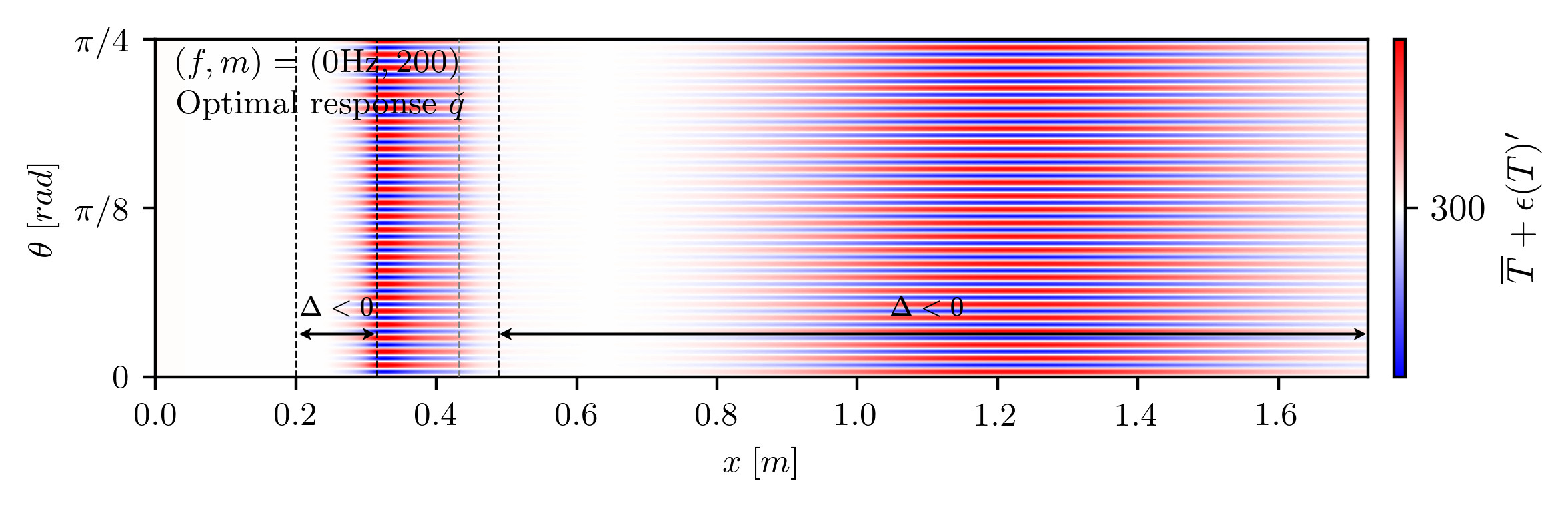}
            \includegraphics[width=0.32\textwidth]{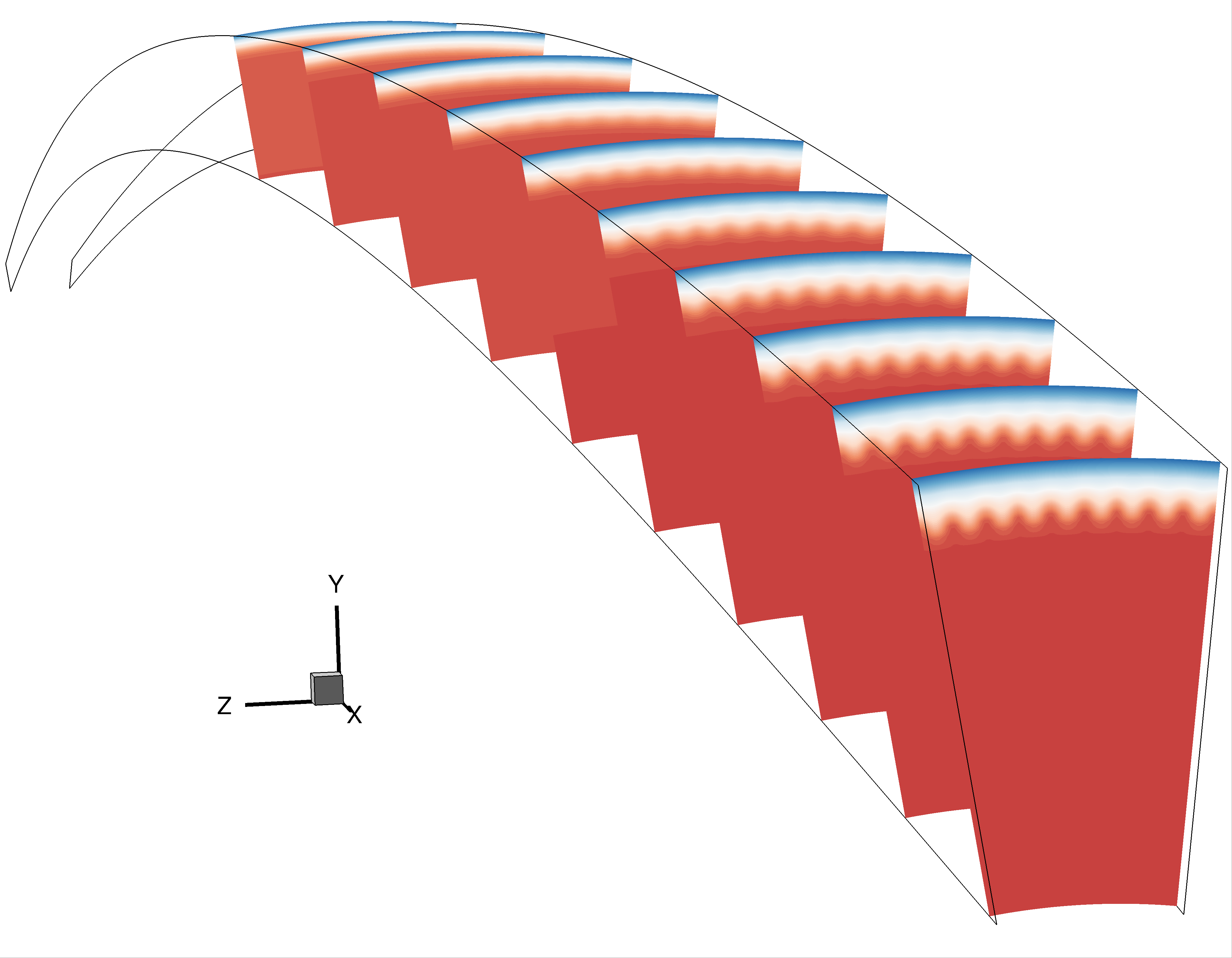}
            \caption{Görtler instability $(f,m) = (0 \, \text{Hz}, 200)$: plot of $q = \overline{q} + \epsilon q'$ with $\epsilon$ equal to 10\% of the maximum between base-flow and fluctuations. Left: one-eighth of the nozzle wall temperature, i.e.\ 25 structures since $m=200$. Right: response $u_{\zeta} = \overline{u_{\zeta}} + \epsilon u'_{\zeta}$ at different $x$-position in the divergent.}
            \label{fig:Plot_baseflow_plus_disturbance}
        \end{figure}

        Figure \ref{fig:Plot_baseflow_plus_disturbance} shows the superposition of the base-flow and the fluctuations. The evolution of the wall temperature reveals the formation of
        streaky
        structures along the wall in the first part of the convergent and the second part of the divergent (i.e.\ in the regions where the wall is concave, figure \ref{fig:Rayleigh_discriminant}). This observation aligns with the results reported by \citet{beckwith1981gortler, beckwith1984nozzle} in hypersonic wind tunnels or by \citet{chen2019hypersonicConcaveWall} on a concave wall. Note that these
        streaky
        structures seem to persist for some distance into the convex region of the convergent, as also noted by \citet{schneider2008development}. 
        Indeed, the Chu energy associated with the Görtler instability is weakly damped after the concave portion of the convergent, as shown in figure \ref{fig:Streamwise_gortler}. Consequently, the characteristic structures of the instability are gradually eliminated and advected downstream of the inflection point on the convergent wall section.
        Figure \ref{fig:Plot_baseflow_plus_disturbance} also reveals that the structures observed at the wall are alternated between the convergent and divergent sections. This variation is likely attributed to the opposite signs of wall curvature in these two regions.
        Figure \ref{fig:Plot_baseflow_plus_disturbance}, showing the streamwise structures at several slices of the nozzle, clearly reveals the first linear structures commonly identified in Görtler instability studies (\citet{chen2019hypersonicConcaveWall} study on a concave wall and \citet{li2010developmentBAM6QT} study on a portion of the BAM6QT nozzle).
        If the nonlinear system is studied, all these structures then evolve through nonlinearities into mushroom-shaped structures usually associated with Görtler vortices.

        Before detailing the other resolvent mode mechanisms highlighted in figures \ref{fig:OptimalGainMap} and \ref{fig:Streamwise_gortler}, we provide further insight into the two distinct bumps of the optimal gain $\mu_0^2$, which are clearly noticeable around $(f,m)=(10^3 \, \text{Hz}, \, 200)$ and $(f,m)=(10^1 \, \text{Hz}, \, 450)$. These peaks suggest the presence of potential physical phenomena occurring in these two specific regions.
        Given that two regions of the flow can develop centrifugal instabilities, as shown in figure \ref{fig:Rayleigh_discriminant}, the Görtler instability can develop in three possible scenarios: growth of the instability in both $\Delta<0$ regions (the case discussed above), growth of the instability exclusively in the divergent (second $\Delta<0$ region) or in the convergent (first $\Delta<0$ region). The last two possibilities are discussed in the following two paragraphs.
        
        \subsubsection{Görtler instability with frequency}\label{subsubsection:Gortler_with_frequency}
            
            \begin{figure}
                \centering
                \includegraphics[width=0.49\textwidth]{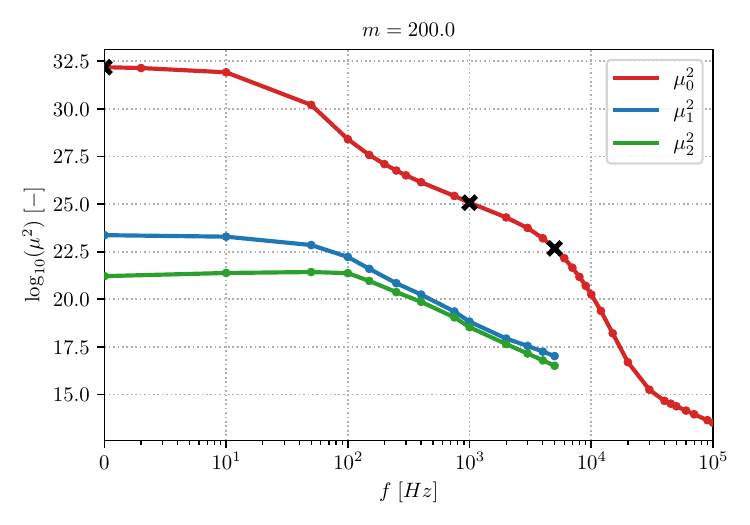}
                \includegraphics[width=0.49\textwidth]{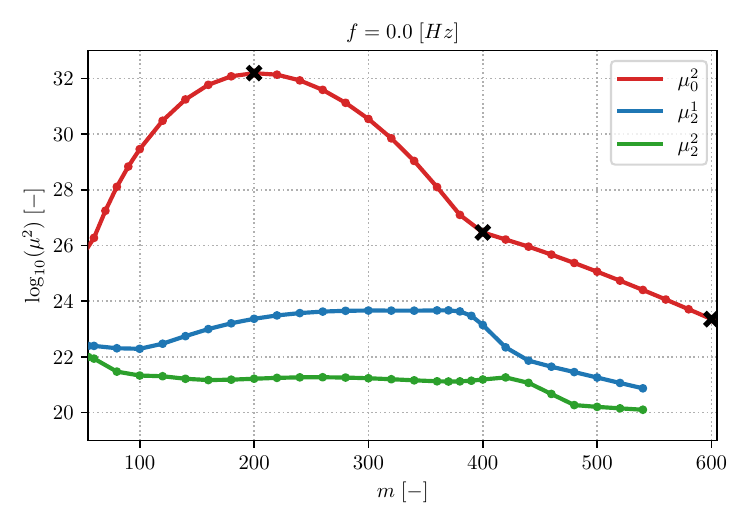}
                \caption{Optimal $\mu_0^2$ and sub-optimal $(\mu_1^2 \; \text{or} \; \mu_2^2)$ gains as a function of: (left) frequency $f$ at a fixed azimuthal wavenumber $m=200$, (right) azimuthal wavenumber $m$ at a fixed frequency $f=0 \; \text{Hz}$. The cross markers ({\includegraphics[width=0.4\baselineskip]{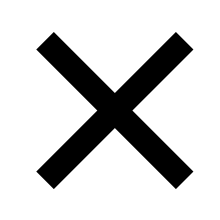}}) indicate the points used for the computations in figures \ref{fig:Streamwise_gortler_m200} and \ref{fig:Streamwise_gortler_f0}.}
                \label{fig:OptimalSuboptimal_GortlerDV}
            \end{figure}
            %
            \begin{figure}
                \vspace{-1.5em}
                \centering
                \includegraphics[width=0.49\textwidth]{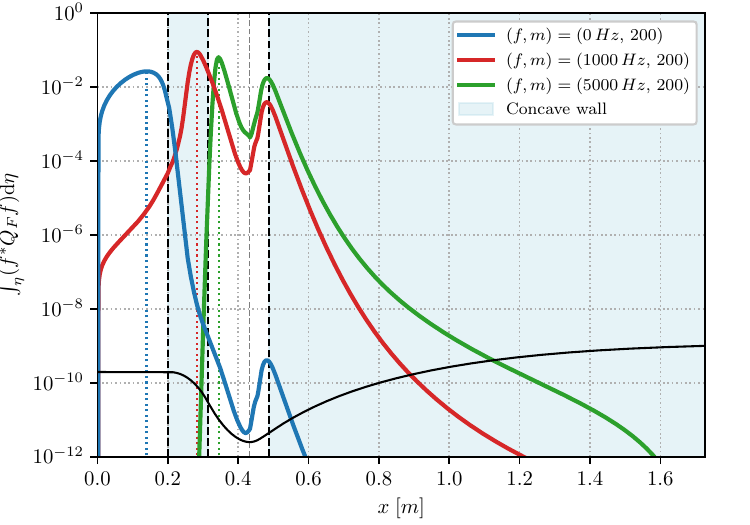}
                \includegraphics[width=0.49\textwidth]{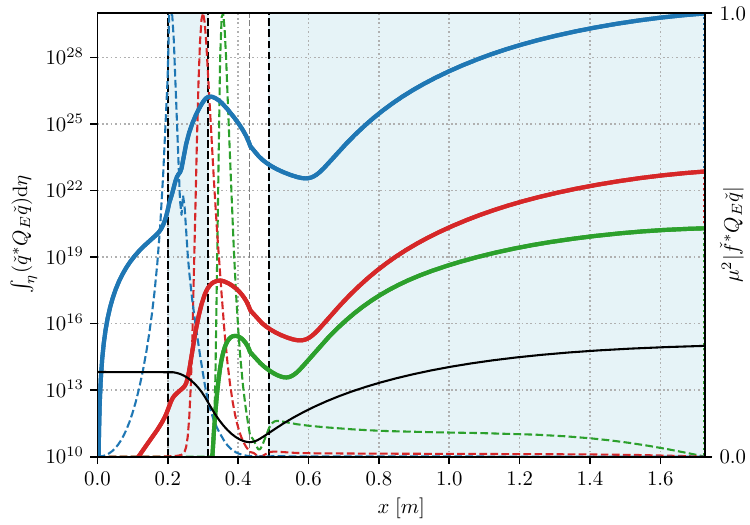}
                \caption{({\includegraphics[width=\baselineskip]{Ligne4.png}}) Evolution of the quantity $E_{Chu}$ at different frequencies $f$ and fixed azimuthal wavenumber $m=200$ of the optimal forcing (left) and response (right) computed along the gridlines in the $r$-direction by integrating the local Chu energy contribution. ({\includegraphics[width=\baselineskip]{Ligne1.png}}) Evolution of the rate of work done by the forcing onto the response mode $\xi = \mu^2 |\check{f}^* Q_E \check{q}|$ computed along the gridlines in the $r$-direction (normalised by the maximum value for each $(f,m)$). Frequencies $f$ used are identified in figure \ref{fig:OptimalSuboptimal_GortlerDV} (left) by the cross markers ({\includegraphics[width=0.4\baselineskip]{Modele_croix.png}}).}
                \label{fig:Streamwise_gortler_m200}
            \end{figure}

            The first bump of the optimal gain $\mu_0^2$, near $(f,m)=(10^3 \, \text{Hz}, \, 200)$, can be detected in the slice of the optimal gain map figure \ref{fig:OptimalSuboptimal_GortlerDV} (left).
            No switch is observed between the optimal gain $\mu_0^2$ and sub-optimal $(\mu_1^2 \; \text{or} \; \mu_2^2)$ gains as frequency increases, indicating that there is no change in the leading mechanism between these two bumps but rather a change in mode behaviour. Therefore, the Görtler instability remains the leading mechanism for these frequencies lower than $f=10 \; \text{kHz}$ and $m=200$.

            Examining the streamwise evolution of $E_{Chu}$ of the forcing and response at different frequencies and fixed $m$, figure \ref{fig:Streamwise_gortler_m200}, it is observed that as the frequency increases the forcing moves downstream in the nozzle. 
            In contrast, the response appears almost unchanged in terms of energy distribution, with only a Chu energy loss in the divergent and almost no energy contribution from the convergent at higher frequencies.
            Thus, as the forcing moves downstream in the nozzle, the region where centrifugal instabilities can develop becomes smaller (see figure \ref{fig:Rayleigh_discriminant}), until the instability can only grow in the divergent part of the nozzle (second region where $\Delta < 0$). This leads to a reduction in the Chu energy of the instability and thus a reduction of the optimal gain, leading to the second bump near $f=10^3 \, \text{Hz}$. Thus, this bump in the optimal gain map $\mu_0^2$ corresponds to the growth of this instability exclusively within the divergent section. 
            This effect will be further validated when studying the restriction of the forcing field in the divergent section (see \S \hspace*{0.01em} \ref{section:forcing_field_restriction}), where a single gain bump is observed (near $log_{10}(\mu_0^2) = 24.3$) matching the gain bump when the Görtler instability grows exclusively within the divergent at higher frequencies. 
            Note that the increase in Chu energy observed in a non-concave region near the throat for the Görtler instability at $(f,m) = (5000 \, \text{Hz}, 200)$ (figure \ref{fig:Streamwise_gortler_m200}) originates from the rate of work done by the forcing onto the response mode $\xi = \mu^2 |\check{f}^* Q_E \check{q}|$, rather than from a contribution of the base-flow (i.e.\ suitable curvature).
            
            

        \subsubsection{Görtler instability with azimuthal wavenumber}\label{subsubsection:Gortler_with_azimuthalwavenumber}

            Next, the second gain bump on the optimal gain map in figure \ref{fig:OptimalGainMap}, located around $(f,m)=(10^1 \, \text{Hz}, \, 450)$, is examined. To better visualise this bump, a slice of the optimal gain map at zero frequency is shown in figure \ref{fig:OptimalSuboptimal_GortlerDV} (right).
            As in the previous case, no change between the optimal and sub-optimal gains is observed, suggesting that there is no change in mode but rather a change in behaviour. This supports that the Görtler instability is still observed here.

            \begin{figure}
                \centering
                \includegraphics[width=0.49\textwidth]{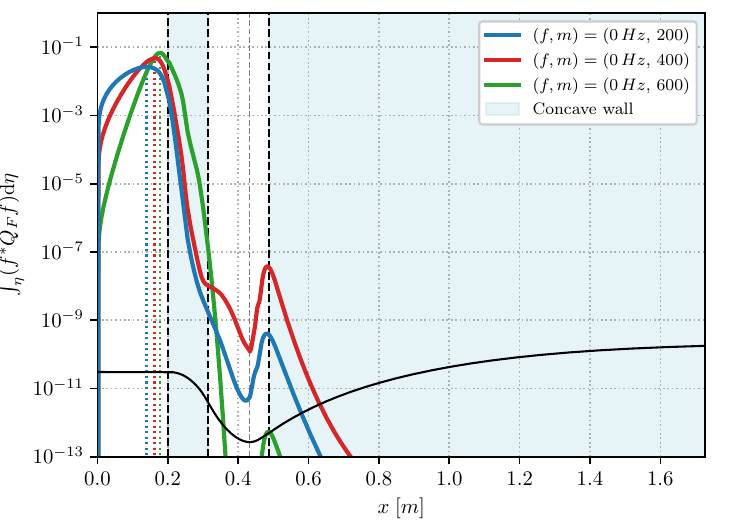}
                \includegraphics[width=0.49\textwidth]{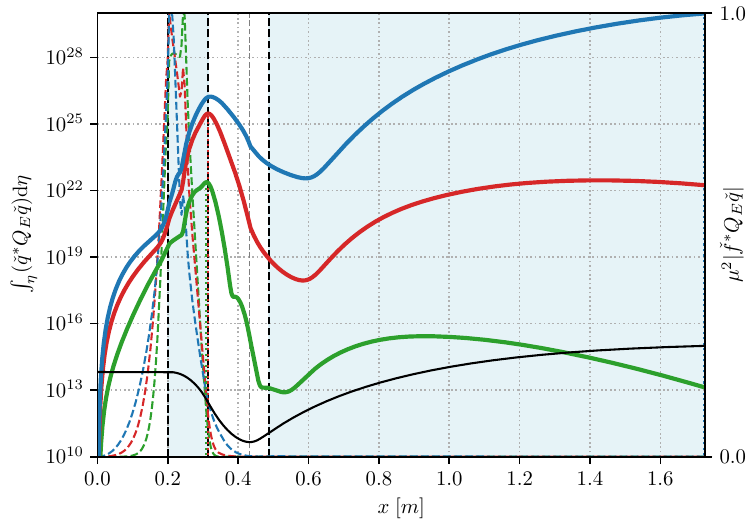}
                \caption{({\includegraphics[width=\baselineskip]{Ligne4.png}}) Evolution of the quantity $E_{Chu}$ at different azimuthal wavenumbers $m$ and fixed frequency $f = 0 \; \text{Hz}$ of the optimal forcing (left) and response (right) computed along the gridlines in the $r$-direction by integrating the local Chu energy contribution. ({\includegraphics[width=\baselineskip]{Ligne1.png}}) Evolution of the rate of work done by the forcing onto the response mode $\xi = \mu^2 |\check{f}^* Q_E \check{q}|$ computed along the gridlines in the $r$-direction (normalised by the maximum value for each $(f,m)$). Azimuthal wavenumbers $m$ used are identified in figure \ref{fig:OptimalSuboptimal_GortlerDV} (right) by the cross markers ({\includegraphics[width=0.4\baselineskip]{Modele_croix.png}}).}
                \label{fig:Streamwise_gortler_f0}
            \end{figure}

            By applying the same approach as in the previous case, but this time examining the streamwise evolution of $E_{Chu}$ as a function of the azimuthal number $m$ at a fixed frequency $f=0 \, \text{Hz}$, as shown in figure \ref{fig:Streamwise_gortler_f0}, it is observed that as $m$ increases, the forcing remains essentially unchanged and localised in the settling chamber, while the response shifts upstream within the nozzle until the maximum of Chu's energy is no longer located in the divergent but rather in the convergent. Specifically focusing on the divergent region, it is also noticeable that the peak disturbance energy moves also upstream, towards the region where the boundary layer is thinner. The following conclusion is drawn: this increase of the azimuthal number $m$ leads the Görtler instability to adapt to the thinner boundary layers localised in the convergent and the beginning of the divergent (figure \ref{fig:Baseflow_wallT_BoundaryLayer}),
            due to the link between the azimuthal wavenumber $m$ and the boundary layer thickness $\delta$. 
            As a result, the instability can almost exclusively grow in the first $\Delta<0$ zone located in the first part of the convergent, with limited growth in the second $\Delta<0$ region as it can no longer adapt to the thicker boundary layer at the end of the divergent section. This adaptation explains the gain reduction observed around $m=400$ and the second bump in the optimal gain map. 

        \subsubsection{Görtler instability conclusion}


            To summarise, two regions of the flow can develop centrifugal instabilities, as depicted in figure \ref{fig:Rayleigh_discriminant}, leading to three scenarios for Görtler instability development. First, the instability grows in both $\Delta<0$ regions (see \S \hspace*{0.01em} \ref{subsection:gortler_instability}). This is the worst-case scenario for the development of this instability. Second, the instability develops exclusively in the divergent section (see \S \hspace*{0.01em} \ref{subsubsection:Gortler_with_frequency}) at higher frequencies, the forcing field shifts downstream, allowing the instability to develop only in the second centrifugal instability zone. Third, the instability grows in the convergent section (see \S \hspace*{0.01em} \ref{subsubsection:Gortler_with_azimuthalwavenumber}), an increase in the azimuthal wavenumber $m$ leads the Görtler instability to adapt to thinner boundary layers, due to the link between the azimuthal wavenumber $m$ and the boundary layer thickness $\delta$, thus preventing its development in the thicker boundary layer at the end of the divergent section. These three different scenarios manifest as three distinct bumps on the optimal gain map.

    \subsection{Second Mack mode instability}

        \begin{figure}
            \centering
            \includegraphics[width=0.49\textwidth]{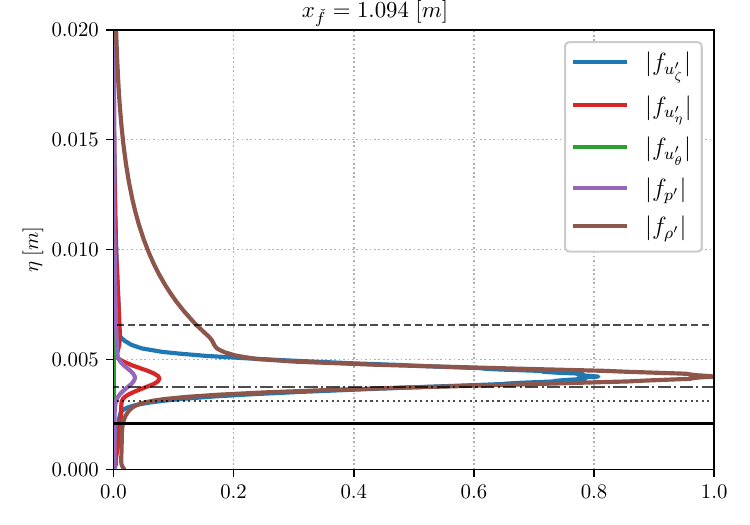}
            \includegraphics[width=0.49\textwidth]{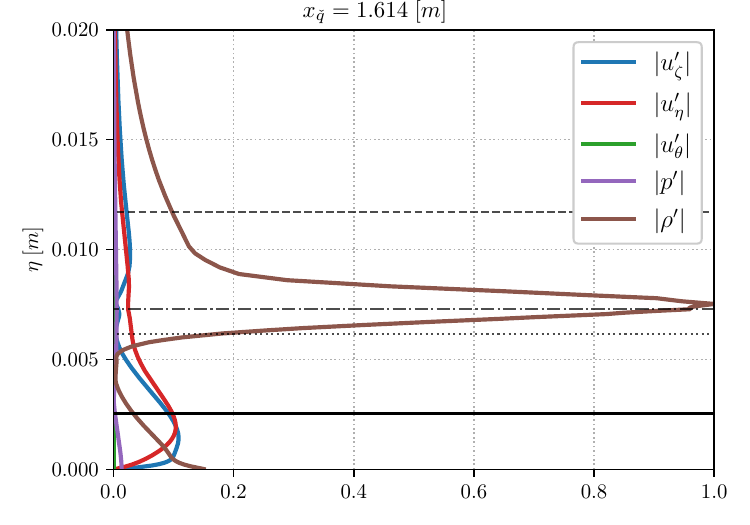}
            \caption{Second Mack Mode $(f,m) = (45 \, \text{kHz}, 0)$: eigenfunctions at the maximum Chu energy. Left: Forcing. Right: Response. ({\includegraphics[width=\baselineskip]{Ligne1.png}}): boundary layer thickness $\delta$. ({\includegraphics[width=\baselineskip]{Ligne2.png}}): displacement thickness $\delta^*$. ({\includegraphics[width=\baselineskip]{Ligne3.png}}): generalised inflection point. ({\includegraphics[width=\baselineskip]{Ligne4.png}}): $\hat{M} = -1$, the supersonic region is located below this line.}
            \label{fig:Eigenfunctions_2ndmode}
        \end{figure}
        %
        \begin{figure}
            \vspace{-1.0em}
            \centering
            \includegraphics[width=0.95\textwidth]{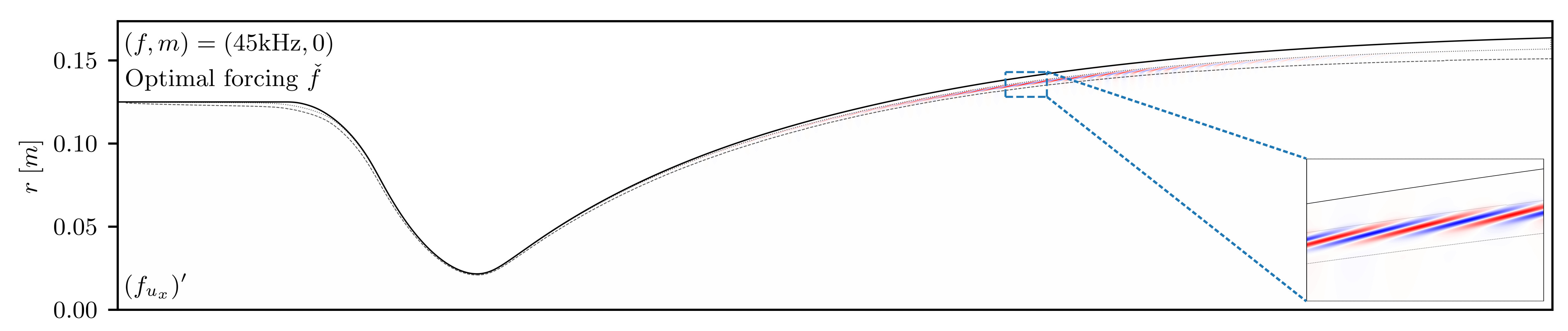}
            \includegraphics[width=0.95\textwidth]{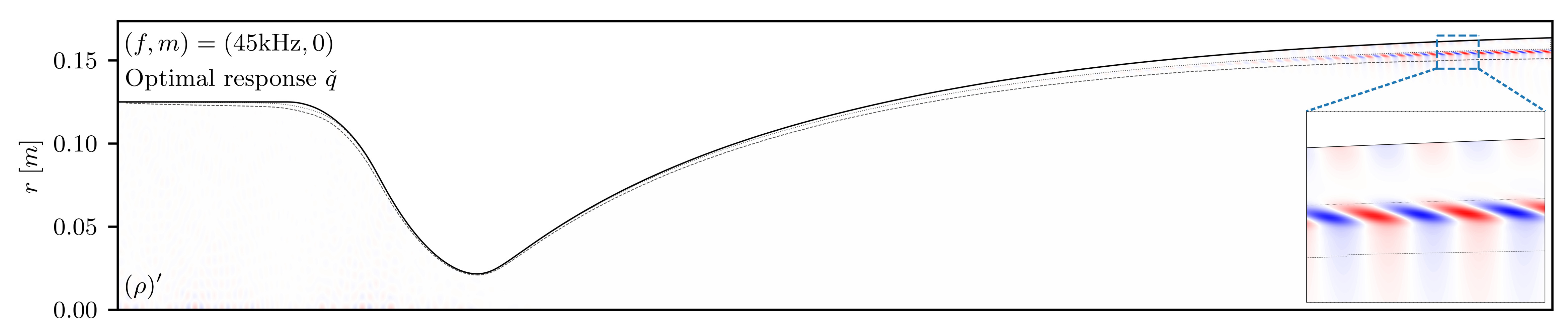}
            \includegraphics[width=0.95\textwidth]{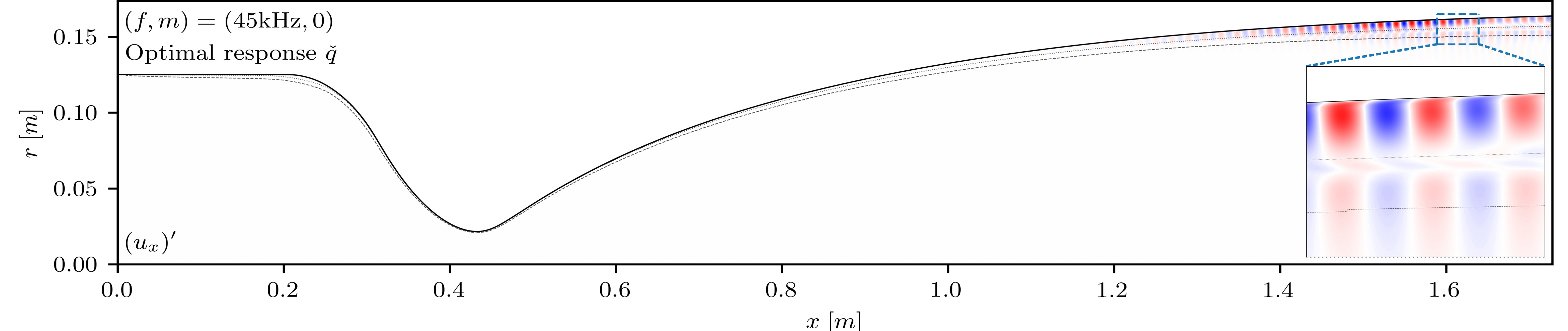}
            \caption{Second Mack Mode $(f,m) = (45 \, \text{kHz}, 0)$: eigenfunctions field in the nozzle. Top: streamwise $f_{u_{x}'}$ optimal forcing. Middle: density $\rho'$ optimal response. Bottom: streamwise $u_{x}'$ optimal response. Zoom: on the area near maximum Chu energy for the forcing and the response. ({\includegraphics[width=\baselineskip]{Ligne1.png}}): boundary layer thickness $\delta$. ({\includegraphics[width=\baselineskip]{Ligne2.png}}): displacement thickness $\delta^*$.}
            \label{fig:Eigenfunctions_2ndmode_plot2d}
        \end{figure}

        Having identified and understood the dominant instability in our flow, attention can now turn to the other resolvent modes, specifically the high-frequency peak observed on the optimal gain map in figure \ref{fig:OptimalGainMap}. This instability, peaking at $m=0$ (i.e.\ axisymmetric) and around $f = 45 \; \text{kHz}$, corresponds to the second Mack mode. The estimated frequency, determined using the Resolvent analysis, aligns with the estimation provided by \citet{mack1984boundary} where $f = U_e/2\delta$. 
        The second Mack mode is the least amplified instability compared to others examined in this study, which is consistent with findings from previous Mach 6 wind tunnel investigations \citep{schneiderENMethode, lakebrink2018optimization, schneider2001shakedown}. 
        Indeed, this mode is known to develop above $M \approx 4$ \citep{mack1984boundary}, a condition reached only in a limited portion of the divergent section due to the short length of the studied nozzle.




        The eigenfunctions obtained at the maximum of the Chu energy for the second Mack mode (figure \ref{fig:Streamwise_gortler}) are presented in figure \ref{fig:Eigenfunctions_2ndmode}. The maximum forcing and response are located in the divergent section. The shape of these eigenfunctions and their interpretation are consistent with the results by \citet{bugeat20193d} on a flat plate.
        To compute the relative Mach number $\hat{M}$ the method proposed by \citet{bugeat20193d} is employed. Supposing a wavelike structure of the perturbation fields $\check{q}(x,r) = \tilde{q}(r) \, e^{i \alpha x} = \tilde{q}(r) \, e^{i \alpha_r x} \, e^{- \alpha_i x}$, thus the streamwise wavenumber $\alpha_r$ can be computed using $\check{q} = \check{u}(x, r=\delta^{*})$.
        Once the streamwise wavenumber is calculated, the phase velocity $c_{\phi}$ of the wave can be determined, enabling the computation of the relative Mach number $\hat{M} = (\overline{u} - c_{\phi})/\overline{c}$, where $\overline{c}$ is the base-flow sound speed. Thus, a supersonic region of relative Mach number
        $\hat{M}<-1$ is detected close to the wall for both forcing and response fields, which is a condition for the existence of additional unstable modes according to \citet{mack1984boundary}.
        
        The optimal response fluctuations $\check{q}$ grow within this near-wall supersonic region, while the density is also amplified near the generalised inflection point. This property of the density is also observed in other instabilities and in other studies \citep{hanifi1996transient, bugeat20193d}. Moreover, the optimal forcing disturbances $\check{f}$ are not localised in this supersonic region but are instead concentrated along the generalised inflection point. 
        \citet{bugeat20193d} concluded that it seems that there are two distinct mechanisms coexisting. The first mechanism is specific to the second Mack mode and corresponds to the growth of hydrodynamic and thermodynamic disturbances in the supersonic region. The second is an inflectional mechanism, where thermodynamic disturbances are also amplified along the generalised inflection point.

        The axisymmetric fluctuation structures of the second Mack mode, illustrated in figure \ref{fig:Eigenfunctions_2ndmode_plot2d}, align with the linear stability results from the Resolvent analyses on a flat plate \citep{poulain2023broadcast, bugeat20193d}, as well as with the results comparing LST and DNS obtained by \citet{egorov2006direct, egorov2007direct, unnikrishnan2020linear}, showing the classic two-lobed structure of second-mode instability. 
        Moreover, figure \ref{fig:Streamwise_gortler} reveals acoustic disturbance structures within the subsonic region of the nozzle (oscillating/noisy curve in the subsonic region). This phenomenon, observed only for the second Mack mode, results in a slight contribution of Chu energy (figure \ref{fig:Streamwise_gortler}) for both the forcing and the response. This energy contribution is relatively small compared to the energy in the divergent section. Consequently, this study has not identified any significant synchronisation between acoustic mode resonances in the subsonic region and the second Mack mode.
        
    \subsection{First Mack mode instability}

\begin{figure}
            \centering
            \includegraphics[width=0.49\textwidth]{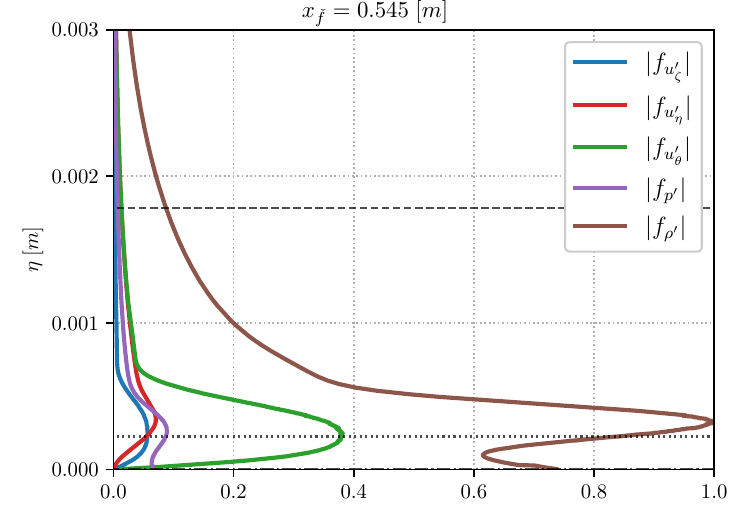}
            \includegraphics[width=0.49\textwidth]{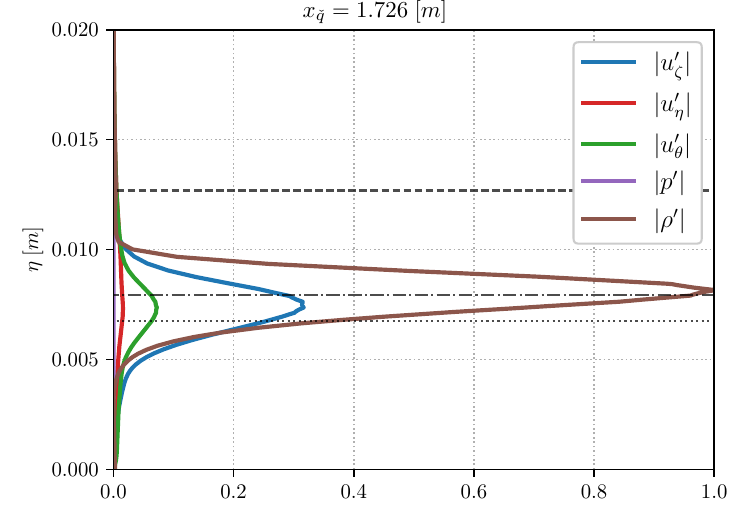}
            \caption{First Mack Mode $(f,m) = (12 \, \text{kHz}, 50)$: eigenfunctions at the maximum Chu energy. Left: Forcing. Right: Response. ({\includegraphics[width=\baselineskip]{Ligne1.png}}): boundary layer thickness $\delta$. ({\includegraphics[width=\baselineskip]{Ligne2.png}}): displacement thickness $\delta^*$. ({\includegraphics[width=\baselineskip]{Ligne3.png}}): generalised inflection point.}
            \label{fig:Eigenfunctions_1stmode}
        \end{figure}

\begin{figure}
            \vspace{-1.0em}
            \centering
            \includegraphics[width=0.95\textwidth]{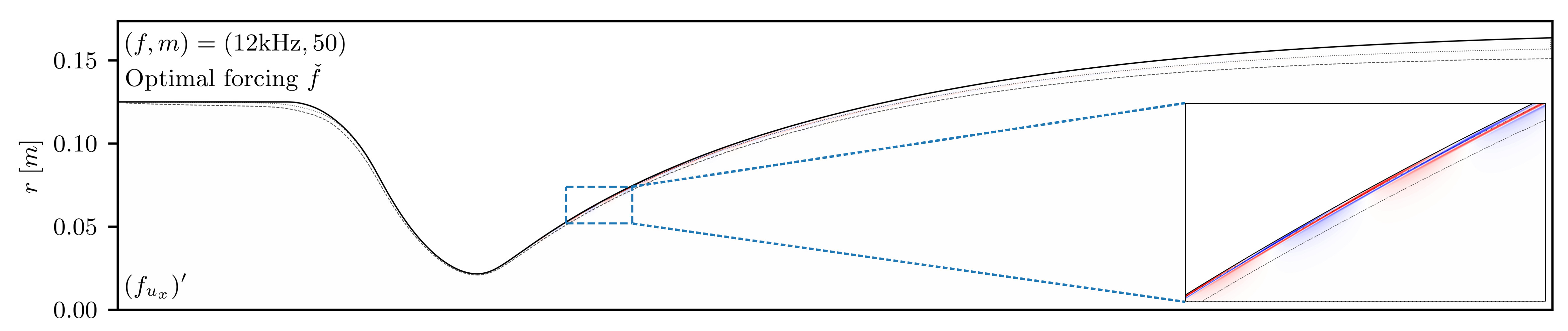}
            \includegraphics[width=0.95\textwidth]{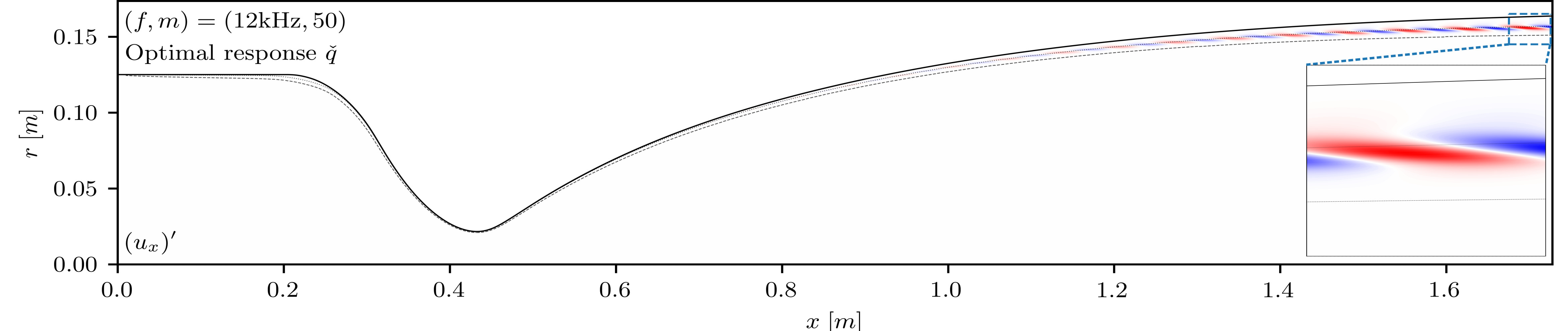}
            \caption{First Mack Mode $(f,m) = (12 \, \text{kHz}, 50)$: eigenfunctions field in the nozzle. Top: streamwise $f_{u'_{x}}$ optimal forcing. Bottom: streamwise $u_{x}'$ optimal response. Zoom on the area near maximum Chu energy for the forcing and the response. ({\includegraphics[width=\baselineskip]{Ligne1.png}}): boundary layer thickness $\delta$. ({\includegraphics[width=\baselineskip]{Ligne2.png}}): displacement thickness $\delta^*$.}
            \label{fig:Eigenfunctions_1ermode_plot2d}
        \end{figure}

        Another gain peak is observed around $(f,m)=(12 \, \text{kHz}, \, 40)$. Given that both the frequency and the azimuthal wavenumber are non-zero, this corresponds to an oblique mode. This instability matches the characteristics of the first Mack mode. The eigenfunctions for this first Mack mode, extracted at the peak Chu energy, are shown in figure \ref{fig:Eigenfunctions_1stmode}. The maximum forcing and response are located in the divergent section in figure \ref{fig:Streamwise_gortler}. These eigenfunctions are consistent with the findings from the flat plate study using the Resolvent analysis by \citet{bugeat20193d}. The forcing exhibits a dominant transverse component, whereas the response is mainly longitudinal with a streamwise velocity response. The optimal response is localised near the generalised inflection point of the boundary layer, which is characteristic of the first Mack mode \citep{mack1984boundary}. Figure \ref{fig:Eigenfunctions_1ermode_plot2d} highlights a reorientation of the disturbance structures between the forcing and the response fields, interpreted as the action of the non-modal Orr mechanism. The numerical estimation of the oblique angle of the perturbation is approximately $\Psi \approx 73$\textdegree, which is consistent with commonly observed waves angles for this instability \citep{mack1984boundary}.


    \subsection{Kelvin-Helmholtz instability}

\begin{figure}
            \centering
            \includegraphics[width=\textwidth]{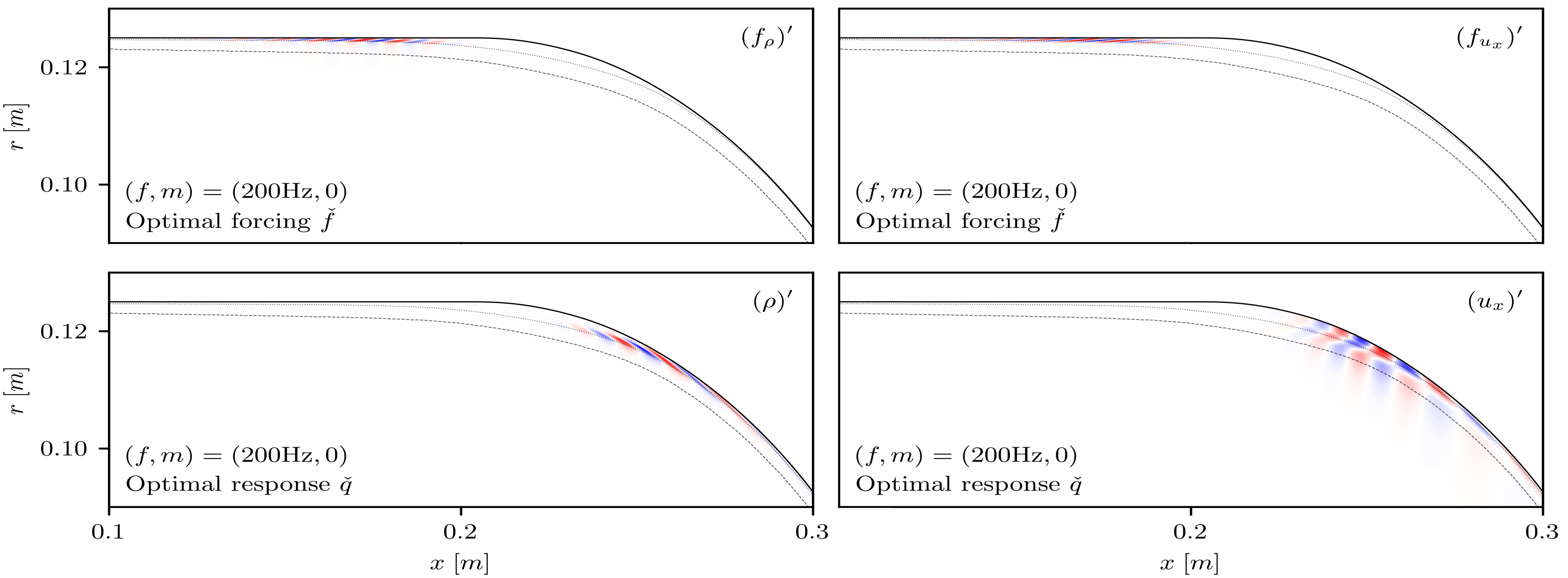}
            \caption{Kelvin-Helmholtz instability, shear mode due to the recirculation bubble, $(f,m) = (200 \, \text{Hz}, 0)$: eigenfunctions field near the convergent inlet. Top: density $f_{\rho'}$ and streamwise $f_{u'_{x}}$ optimal forcing. Bottom: density $\rho'$ and streamwise $u_{x}'$ optimal response. ({\includegraphics[width=\baselineskip]{Ligne1.png}}): boundary layer thickness $\delta$. ({\includegraphics[width=\baselineskip]{Ligne2.png}}): displacement thickness $\delta^*$.}
            \label{fig:Eigenfunctions_KH_plot2d}
        \end{figure}

        The last mechanism identified in this nozzle is a low-frequency instability at $f = 200 \; \text{Hz}$, which peaks at zero azimuthal wavenumber for the isothermal case. This instability is attributed to the Kelvin-Helmholtz instability arising from the recirculation bubble at the convergent inlet. Figure \ref{fig:Streamwise_gortler} shows that the maximum optimal forcing occurs in the settling chamber just before the boundary layer separation, while the response is located in the convergent section immediately downstream of the recirculation bubble. Examination of the disturbance structures associated with this mode, figure \ref{fig:Eigenfunctions_KH_plot2d}, reveals that the instability emanates from the recirculation bubble 
        displaying the characteristic structures of the Kelvin-Helmholtz instability. 
        The results obtained regarding the linear structures of the Kelvin-Helmholtz instability, which emerge from a recirculation bubble, are consistent with studies using resolvent analysis in backward-facing step flows \citep{dergham2013stochastic, barbagallo2012closed} and laminar separation bubbles around airfoils \citep{thomareis2018resolvent}.

        This low-frequency instability observed here may explain the low-frequency instability that persisted in the Mach-6 NASA Langley nozzle \citep{wilkinson1997review}, because as noticed by \citet{schneider2001shakedown} the separation bubble that could appear in wind tunnel contractions are usually unstable and thus would transmit noise downstream. 



\vspace{-1.0em}
\section{Isothermal and adiabatic walls effects on resolvent modes}\label{section:Comp_isoT_adiaT}

    In this section, linear stability results obtained through resolvent analysis are compared for the isothermal and adiabatic cases. Such a comparison is valuable because the wall temperature distribution, and consequently the boundary layer properties, differ between the two cases (see figures \ref{fig:Baseflow_wallT_BoundaryLayer} and \ref{fig:BL_comp}). This analysis highlights the significant role of wall temperature in the development of instabilities.

\begin{figure}
        \centering
        \includegraphics[width=\textwidth]{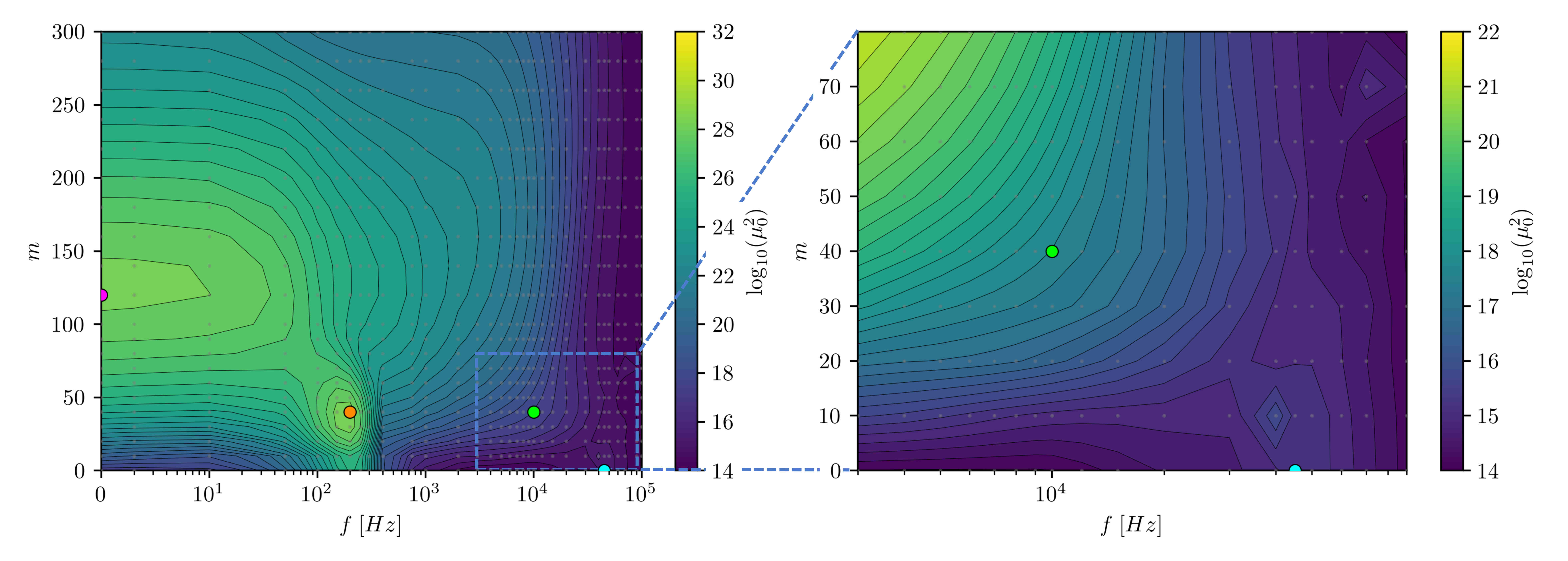}
        \caption{Optimal gain map $\mu_0^2$ in frequency $f$ and azimuthal wavenumber $m$ space in the adiabatic case. The right plot shows the optimal gain map zoom on the area near the first and second Mack modes. The grey dots ({\includegraphics[width=0.4\baselineskip]{Modele_rond5.png}}) show the different computation locations. ({\includegraphics[width=0.4\baselineskip]{Modele_rond4.png}}): Görtler instability peak. ({\includegraphics[width=0.4\baselineskip]{Modele_rond3.png}}): Kelvin-Helmholtz instability peak. ({\includegraphics[width=0.4\baselineskip]{Modele_rond2.png}}): first Mack mode peak. ({\includegraphics[width=0.4\baselineskip]{Modele_rond1.png}}): second Mack mode peak.}
        \label{fig:OptimalGainMap_adiaT}
    \end{figure}

    By comparing the two gain maps in Figures \ref{fig:OptimalGainMap} and \ref{fig:OptimalGainMap_adiaT}, differences in the amplification of various instabilities can be observed between the two wall boundary conditions. When an adiabatic boundary condition is applied, the optimal gain of all instabilities decreases (see tables \ref{tab:Instability_Tiso} and \ref{tab:Instability_Tadia}), however this condition is less representative of the physical reality of the nozzle wall compared to an isothermal wall \citep{schneider1995quiet}. 
    This reduction in flow instability when the wall temperature is higher at the throat and cooled downstream ($T_{wall,adiabatic} > T_{wall,isothermal}$ and $T_{wall,adia}$ gradually "cools" along the nozzle wall, see figure \ref{fig:Baseflow_wallT_BoundaryLayer}) has also been noted by \citet{schneiderENMethode}.
    This behaviour is particularly interesting because previous studies on flat plates by \citet{mack1993effect} have shown that wall heating and cooling influence Mack’s modes: a uniformly cooled wall stabilises the first Mack mode but destabilizes the second. 
    Here, a reduction in the second Mack mode is also observed, thus its behaviour in nozzle flow remains to be fully explained, as noted by \citet{schneiderENMethode}.
    Therefore, this highlights the importance of studying wall temperature effects in nozzle flows for delaying boundary layer transition by limiting the growth of instabilities. Such considerations are crucial for the development of quiet hypersonic wind tunnels.
    This aspect has been explored by \citet{schneiderENMethode} by increasing the throat temperature while decreasing the wall temperature downstream to reduce the growth of instabilities (noted that throat heating is essential for proper nozzle operation \citep{demetriades1996stabilization}). More recently, optimisation studies for the Notre Dame large hypersonic quiet tunnel \citep{lakebrink2018optimization} found that stabilizing the T-S and the second Mack mode through optimisation of the wall temperature could reduce the $N_{tot}$ factors envelope by as much as $1.5$. 
    
    {\renewcommand{\arraystretch}{1.25} 
    {\setlength{\tabcolsep}{0.18cm}
        \begin{table}
            \caption{\label{tab:Instability_Tadia} Resolvent modes in the adiabatic case.}
            \centering
            \begin{tabular}{c|cccc}
                \hline
                Instability & frequency $f$ [Hz] & Wavenumber $m$ & Optimal gain $\log_{10}(\mu_0^2)$ & $\frac{\log_{10}(\mu_0^2)_{\text{adiabatic}}}{ \log_{10}(\mu_0^2)_{\text{isothermal}}}$  \\ 
                \hline
                Görtler & 0 & 120 & 28.12 & 0.87 \\ 
                Kelvin-Helmholtz & 200 & 40 & 28.35 & 0.95 \\ 
                First Mack mode & $\approx$ 10 000 & $\approx$ 40 & $\approx$ 17.61 & $\approx$ 0.92 \\ 
                Second Mack mode & 45 000 & 0 & 15.26 & 0.94 \\ 
                \hline
            \end{tabular}
    \end{table}}}
    
    For frequencies $f$ and azimuthal wavenumbers $m$ where instabilities are amplified, the following observations can be made. For the Görtler instability, the peak amplification occurs at zero frequency (stationary) under both boundary conditions. However, it does not appear at the same azimuthal wavenumber $m$ due to differences in boundary layer thickness (figures \ref{fig:Baseflow_wallT_BoundaryLayer} and \ref{fig:BL_comp}). The Kelvin-Helmholtz instability resulting from the recirculation bubble occurs within the same frequency range for both cases. Nevertheless, a three-dimensionalisation of the instability is observed in the adiabatic case, whereas it remains strictly two-dimensional $(m=0)$ in the isothermal case. The frequency ranges for the first and second Mack modes do not show significant differences between the two boundary conditions. However, the first Mack mode is less easily identifiable in the adiabatic case as the instability is more reduced in gain and the Görtler instability becomes more amplified at low azimuthal wavenumbers (compared to the first mode), thus covering a larger portion of the amplification region associated with the first Mack mode.

\begin{figure}
        \centering
        \includegraphics[width=0.49\textwidth]{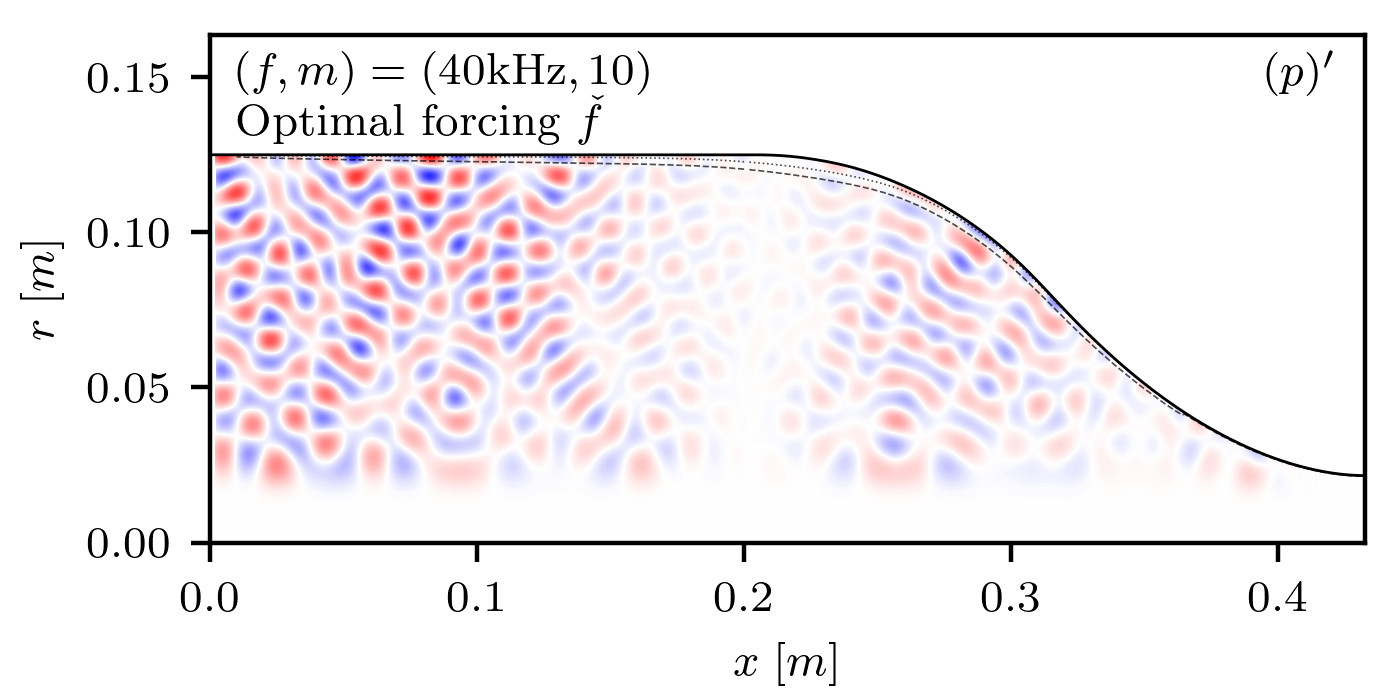}
        \includegraphics[width=0.49\textwidth]{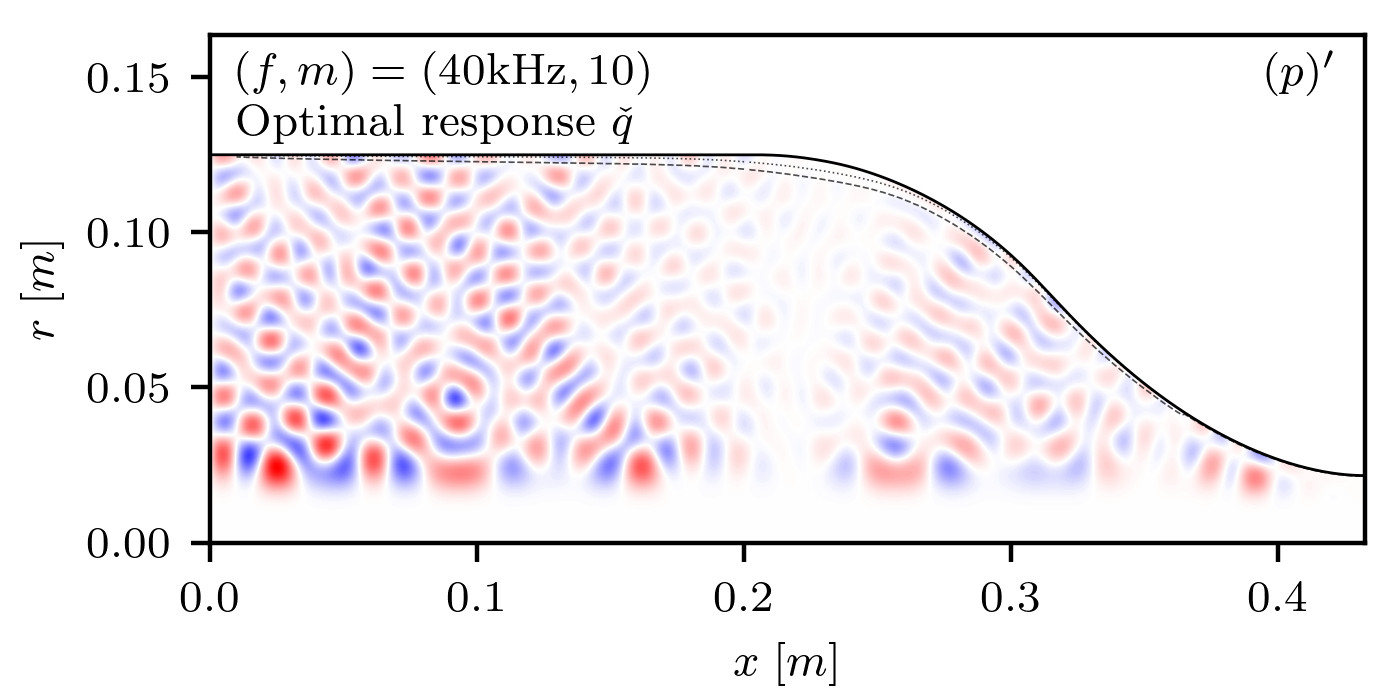}
        \caption{Example of disturbance $p'$ associated with an acoustic mode in the adiabatic case, $(f,m) = (40\; \text{kHz}, \, 10)$.}
        \label{fig:Resolvent_acoustic_mode}
    \end{figure}

    The gain peaks observed in the adiabatic case at $(f,m)=(40 \; \text{kHz}, 10)$ (close to the second Mack mode) and also at $(f,m)=(70 \; \text{kHz}, 70)$ (the second peak is also observable in the isothermal case), correspond to acoustic modes that can exist in the settling chamber and convergent regions due to the confined subsonic flow. These modes are illustrated in figure \ref{fig:Resolvent_acoustic_mode}.
    Typically, such acoustic modes may appear in various locations on the optimal gain map. However, their detection has been limited in this study due to the Chu energy restriction near the wall to only study resolvent modes that develop within the boundary layer. Further investigation is required to determine whether these acoustic modes play a significant role in the transition process. However, this study has not identified any dominant synchronisation between the acoustic mode resonances in the subsonic region and the boundary layer modes (see figure \ref{fig:Streamwise_gortler}). 
    Moreover, given their low gain it can be assumed that they will exhibit only limited growth.

    For the sake of conciseness, the details of the various resolvent modes observed in the adiabatic case have not been presented.
    The only differences between the two boundary conditions would lie in the optimal gain value, the size, and the periodicity of the instability fluctuation structures. 
    However, these variations would not alter the observations made previously in the isothermal case.



\vspace{-1.0em}
\section{Forcing field restricted to the divergent section only}\label{section:forcing_field_restriction}

    This section aims to model the influence of a boundary layer suction slot.
    Such a suction slot, present in all quiet hypersonic wind tunnel nozzles, is used to bleed off the instability or the turbulent boundary layer that may develop in the settling chamber and/or convergent parts of the nozzle. It is an extremely effective device that has enabled the achievement of quiet wind tunnels \citep{schneider2008development, beckwith1975development, chen1993gortler}. If the boundary layer suction slot upstream of the throat is closed, the flow returns to 
    a noisy state
    \citep{chen1993gortler} (see also Fig.\ 3 in \citet{liu2024experimental} for a clear experimental visualization, which highlights the crucial role of this suction slot on the boundary layer transition). This effect is illustrated by examining the TSP results in the experimental studies in quiet (bleed slot open) and noisy (bleed slot closed) wind tunnels by \citet{durant2015mach} (Fig.\ 12) and \citet{mckiernan2021boundary} (Figs.\ 12 and 13), with an early transition on the model when the bleed slot is closed indicating a noisy freestream flow.

    The restriction matrix $P$ in the eigenvalue problem (\ref{eq:eigenvalueproblem_resolvent}) is used to model this suction slot. The forcing field is restricted to the divergent section of the nozzle (from the throat to the outlet), meaning that all disturbances originating from the settling chamber and/or the convergent parts are eliminated, which is precisely the goal of the suction slot upstream of the throat. This setup is a simplified model as in reality the suction slot would generate a new boundary layer, which is not taken into account in our case. However, due to the thin boundary layer in the throat area, resulting from the strong contraction, this assumption is not far from physical reality. Moreover, this simple model also eliminates the necessity of including the suction lip in the simulation with all the complexities it involves \citep{taskinoglu2005numerical, benay2004design}: boundary layer separation, meshing, etc.\ It is therefore a relatively simple model to implement,
    allowing us to draw conclusions.

    \begin{figure}
        \centering
        \subfigure[Isothermal case restricted to the divergent section.]
        {\includegraphics[width=0.49\textwidth]{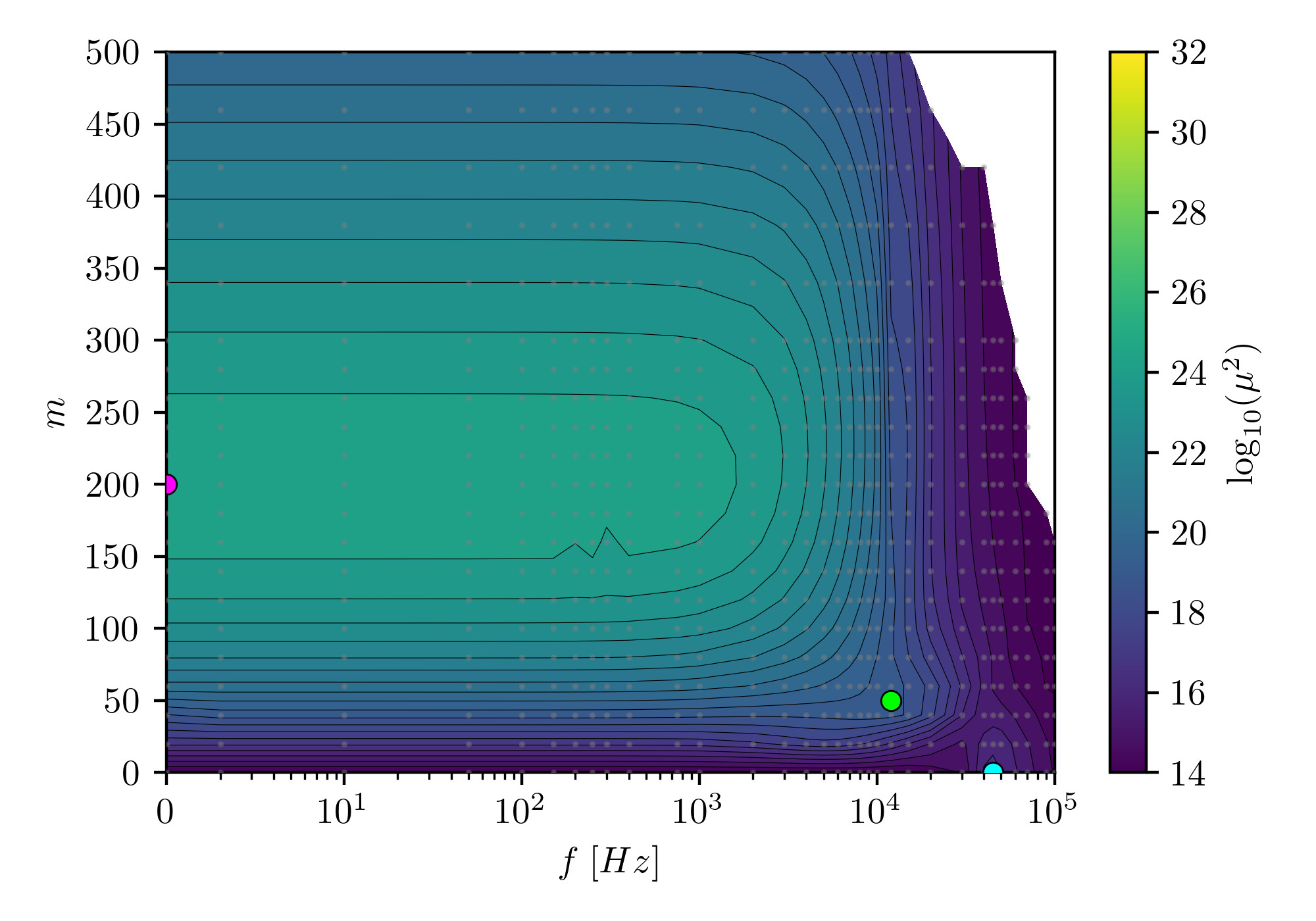}}
        \subfigure[Difference in the optimal gain with and without restrictions.]
        {\includegraphics[width=0.49\textwidth]{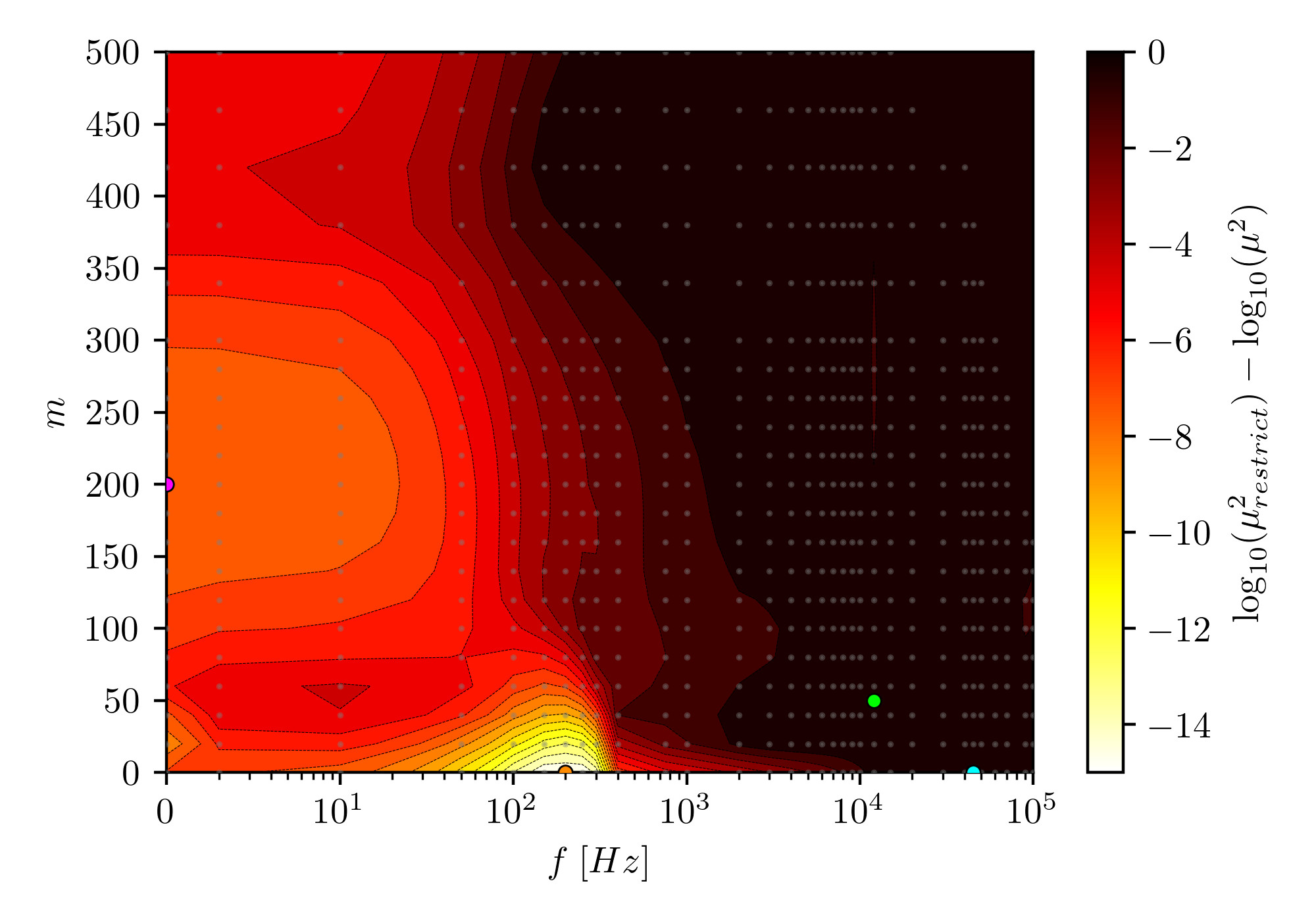}}
        \caption{Left: Optimal gain map $\mu_0^2$ in frequency $f$ and azimuthal wavenumber $m$ space with the forcing field restricted to the divergent section. Right: Difference between the optimal gain map with the forcing field restricted to the divergent section and the optimal gain map in figure \ref{fig:OptimalGainMap} without restriction of the forcing field, in the isothermal case. The grey dots ({\includegraphics[width=0.4\baselineskip]{Modele_rond5.png}}) show the different computation locations. ({\includegraphics[width=0.4\baselineskip]{Modele_rond4.png}}): Görtler instability peak. ({\includegraphics[width=0.4\baselineskip]{Modele_rond3.png}}): Kelvin-Helmholtz instability peak. ({\includegraphics[width=0.4\baselineskip]{Modele_rond2.png}}): first Mack mode peak. ({\includegraphics[width=0.4\baselineskip]{Modele_rond1.png}}): second Mack mode peak.}
        \label{fig:OptimalGainMap_restrict}
    \end{figure}
    \begin{figure}
        \vspace{-1.5em}
        \centering
        \includegraphics[width=0.49\textwidth]{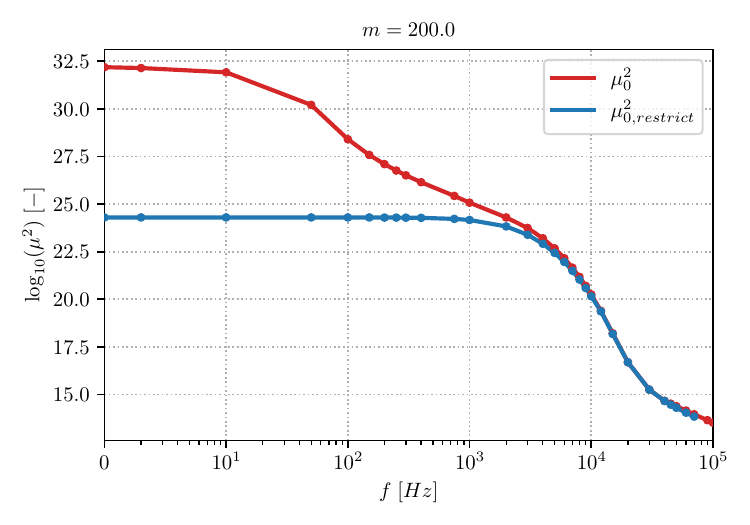}
        \includegraphics[width=0.49\textwidth]{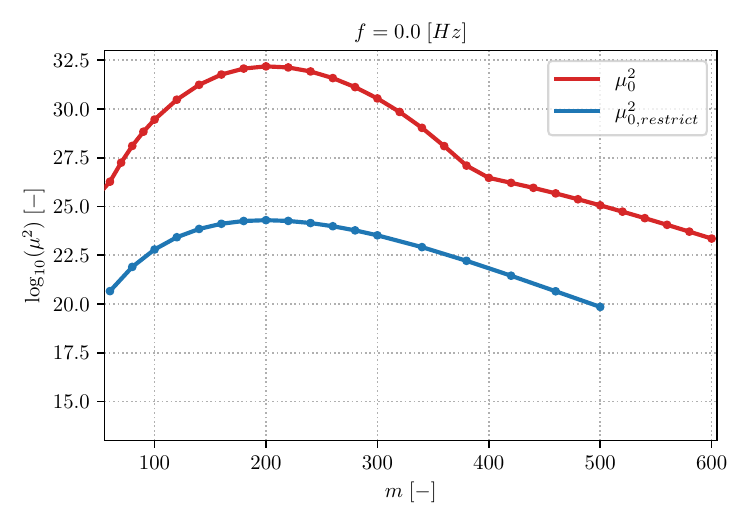}
        \caption{Comparison of the optimal gain at $f=0 \; \text{Hz}$ and $m=200$, with and without forcing field restriction, in the isothermal case.}
        \label{fig:OptimalGainMap_restrict_f0_m200}
    \end{figure}


    The restricted gain map, figure \ref{fig:OptimalGainMap_restrict}, shows significant differences compared to the unrestricted case in figure \ref{fig:OptimalGainMap} (a). With this restriction, the Görtler instability now only displays a single bump on the gain map. It corresponds to the growth of the Görtler instability exclusively in the divergent section of the nozzle, as the instability can now develop only within a single negative Rayleigh discriminant zone ($\Delta < 0$) located in the divergent section (see figure \ref{fig:Rayleigh_discriminant}). This result aligns more closely with those obtained through Resolvent analysis by 
    \citet{hao2023response, cao2023stability}. 
    These studies displayed only a single bump in their optimal gain maps for the Görtler instability
    because in these different cases there is only one zone of $\Delta < 0$. 
    This behaviour had already been observed earlier during the study of Görtler instability variations with frequency (see \S \hspace*{0.01em} \ref{subsubsection:Gortler_with_frequency}). 
    Figure \ref{fig:OptimalGainMap_restrict_f0_m200} (a) illustrates that the new single gain bump obtained aligns with the second bump identified in the unrestricted case when the frequency increases (near $log_{10}(\mu_0^2) = 24.3$), thereby supporting earlier observations.
    Additionally, figure \ref{fig:OptimalGainMap_restrict_f0_m200} (b) shows the decrease in optimal gain with the azimuthal wavenumber $m$, already identified earlier, as the instability scales to the thinner boundary layer located upstream in the nozzle.


    Figure \ref{fig:OptimalGainMap_restrict} (b), which compares optimal gains with and without the restriction, illustrates the impact of the convergent section on the development of resolvent modes. The Görtler instability exhibits non-negligible growth in the convergent section. The Kelvin-Helmholtz instability, associated to the recirculation bubble, has completely disappeared in the restricted case, as expected since the restriction eliminates all disturbances originating from the settling chamber or convergent section. However, for the first and second Mack modes no difference in optimal gain is observed, which is not surprising as the forcing and response fields for these instabilities are located mainly in the divergent section, figure \ref{fig:Streamwise_gortler}. 

    This simple model highlights the crucial role of a boundary layer suction slot upstream of the nozzle throat, as a significant difference in optimal gain is observed between the cases with and without forcing field restriction. 
    When effectively implemented, such a suction slot can eliminate all instabilities originating from the settling chamber or convergent section. By inhibiting these instabilities, this device effectively delays the growth of instabilities within the boundary layer, thereby postponing the onset of a turbulent boundary layer on the nozzle wall. 



\vspace{-1.0em}
\section{Conclusion and future works}

    \begin{figure}
        \centering
        \subfigure[Isothermal case.]
        {\includegraphics[width=0.49\textwidth]{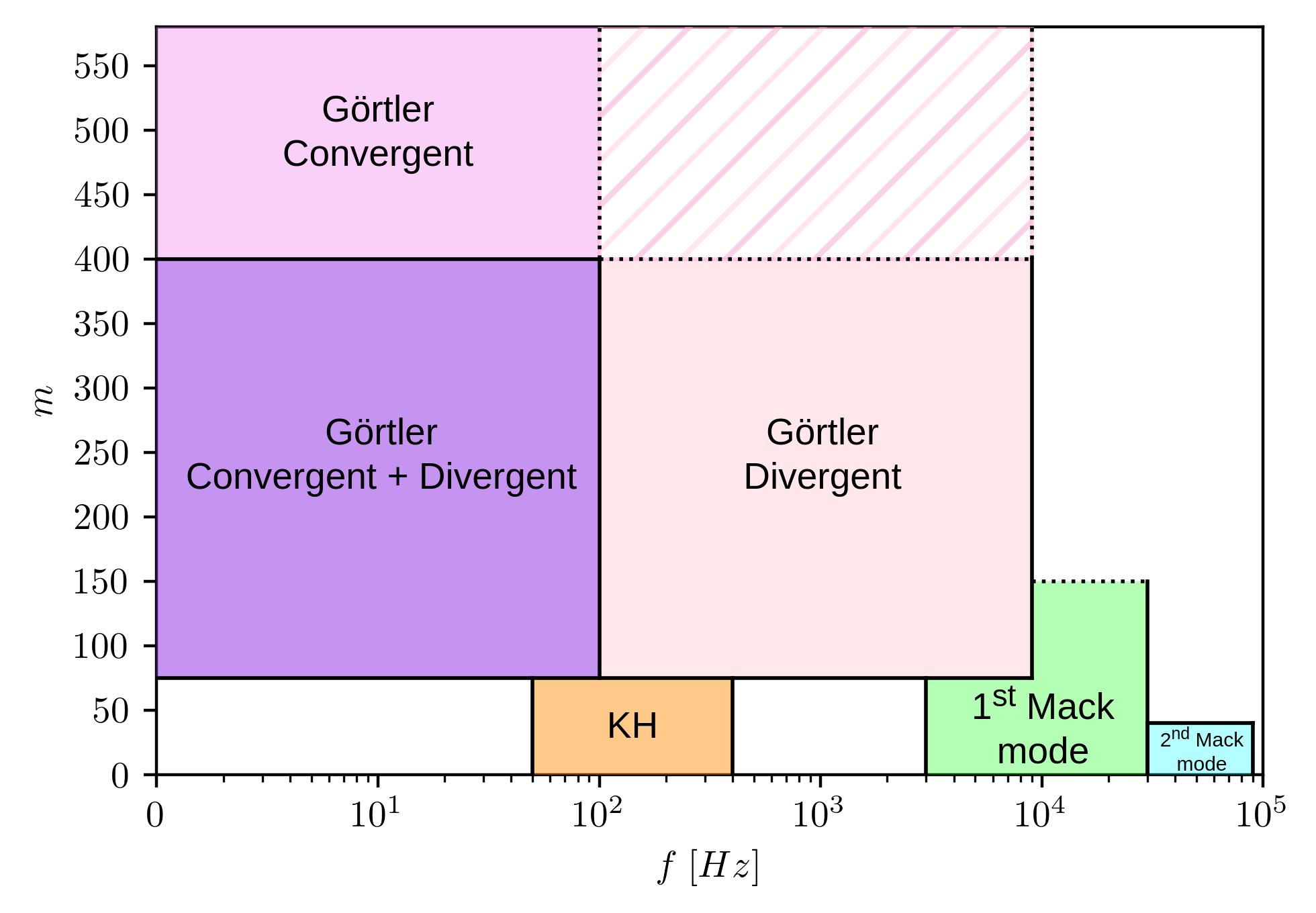}}
        \subfigure[Isothermal case restricted to the divergent.]
        {\includegraphics[width=0.49\textwidth]{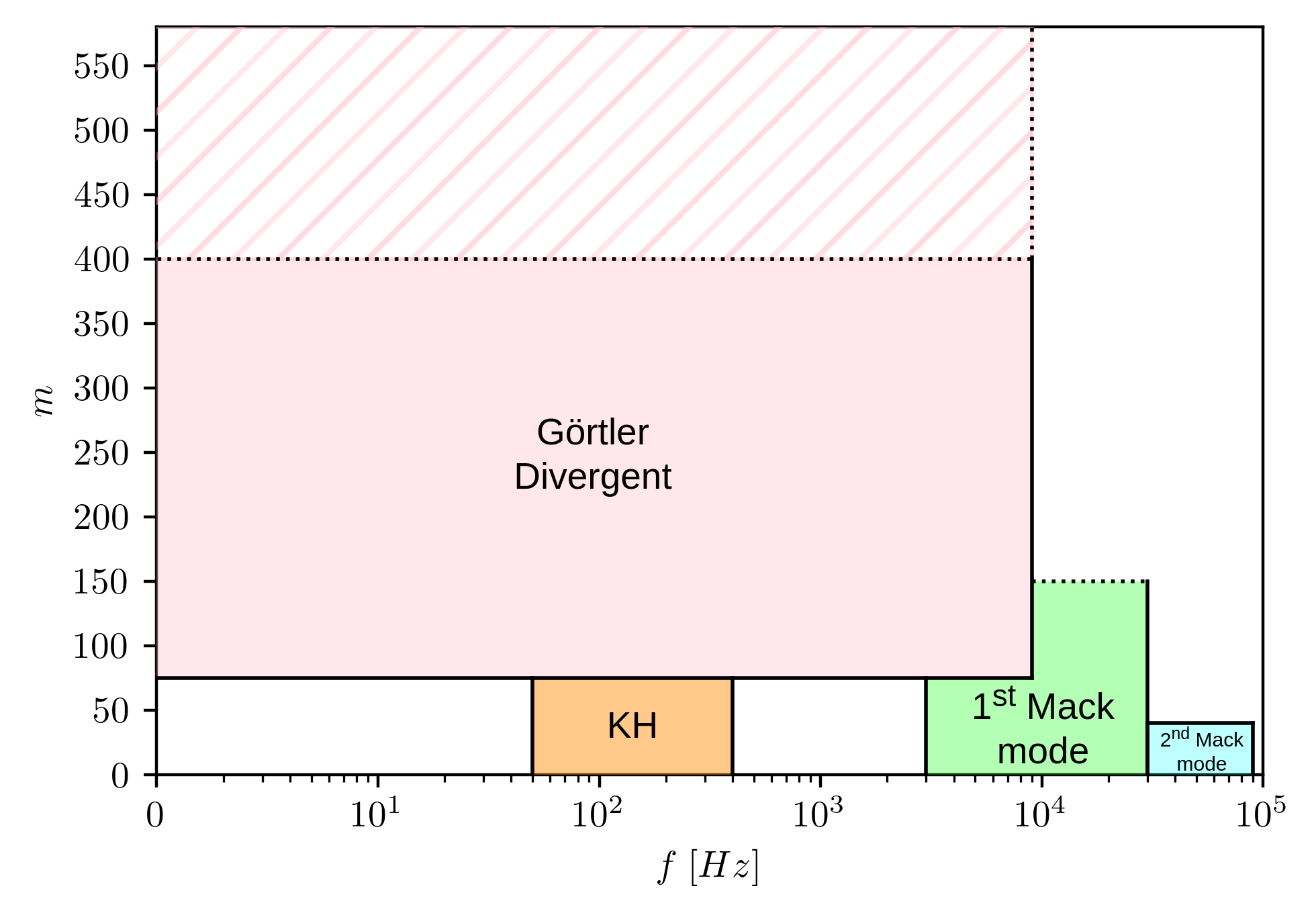}}
        \caption{Simplified overview of figures \ref{fig:OptimalGainMap} and \ref{fig:OptimalGainMap_restrict}, showing the optimal gain $\mu_0^2$ in frequency $f$ and azimuthal wavenumber $m$ space.}
        \label{fig:OptimalGainMap_resume}
    \end{figure}

    In this paper, all linear instabilities that may develop within the boundary layer of hypersonic wind tunnel nozzles (see figure \ref{fig:SchemaInstabNozzle}), including both global modes (recirculation bubble mode) and resolvent modes (Görtler, Kelvin-Helmholtz, first and second Mack modes) are characterised and identified using global linear stability analysis. A summary of all identified resolvent modes and their corresponding locations on the optimal gain map, for both restricted and unrestricted cases, is provided in figure \ref{fig:OptimalGainMap_resume}.

    Firstly, the eigenvalue analysis of the Jacobian operator has highlighted global modes that may develop within recirculation bubbles. These recirculation bubbles can form at the convergent inlet and on the inner or outer walls of the bleed slot upstream of the throat. Such separations may compromise the proper operation of the wind tunnel by introducing new instabilities that could lead to an early transition of the boundary layer. 
    
    Secondly, resolvent analysis was performed on the base-flow, enabling the identification and characterisation of resolvent modes. The Görtler instability was identified as the dominant instability. It can develop in two distinct zones suitable for the development of centrifugal instability, located where there is a concave wall i.e.\ in the first part of the convergent section and the second part of the divergent section. The peak of the Görtler instability is characterised by a streamwise mode at zero frequency (i.e.\ stationary) and a non-zero azimuthal number (i.e.\ three-dimensional). For this instability, it was observed that as the frequency increases, the optimal forcing shifts downstream in the nozzle until the instability only develops in the divergent section.
    On the other hand, as the azimuthal wavenumber increases, the instability scales to thinner boundary layers and can no longer develop in the thicker boundary layer at the end of the divergent section.
    The first and second Mack modes were also identified in the nozzle. Both of these instabilities develop in the divergent section, and their characteristics are consistent with previous state-of-the-art results. The last identified instability is the Kelvin-Helmholtz instability, which arises from the recirculation bubble. This instability can be controlled in such flows by ensuring the suppression of recirculation bubbles. 
    
    Finally, the restriction of the forcing field to the divergent section provided a simple model for the
    implementation of a bleed slot upstream of the nozzle throat. This device enables the suppression of the instability or the turbulent boundary layer that may develop in the settling chamber and/or convergent parts of the nozzle by creating a new boundary layer at the lip. It was found that the Görtler instability 
    grows only in the concave part of the divergent section and is strongly damped compared to the case without restriction, as the instability cannot develop in the convergent section of the nozzle. This study illustrates the impact of the convergent section on the development of resolvent modes. As a result, the optimal gain for the Görtler instability is significantly reduced, and the Kelvin-Helmholtz instability is completely eliminated. However, the first and second Mack modes remain unaffected by this device, as their energy is located almost exclusively in the divergent section of the nozzle. 
    
    The prospect of this work is to take advantage of
    the global linear stability framework
    to conduct optimisation studies for these nozzles. For example, discrete sensitivity analysis based on the Hessian of the Navier-Stokes equation around the base-flow provides insights into the regions of the flow where small modifications to the base-flow have the greatest impact on its stability \citep{metto2013, marquet2008sensitivity, poulain2024adjoint}. Using the Hessian operator, optimisation strategies such as volume control, wall temperature, wall blowing/suction, and so on could then be considered to optimally reduce instability growth. Additionally, sensitivity studies focused on nozzle geometry could also be performed using the formalism proposed by \citet{kitzinger2023receptivity} to find an optimal wall contour in order to delay the transition. 





\begin{Backmatter}

\paragraph{Acknowledgements}
The author thanks Clément Caillaud and Cédric Content for implementing the 3D contributions of the residual in cylindrical coordinates in BROADCAST (see Table \ref{tab:Resume_3D}). We are also grateful to Arthur Poulain for his development of the open-source CFD code BROADCAST. 

\paragraph{Funding Statement}
This study was supported by funding from ONERA and CEA-CESTA.

\paragraph{Competing Interests}
The authors declare no conflict of interest.

\paragraph{Data availability statement}
Data that support the findings of this study are available by the authors upon reasonable request.





\end{Backmatter}


\newpage

\appendix

    \section{Chu energy matrix definition}\label{appendix:ChuDefinition}

    For the definition of the Chu energy, we rely on the expression derived in the studies by \citet{chu1965energy, hanifi1996transient, bugeat20193d, poulain2024adjoint}. Chu's energy can be expressed using dimensionless variables of density $\check{\rho}$, velocity $\check{u} = (\check{u}_x,\check{u}_{r},\check{u}_{\theta})$, and temperature $\check{T}$ as follows:
    \begin{equation} 
        E_{Chu} = \check{q}^{*} Q_E \check{q} = \frac{1}{2} \int_{\Omega} \left( \overline{\rho} \lVert \check{u} \rVert^2 + \frac{\overline{T}}{\overline{\rho} \gamma M^2} \check{\rho}^2 + \frac{\overline{\rho}}{(\gamma-1)\gamma M^2 \overline{T}} \check{T}^2 \right) \mathrm{d}\Omega.
        \label{eq:Echu_definition} 
    \end{equation}
    The quantities denoted by $\overline{q}$ correspond to the base-flow quantities. Note that Chu's energy can also be expressed using the forcing terms $\check{f}$. The previous expression of the Chu energy (\ref{eq:Echu_definition}) is not yet suitable for our study as it is written with primitive variables, whereas the Navier-Stokes equations used are written with conservative variables (equation (\ref{eq:NSequation})). To switch from conservative variables to primitive variables, the following expressions are used: 
    \begin{equation} 
        \check{u}_i = \frac{1}{\overline{\rho}} \left( \check{(\rho u_i)} - \overline{u}_i \check{\rho} \right),
        \label{eq:primitive_to_conservative_u} 
    \end{equation}
    \begin{equation} 
        \check{T} = \frac{(\gamma-1)\gamma M^2}{\overline{\rho}} \left[ \left( \frac{1}{2} \lVert \overline{u} \rVert^2 - \overline{e} \right) \check{\rho} - \overline{u}_i \check{(\rho u_i)} + \check{(\rho E)} \right],
        \label{eq:primitive_to_conservative_T} 
    \end{equation}
    where $\overline{e}$ is the internal energy. For the matrix form of Chu's energy, we follow the matrix introduced by \citet{bugeat20193d}. Thus, the Chu energy matrix is written as follows:
    \begin{equation} 
        Q_{Chu} = \frac{1}{2} \mathrm{d}\Omega 
        \renewcommand{\arraystretch}{1.75}  
        \setlength{\arraycolsep}{12pt}     
        \begin{pmatrix}
        \frac{\lVert \overline{u} \rVert^2}{ \overline{\rho}} + \frac{\overline{T}}{\overline{\rho} \gamma M^2} + a_1 a_2^2 & -\frac{\overline{u} \left( 1 + a_1 a_2 \right)}{\overline{\rho}} & -\frac{\overline{v} \left( 1 + a_1 a_2 \right)}{\overline{\rho}} & 0 & \frac{a_1 a_2}{\overline{\rho}} \\
        -\frac{\overline{u} \left( 1 + a_1 a_2 \right)}{\overline{\rho}} & \frac{1}{\overline{\rho}} + \frac{\overline{u}^2 a_1}{\overline{\rho}^2} & \frac{\overline{u} \, \overline{v} a_1}{\overline{\rho}^2} & 0 & -\frac{\overline{u} a_1}{\overline{\rho}^2} \\
        -\frac{\overline{v} \left( 1 + a_1 a_2 \right)}{\overline{\rho}} & \frac{\overline{u} \, \overline{v} a_1}{\overline{\rho}^2} & \frac{1}{\overline{\rho}} + \frac{\overline{v}^2 a_1}{\overline{\rho}^2} & 0 & -\frac{\overline{v} a_1}{\overline{\rho}^2} \\
        0 & 0 & 0 & \frac{1}{\overline{\rho}} & 0 \\
        \frac{a_1 a_2}{\overline{\rho}} & -\frac{\overline{u} a_1}{\overline{\rho}^2} & -\frac{\overline{v} a_1}{\overline{\rho}^2} & 0 & \frac{a_1}{\overline{\rho}^2}
        \end{pmatrix},
        \label{eq:Qchu_matrix_definition} 
    \end{equation}
    where:
    \begin{equation} 
        a_1 = \frac{(\gamma-1) \gamma M^2 \overline{\rho}}{\overline{T}},
        \label{eq:a1_definition} 
    \end{equation}  
    \begin{equation} 
        a_2 = \frac{\left( \frac{1}{2} \lVert \overline{u} \rVert^2 - \overline{e} \right)}{\overline{\rho}}.
        \label{eq:a2_definition} 
    \end{equation}

    \section{Extension to 3D perturbations in axisymmetric case}\label{appendix:Extension_3D}

        \subsection{Navier-Stokes equations}\label{appendix:NavierStokesEquation}

        The governing equations are the compressible Navier-Stokes equations:
        \begin{equation} 
            \left\{\begin{aligned}
                & \frac{\partial \rho}{\partial t} + \nabla \cdot (\rho \mathbf{u}) = 0, \\
                & \frac{\partial (\rho \mathbf{u})}{\partial t} + \nabla \cdot \Bigr[ \rho \mathbf{u} \otimes \mathbf{u} + p \mathbf{I} - \bm{\tau} \Bigr] = 0, \\
                & \frac{\partial (\rho E)}{\partial t} + \nabla \cdot \Bigr[ (\rho E + p)\mathbf{u} - \bm{\tau} \cdot \mathbf{u} - \lambda \nabla T \Bigr] = 0,
    	\end{aligned}\right.            
            \label{eq:NSequation_detail} 
        \end{equation}
        with, $\mathbf{u}=(u_r, u_{\theta}, u_z = u_x)^{T}$ the velocity vector written in cylindrical coordinates (for an axisymmetric nozzle), $E = \frac{p}{\rho (\gamma-1)} + \frac{1}{2} \mathbf{u} \cdot \mathbf{u}$ the total energy, $\lambda = \frac{\mu c_p}{Pr}$ the isobaric heat capacity, $Pr = 0.72$ the Prandtl number, and $\mathbf{I}$ the identity matrix. Note that all variables used are non-dimensional. The viscous stress tensor of a Newtonian fluid is given by:
        \begin{equation} 
            \bm{\tau} = \mu \Bigr[ \nabla \otimes \mathbf{u} + (\nabla \otimes \mathbf{u})^T  \Bigr] - \frac{2}{3} \mu (\nabla \cdot \mathbf{u})\mathbf{I}.
            \label{eq:NSequation_StressTensor} 
        \end{equation}
        Note that all operators must be written in cylindrical coordinates as we are studying an axisymmetric nozzle. In order to close the system, two additional equations are required. First, for a thermally and calorically perfect gas, the non-dimensional pressure is written using the perfect gas law:
        
        \begin{equation} 
             P = \frac{1}{\gamma M^2} \rho T
            \label{eq:PerfectGasLaw} 
        \end{equation}
        For the second equation, Sutherland's law \citep{sutherland1893lii} is selected for the viscosity:
        \begin{equation} 
             \mu (T) = \mu_{ref} \left( \frac{T}{T_{ref}} \right)^{3/2} \frac{T_{ref}+S}{T+S},
            \label{eq:NSequation_Sutherland} 
        \end{equation}
        with, $S=110.4 \; K$ the Sutherland's temperature, $\mu_{ref}=1.716 \times 10^{-5} \; kg.m^{-1}.s^{-1}$ and $T_{ref}=273.15 \; K$.

        \subsection{Extension to 3D stability of the base-flow}\label{subsubsection:Extension3Dstability}
            
            In order to extend the global linear stability analysis of the 2D base-flow to 3D linear instabilities, \citet{poulain2023broadcast} used in BROADCAST the method proposed by \citet{metto2013, bugeat20193d}, except here linearisation is not performed using a finite difference method but using AD. A summary of this method for axisymmetric flow is given below. 

            Given the form of the perturbations introduced in equations (\ref{eq:disturbance_globalmode}) (global modes) and (\ref{eq:equation_disturbances}) (resolvent modes), the eigenmodes can be studied in the 2D base-flow, with the 3D dependence along the $\theta$-direction computed analytically in Fourier space. Thus, the discrete residual can be separated into two components: a strictly 2D part computed from the 2D base-flow and a 3D part accounting for the derivatives in the $\theta$-direction within the Navier-Stokes operator:
            \begin{equation}
                R_{3D} = R_{2D} + R_{\theta}
                \label{eq:Residual3D_twocomponents}.
            \end{equation}
            The 3D part of the residual $R_{\theta}$, accounting for the derivatives in $\theta$-direction, can be decomposed as a sum of three functions:
            \begin{equation}
                R_{\theta} (q) = \sum_{k} \alpha_k(q) \frac{\partial a_k (q)}{\partial \theta} + \sum_{l} \lambda_{l}(q) \frac{\partial^2 b_l (q)}{\partial \theta^2} + \sum_m \gamma_m (q) \frac{\partial c_m (q)}{\partial \theta} \odot \frac{\partial d_m (q)}{\partial \theta} 
                \label{eq:Residual_3fonctions},
            \end{equation}
            where the notation $\odot$ refers to the element-wise product of two vectors or matrices.
            The definition of the various coefficients ($\alpha_k$, $a_k$, $\lambda_{l}$, $b_l$, $\gamma_m$, $c_m$, $d_m$) in the above equation is proposed by \citet{poulain2023broadcast} for flow in Cartesian coordinates, these coefficients must be reformulated for flow in cylindrical coordinates. In appendix \ref{appendix:3DResidual}, the 3D residual and the definitions of the various coefficients in cylindrical coordinates (table \ref{tab:Resume_3D}) are provided. Note that the implementation in BROADCAST of the axisymmetric case was validated by \citet{caillaud2024separation} through a study of a Cone-Cylinder-Flare (CCF) geometry. 
            
            To achieve the 3D stability analysis, the 3D Jacobian must be derived from the previously defined 3D residual. \citet{poulain2023broadcast} propose to linearise equation (\ref{eq:Residual_3fonctions}) ($R_{\theta} (\overline{q} + q') = \bm{\mathcal{J}}_{\theta} q'$), taking into account that the base-flow is 2D (i.e. homogeneous in the $\theta$-direction) and keeping only the first-order terms:
            \begin{equation}
                \bm{\mathcal{J}}_{3D} q' = \left( \bm{\mathcal{J}}_{2D} + \bm{\mathcal{J}}_{\theta} \right) q' = \left( \bm{\mathcal{J}}_{2D} + i m D_{\theta} - m^2 D_{\theta \theta} \right) q'
                \label{eq:Jacobian_3D_cylindrical},
            \end{equation}
            with:
            \begin{equation} 
                D_{\theta} = \sum_{k} \alpha_k(\overline{q}) \frac{\partial a_k}{\partial q}\Bigg|_{\overline{q}} \; \text{ and } \;
                D_{\theta \theta} = \sum_{l} \lambda_l(\overline{q}) \frac{\partial b_l}{\partial q}\Bigg|_{\overline{q}}.
                \label{eq:Jacobian_3D_cylindrical_Dz_Dzz_component} 
            \end{equation}
        

            Thus, the 3D stability of the 2D base-flow can be analysed by replacing the Jacobian $\bm{\mathcal{J}}$ in eigenvalue problems (\ref{eq:globalstability}) and (\ref{eq:eigenvalueproblem_resolvent}) with the previously derived 3D Jacobian $\bm{\mathcal{J}}_{3D}$. Note that the $D_{\theta}$ and $D_{\theta \theta}$ matrices are computed using Algorithmic Differentiation.
            

        \subsection{3D contributions of the residual}\label{appendix:3DResidual}

        In the cylindrical coordinate system, the fluxes terms in equations (\ref{eq:NSequation}) and (\ref{eq:NSequation_detail}) can be separated into the sum of four functions $(\mathbf{G},\mathbf{H},\mathbf{F},\mathbf{K})$ as follows:

        \begin{equation}
            R(q) = - \nabla \cdot F(q) = \frac{1}{r}\frac{\partial (r \mathbf{G})}{\partial r} + \frac{1}{r}\frac{\partial \mathbf{H}}{\partial \theta} + \frac{\partial \mathbf{F}}{\partial z} + \frac{\mathbf{K}}{r}
            \label{eq:Flux_cylindrical} 
        \end{equation}

        To extend the analysis to 3D stability, we follow the method introduced by \citet{bugeat20193d} for studies in Cartesian coordinates. Similarly to the Cartesian case, the 3D component of the flux terms, containing the $\partial_{\theta}$ derivative, is separated from the 2D component (equation (\ref{eq:decomposition3D})). This 3D component is useful for deriving the 3D analytical residual by applying a Fourier transform in the $\theta$-direction.

        \begin{equation}
            F = F_{2D} + F_{\theta}; \;\;\;
            G = G_{2D} + G_{\theta}; \;\;\;
            H = H_{2D} + H_{\theta}; \;\;\;
            K = K_{2D} + K_{\theta}
            \label{eq:decomposition3D} 
        \end{equation}

        The objective is to express the various equations (continuity, momentum, energy) by separating the 3D components from the 2D ones, as outlined above. In other words, the aim is to reformulate all equations in the form presented in equation (\ref{eq:Flux_cylindrical}). This approach enables the derivation of the 3D analytical residual, which can be expressed as follows:
        \begin{equation}
            R_{3D} = R' + R_{\theta}
            \label{eq:Residual3D} 
        \end{equation}
        With (given that $\partial_{\theta} \overline{q} = 0$ and $\overline{u}_{\theta} = 0$ for the base-flow),
        \begin{equation}
            R_{\theta} = \frac{1}{r}\frac{\partial (r \mathbf{G_{\theta}})}{\partial r} + \frac{1}{r}\frac{\partial (\mathbf{H}' + \mathbf{H_{\theta}})}{\partial \theta} + \frac{\partial \mathbf{F'}}{\partial z} + \frac{\mathbf{K'}}{r} 
            \label{eq:ResidualTheta} 
        \end{equation}

            \subsubsection{3D components of the Navier-Stokes flux terms}\label{appendix:3DFluxComponents}

            The separation of the 3D components from the 2D components in cylindrical coordinates is achieved using the following expressions:

        \begin{equation}
            F' =
            \renewcommand{\arraystretch}{1.5}  
            \begin{pmatrix}
                \rho u_z \\
                \rho u_r u_z - \mu \left( \frac{\partial u_r}{\partial z} + \frac{\partial u_z}{\partial z} \right) \\
                \rho u_{\theta} u_z - \mu \left( \frac{\partial u_{\theta}}{\partial z} \right) \\
                \rho u_z^2 + p - \mu \left[ 2 \frac{\partial u_z}{\partial z} - \frac{2}{3} \left( \frac{1}{r}\frac{\partial (r u_r)}{\partial r} + \frac{\partial u_z}{\partial z} \right) \right] \\
                \left( \rho E + p \right) u_z - \mu \left \{ u_r \left[ \frac{\partial u_z}{\partial r} + \frac{\partial u_r}{\partial z} \right] + u_{\theta} \left[ \frac{\partial u_{\theta}}{\partial z} \right] + u_z \left[ 2 \frac{\partial u_z}{\partial z} - \frac{2}{3} \left( \frac{1}{r}\frac{\partial (r u_r)}{\partial r} + \frac{\partial u_z}{\partial z} \right) \right] \right \} - \lambda \frac{\partial T}{\partial z}
            \end{pmatrix}; \;\;\;
            \label{eq:Fcomponent2D} 
        \end{equation}
        \begin{equation}
            F_{\theta} = 
            \renewcommand{\arraystretch}{1.5}  
            \begin{pmatrix}
                0 \\
                0 \\
                -\frac{1}{r} \mu \frac{\partial u_z}{\partial \theta} \\
                \frac{2}{3}\frac{1}{r} \mu \frac{\partial u_{\theta}}{\partial \theta} \\
                - \mu \left[ \frac{1}{r} u_{\theta} \frac{\partial u_z}{\partial \theta} - \frac{2}{3}\frac{1}{r} u_z \frac{\partial u_{\theta}}{\partial \theta} \right]
            \end{pmatrix}.
            \label{eq:Fcomponent3D} 
        \end{equation}
    
        \begin{equation}
            G' =
            \renewcommand{\arraystretch}{1.5}  
            \begin{pmatrix}
                \rho u_r \\
                \rho u_r^2 + p - \mu \left[ 2 \frac{\partial u_r}{\partial r} - \frac{2}{3} \left( \frac{1}{r}\frac{\partial (r u_r)}{\partial r} + \frac{\partial u_z}{\partial z} \right) \right] \\
                \rho u_{\theta} u_r - \mu \left( \frac{\partial u_{\theta}}{\partial r} - \frac{ u_{\theta}}{r} \right) \\
                \rho u_z u_r - \mu \left( \frac{\partial u_z}{\partial r} + \frac{\partial u_r}{\partial z} \right) \\
                \left( \rho E + p \right) u_r - \mu \left \{ u_r \left[ 2\frac{\partial u_r}{\partial r} - \frac{2}{3} \left( \frac{1}{r}\frac{\partial (r u_r)}{\partial r} + \frac{\partial u_z}{z} \right) \right] + u_{\theta} \left[ -\frac{u_{\theta}}{r} + \frac{\partial u_{\theta}}{\partial r} \right] + u_z \left[ \frac{\partial u_r}{\partial z} + \frac{\partial u_z}{\partial r} \right] \right \} - \lambda \frac{\partial T}{\partial r}
            \end{pmatrix}; \;\;\;
            \label{eq:Gcomponent2D} 
        \end{equation}
        \begin{equation}
            G_{\theta} = 
            \renewcommand{\arraystretch}{1.5}  
            \begin{pmatrix}
                0 \\
                \frac{2}{3} \frac{1}{r} \mu \frac{\partial u_{\theta}}{\partial \theta} \\
                -\frac{1}{r} \mu \frac{\partial u_r}{\partial \theta} \\
                0 \\
                - \mu \left[ -\frac{2}{3} \frac{1}{r} u_r \frac{\partial u_{\theta}}{\partial \theta} + \frac{1}{r} u_{\theta} \frac{\partial u_r}{\partial \theta} \right]
            \end{pmatrix}.
            \label{eq:Gcomponent3D} 
        \end{equation}
    
        \begin{equation}
            H' =
            \renewcommand{\arraystretch}{1.5}  
            \begin{pmatrix}
                \rho u_{\theta} \\
                \rho u_r u_{\theta} - \mu \left( -\frac{u_{\theta}}{r} + \frac{\partial u _{\theta}}{\partial r} \right) \\
                \rho u_{\theta}^2 + p - \mu \left[ 2 \frac{u_r}{r} - \frac{2}{3} \left( \frac{1}{r}\frac{\partial (r u_r)}{\partial r} + \frac{\partial u_z}{\partial z} \right) \right] \\
                \rho u_z u_{\theta} - \mu \left( \frac{\partial u_{\theta}}{\partial z} \right) \\
                \left( \rho E + p \right) u_{\theta} - \mu \left \{ u_r \left[ \frac{\partial u_{\theta}}{\partial r} - \frac{u_{\theta}}{r} \right] + u_{\theta} \left[ 2 \frac{u_r}{r} - \frac{2}{3} \left( \frac{1}{r}\frac{\partial (r u_r)}{\partial r} + \frac{\partial u_z}{\partial z} \right) \right] + u_z \left[ \frac{\partial u_{\theta}}{\partial z} \right] \right \}
            \end{pmatrix}; \;\;\;
            \label{eq:Hcomponent2D} 
        \end{equation}
        \begin{equation}
            H_{\theta} = 
            \renewcommand{\arraystretch}{1.5}  
            \begin{pmatrix}
                0 \\
                -\frac{1}{r} \mu \frac{\partial u_{r}}{\partial \theta} \\
                -\frac{4}{3}\frac{1}{r} \mu \frac{\partial u_{\theta}}{\partial \theta}\\
                -\frac{1}{r} \mu \frac{\partial u_z}{\partial \theta} \\
                - \mu \left[ \frac{1}{r} u_r \frac{\partial u_r}{\partial \theta} + \frac{4}{3}\frac{1}{r} u_{\theta} \frac{\partial u_{\theta}}{\partial \theta} + \frac{1}{r} u_z \frac{\partial u_z}{\partial \theta} \right] - \lambda \frac{1}{r}\frac{\partial T}{\partial \theta}
            \end{pmatrix}.
            \label{eq:Hcomponent3D} 
        \end{equation}
    
        \begin{equation}
            K' =
            \renewcommand{\arraystretch}{1.5}  
            \begin{pmatrix}
                0 \\
                -\left( \rho u_{\theta}^2 + p - \mu \left[ 2 \frac{u_r}{r} - \frac{2}{3} \left( \frac{1}{r}\frac{\partial (r u_r)}{\partial r} + \frac{\partial u_z}{\partial z} \right) \right] \right) \\
                 \rho u_r u_{\theta} - \mu \left( -\frac{u_{\theta}}{r} + \frac{\partial u _{\theta}}{\partial r} \right) \\
                 0 \\
                 0
            \end{pmatrix}; \;\;\;
            \label{eq:Kcomponent2D} 
        \end{equation}
        \begin{equation}
            K_{\theta} = 
            \renewcommand{\arraystretch}{1.5}  
            \begin{pmatrix}
                0 \\
                \frac{4}{3}\frac{1}{r} \mu \frac{\partial u_{\theta}}{\partial \theta} \\
                -\frac{1}{r} \mu \frac{\partial u_{r}}{\partial \theta} \\
                0 \\
                0
            \end{pmatrix}.
            \label{eq:Kcomponent3D} 
        \end{equation}
        
        NB : $\left[ 2\frac{u_r}{r} - \frac{2}{3} \left( \frac{1}{r}\frac{\partial (r u_r)}{\partial r} + \frac{\partial u_z}{\partial z} \right) \right] = - \frac{2}{3}\left[ \frac{\partial u_r}{\partial r} + \frac{\partial u_z}{\partial z} - 2\frac{u_r}{r} \right]$.

            \subsubsection{Summary of the component}\label{appendix:3DFluxComponents_summary}

            Using the formulation provided by \citet{poulain2023broadcast} for Cartesian coordinates, the four functions ($\mathbf{G}$, $\mathbf{H}$, $\mathbf{F}$, $\mathbf{K}$) describing the fluxes can be expressed as the sum of three functions (see equation (\ref{eq:Residual_3fonctions})), which are computed in the BROADCAST solver), summarised in the table below:

{\renewcommand{\arraystretch}{1.35} 
{\setlength{\tabcolsep}{0.07cm} 
    \begin{table}[H]
        \caption{\label{tab:Resume_3D} 3D contributions to the Navier-Stokes equations in cylindrical coordinates, following the formulation by \citet{poulain2023broadcast}.}
        \centering
        \begin{tabular}{lll|lll|llll}
            \hline
            \multicolumn{10}{c}{Continuity equation} \\
            \hline
            $k$ & $\alpha_k$ & $a_k$ & $l$ & $\lambda_l$ & $b_l$ & $m$ & $\gamma_m$ & $c_m$ & $d_m$ \\ 
            1 & $\frac{1}{r}$ & $\rho u_{\theta}$ &  &  &  &  &  &  &  \\ 
            \hline
            \multicolumn{10}{c}{Momentum equation in $z$-direction ($=x$-direction)} \\
            \hline
            $k$ & $\alpha_k$ & $a_k$ & $l$ & $\lambda_l$ & $b_l$ & $m$ & $\gamma_m$ & $c_m$ & $d_m$ \\ 
            1 & $\frac{1}{r}$ & $\rho u_z u_{\theta} - \mu \frac{\partial u_{\theta}}{\partial z}$ & 1 & $-\frac{1}{r^2}\mu$ & $u_z$ & 1 & $-\frac{1}{r^2}$ & $\mu$ & $u_z$ \\
            2 & $\frac{2}{3} \frac{1}{r} \frac{\partial \mu}{\partial z}$ & $u_{\theta}$ &  &  &  &  &  &  &  \\
            3 & $\frac{2}{3} \frac{1}{r} \mu$ & $\frac{\partial u_{\theta}}{\partial z}$ &  &  &  &  &  &  &  \\
            \hline
            \multicolumn{10}{c}{Momentum equation in $r$-direction} \\
            \hline
            $k$ & $\alpha_k$ & $a_k$ & $l$ & $\lambda_l$ & $b_l$ & $m$ & $\gamma_m$ & $c_m$ & $d_m$ \\
            1 & $\frac{1}{r}$ & $\rho u_r u_{\theta} - \mu \left( -\frac{u_{\theta}}{r} + \frac{\partial u_{\theta}}{\partial r} \right)$ & 1 & $-\frac{1}{r^2} \mu$  & $u_r$ & 1 & $-\frac{1}{r^2}$ & $\mu$ & $u_r$ \\
            2 & $\frac{2}{3} \frac{1}{r} \frac{\partial \mu}{\partial r} + \frac{4}{3} \frac{1}{r^2} \mu$ & $u_{\theta}$ &  &  &  &  &  &  &  \\
            3 & $\frac{2}{3} \frac{1}{r} \mu$ & $\frac{\partial u_{\theta}}{\partial r}$ &  &  &  &  &  &  &  \\
            \hline
            \multicolumn{10}{c}{Momentum equation in $\theta$-direction} \\
            \hline
            $k$ & $\alpha_k$ & $a_k$ & $l$ & $\lambda_l$ & $b_l$ & $m$ & $\gamma_m$ & $c_m$ & $d_m$ \\ 
            1 & $\frac{1}{r}$ & $\rho u_{\theta}^2 + p + \frac{2}{3} \mu \left[ \frac{\partial u_r}{\partial r} + \frac{\partial u_z}{\partial z} - 2\frac{u_r}{r} \right]$ & 1 & $-\frac{4}{3}\frac{1}{r^2}\mu$ & $u_{\theta}$ & 1 & $-\frac{4}{3}\frac{1}{r^2}$ & $\mu$ & $u_{\theta}$ \\
            2 & $-\frac{1}{r} \frac{\partial \mu}{\partial z}$ & $u_z$ &  &  &  &  &  &  &  \\
            3 & $-\frac{1}{r} \frac{\partial \mu}{\partial r} -\frac{1}{r^2}\mu$ & $u_r$ &  &  &  &  &  &  &  \\
            4 & $-\frac{1}{r} \mu$ & $\frac{\partial u_z}{\partial z} + \frac{\partial u_r}{\partial r}$ &  &  &  &  &  &  &  \\
            \hline
            \multicolumn{10}{c}{Energy equation} \\
            \hline
            $k$ & $\alpha_k$ & $a_k$ & $l$ & $\lambda_l$ & $b_l$ & $m$ & $\gamma_m$ & $c_m$ & $d_m$ \\ 
            1 & $\frac{1}{r}$ & $(\rho E + p) u_{\theta}$ & 1 & $-\frac{1}{r^2}\lambda$ & $T$ & 1 & $-\frac{1}{r^2}$ & $\lambda$ & $T$ \\
            2 & $\frac{2}{3}\frac{1}{r}\frac{\partial (\mu u_z)}{\partial z}$ & $u_{\theta}$ & 2 & $-\frac{1}{r^2}\mu u_z$ & $u_z$ & 2 & $-\frac{1}{r^2}\mu$ & $u_z$ & $u_z$ \\
            3 & $\frac{2}{3}\frac{1}{r}\mu u_z - \frac{1}{r}\mu u_z$ & $\frac{\partial u_{\theta}}{\partial z}$ & 3 & $-\frac{1}{r^2}\mu u_r$ & $u_r$ & 3 & $-\frac{1}{r^2}u_z$ & $\mu$ & $u_z$ \\
            4 & $-\frac{1}{r}\frac{\partial (\mu u_{\theta})}{\partial z}$ & $u_z$ & 4 & $-\frac{4}{3}\frac{1}{r^2}\mu u_{\theta}$ & $u_{\theta}$ & 4 & $-\frac{1}{r^2}\mu$ & $u_r$ & $u_r$ \\
            5 & $-\frac{1}{r}\mu u_{\theta}$ & $\frac{\partial u_z}{\partial z}$ &  &  &  & 5 & $-\frac{1}{r^2}u_r$ & $\mu$ & $u_r$ \\
            6 & $\frac{2}{3}\frac{1}{r}\frac{\partial (\mu u_r)}{\partial r}$ & $u_{\theta}$ &  &  &  & 6 & $-\frac{4}{3}\frac{1}{r^2}\mu$ & $u_{\theta}$ & $u_{\theta}$ \\
            7 & $\frac{2}{3}\frac{1}{r}\mu u_r$ & $\frac{\partial u_{\theta}}{\partial r}$ &  &  &  & 7 & $-\frac{4}{3}\frac{1}{r^2}u_{\theta}$ & $\mu$ & $u_{\theta}$ \\
            8 & $-\frac{1}{r}\frac{\partial (\mu u_{\theta})}{\partial r}$ & $u_r$ &  &  &  &  &  &  &  \\
            9 & $-\frac{1}{r}\mu u_{\theta}$ & $\frac{\partial u_r}{\partial r}$ &  &  &  &  &  &  &  \\
            10 & $-\frac{1}{r}\frac{\partial u_{\theta}}{\partial z}$ & $\mu u_z$ &  &  &  &  &  &  &  \\
            11 & $-\frac{1}{r} \left( -\frac{u_{\theta}}{r} + \frac{\partial u_{\theta}}{\partial r} \right) $ & $\mu u_r$ &  &  &  &  &  &  &  \\
            12 & $-\frac{1}{r} \mu u_r$ & $\left( -\frac{u_{\theta}}{r} + \frac{\partial u_{\theta}}{\partial r} \right)$ &  &  &  &  &  &  &  \\
            13 & $\frac{2}{3}\frac{1}{r}\left[ \frac{\partial u_r}{\partial r} + \frac{\partial u_z}{\partial z} - 2\frac{u_r}{r} \right]$ & $\mu u_{\theta}$ &  &  &  &  &  &  &  \\
            14 & $\frac{2}{3}\frac{1}{r} \mu u_{\theta}$ & $\left[ \frac{\partial u_r}{\partial r} + \frac{\partial u_z}{\partial z} - 2\frac{u_r}{r} \right]$ &  &  &  &  &  &  &  \\
            \hline
        \end{tabular}
\end{table}}}

\bibliographystyle{apalike}
\bibliography{biblio}


\end{document}